\documentclass[twocolumn,traditabstract,longauth]{aa}

\usepackage{amsmath,amsfonts,amssymb}
\usepackage{txfonts}
\usepackage{graphicx}
\usepackage{fixltx2e}
\usepackage{natbib}
\usepackage{epstopdf}
\usepackage{epsf,color}
\usepackage{ifthen}

\usepackage[breaklinks, colorlinks, citecolor=blue]{hyperref}

\usepackage{mathrsfs}

\bibpunct{(}{)}{;}{a}{}{,} 

\def\setsymbol#1#2{\expandafter\def\csname #1\endcsname{#2}}
\def\getsymbol#1{\csname #1\endcsname}

\def\Planck{\textit{Planck}}

\newbox\tablebox    \newdimen\tablewidth
\def\leaderfil{\leaders\hbox to 5pt{\hss.\hss}\hfil}
\def\endPlancktable{\tablewidth=\columnwidth 
    $$\hss\copy\tablebox\hss$$
    \vskip-\lastskip\vskip -2pt}

\def\tablenote#1 #2\par{\begingroup \parindent=0.8em
    \abovedisplayshortskip=0pt\belowdisplayshortskip=0pt
    \noindent
    $$\hss\vbox{\hsize\tablewidth \hangindent=\parindent \hangafter=1 \noindent
    \hbox to \parindent{$^#1$\hss}\strut#2\strut\par}\hss$$
    \endgroup}
\def\doubleline{\vskip 3pt\hrule \vskip 1.5pt \hrule \vskip 5pt}

\def\L2{\ifmmode L_2\else $L_2$\fi}

\def\DeltaT{\ifmmode \Delta T\else $\Delta T$\fi}
\def\deltat{\ifmmode \Delta t\else $\Delta t$\fi}
\def\fknee{\ifmmode f_{\rm knee}\else $f_{\rm knee}$\fi}
\def\Fmax{\ifmmode F_{\rm max}\else $F_{\rm max}$\fi}
\def\solar{\ifmmode{\rm M}_{\mathord\odot}\else${\rm M}_{\mathord\odot}$\fi}
\def\Msolar{\ifmmode{\rm M}_{\mathord\odot}\else${\rm M}_{\mathord\odot}$\fi}
\def\Lsolar{\ifmmode{\rm L}_{\mathord\odot}\else${\rm L}_{\mathord\odot}$\fi}
\def\inv{\ifmmode^{-1}\else$^{-1}$\fi}
\def\mo{\ifmmode^{-1}\else$^{-1}$\fi}
\def\sup#1{\ifmmode ^{\rm #1}\else $^{\rm #1}$\fi}
\def\expo#1{\ifmmode \times 10^{#1}\else $\times 10^{#1}$\fi}
\def\,{\thinspace}
\def\lsim{\mathrel{\raise .4ex\hbox{\rlap{$<$}\lower 1.2ex\hbox{$\sim$}}}}
\def\gsim{\mathrel{\raise .4ex\hbox{\rlap{$>$}\lower 1.2ex\hbox{$\sim$}}}}

\def\simprop{\mathrel{\raise .4ex\hbox{\rlap{$\propto$}\lower 1.2ex\hbox{$\sim$}}}}
\def\deg{\ifmmode^\circ\else$^\circ$\fi}
\def\pdeg{\ifmmode $\setbox0=\hbox{$^{\circ}$}\rlap{\hskip.11\wd0 .}$^{\circ}
          \else \setbox0=\hbox{$^{\circ}$}\rlap{\hskip.11\wd0 .}$^{\circ}$\fi}
\def\arcs{\ifmmode {^{\scriptstyle\prime\prime}}
          \else $^{\scriptstyle\prime\prime}$\fi}
\def\arcm{\ifmmode {^{\scriptstyle\prime}}
          \else $^{\scriptstyle\prime}$\fi}
\newdimen\sa  \newdimen\sb
\def\parcs{\sa=.07em \sb=.03em
     \ifmmode \hbox{\rlap{.}}^{\scriptstyle\prime\kern -\sb\prime}\hbox{\kern -\sa}
     \else \rlap{.}$^{\scriptstyle\prime\kern -\sb\prime}$\kern -\sa\fi}
\def\parcm{\sa=.08em \sb=.03em
     \ifmmode \hbox{\rlap{.}\kern\sa}^{\scriptstyle\prime}\hbox{\kern-\sb}
     \else \rlap{.}\kern\sa$^{\scriptstyle\prime}$\kern-\sb\fi}
\def\ra[#1 #2 #3.#4]{#1\sup{h}#2\sup{m}#3\sup{s}\llap.#4}
\def\dec[#1 #2 #3.#4]{#1\deg#2\arcm#3\arcs\llap.#4}
\def\deco[#1 #2 #3]{#1\deg#2\arcm#3\arcs}
\def\rra[#1 #2]{#1\sup{h}#2\sup{m}}

\def\dots{\relax\ifmmode \ldots\else $\ldots$\fi}
\def\WHzsr{\ifmmode $W\,Hz\mo\,sr\mo$\else W\,Hz\mo\,sr\mo\fi}
\def\mHz{\ifmmode $\,mHz$\else \,mHz\fi}
\def\GHz{\ifmmode $\,GHz$\else \,GHz\fi}
\def\mKs{\ifmmode $\,mK\,s$^{1/2}\else \,mK\,s$^{1/2}$\fi}
\def\muKs{\ifmmode \,\mu$K\,s$^{1/2}\else \,$\mu$K\,s$^{1/2}$\fi}
\def\muKRJs{\ifmmode \,\mu$K$_{\rm RJ}$\,s$^{1/2}\else \,$\mu$K$_{\rm RJ}$\,s$^{1/2}$\fi}
\def\muKHz{\ifmmode \,\mu$K\,Hz$^{-1/2}\else \,$\mu$K\,Hz$^{-1/2}$\fi}
\def\MJysr{\ifmmode \,$MJy\,sr\mo$\else \,MJy\,sr\mo\fi}
\def\MJysrmK{\ifmmode \,$MJy\,sr\mo$\,mK$_{\rm CMB}\mo\else \,MJy\,sr\mo\,mK$_{\rm CMB}\mo$\fi}
\def\microns{\ifmmode \,\mu$m$\else \,$\mu$m\fi}
\def\micron{\microns}
\def\muK{\ifmmode \,\mu$K$\else \,$\mu$\hbox{K}\fi}
\def\microK{\ifmmode \,\mu$K$\else \,$\mu$\hbox{K}\fi}
\def\muW{\ifmmode \,\mu$W$\else \,$\mu$\hbox{W}\fi}
\def\kms{\ifmmode $\,km\,s$^{-1}\else \,km\,s$^{-1}$\fi}
\def\kmsMpc{\ifmmode $\,\kms\,Mpc\mo$\else \,\kms\,Mpc\mo\fi}
\providecommand{\sorthelp}[1]{}

\def\ben{\begin{enumerate}}
\def\een{\end{enumerate}}
\def\bi{\begin{itemize}}
\def\ei{\end{itemize}}
\def\be{\begin{equation}}
\def\ee{\end{equation}}
\def\bea{\begin{eqnarray}}
\def\eea{\end{eqnarray}}
\def\ba{\begin{align}}
\def\ea{\end{align}}

\def\bdx{\boldsymbol{x }}

\def\bda{\boldsymbol{a }}

\def\bdn{\boldsymbol{n }}
\def\bds{\boldsymbol{s }}
\def\bdf{\boldsymbol{f }}

\def\bdt{\boldsymbol{t }}

\newcommand{\hi}{\ion{H}{i}}

\newcommand{\healpix}{{\tt HEALPix}}

\def\setsymbol#1#2{\expandafter\def\csname #1\endcsname{#2}}
\def\getsymbol#1{\csname #1\endcsname}

\begin{document}


\title{\Planck\ intermediate results. XLVIII.\\ Disentangling Galactic dust emission and cosmic infrared background anisotropies}

\date{Received \today \ / Accepted --}

\authorrunning{Planck Collaboration}
\titlerunning{Disentangling dust and CIB in \Planck\ observations}
\author{\small
Planck Collaboration: N.~Aghanim\inst{47}
\and
M.~Ashdown\inst{57, 4}
\and
J.~Aumont\inst{47}
\and
C.~Baccigalupi\inst{68}
\and
M.~Ballardini\inst{24, 39, 42}
\and
A.~J.~Banday\inst{78, 7}
\and
R.~B.~Barreiro\inst{52}
\and
N.~Bartolo\inst{23, 53}
\and
S.~Basak\inst{68}
\and
K.~Benabed\inst{48, 77}
% ...
}
\institute{\small
APC, AstroParticule et Cosmologie, Universit\'{e} Paris Diderot, CNRS/IN2P3, CEA/lrfu, Observatoire de Paris, Sorbonne Paris Cit\'{e}, 10, rue Alice Domon et L\'{e}onie Duquet, 75205 Paris Cedex 13, France\goodbreak
\and
African Institute for Mathematical Sciences, 6-8 Melrose Road, Muizenberg, Cape Town, South Africa\goodbreak
\and
Agenzia Spaziale Italiana Science Data Center, Via del Politecnico snc, 00133, Roma, Italy\goodbreak
\and
Astrophysics Group, Cavendish Laboratory, University of Cambridge, J J Thomson Avenue, Cambridge CB3 0HE, U.K.\goodbreak
\and
Astrophysics \& Cosmology Research Unit, School of Mathematics, Statistics \& Computer Science, University of KwaZulu-Natal, Westville Campus, Private Bag X54001, Durban 4000, South Africa\goodbreak
\and
CITA, University of Toronto, 60 St. George St., Toronto, ON M5S 3H8, Canada\goodbreak
\and
CNRS, IRAP, 9 Av. colonel Roche, BP 44346, F-31028 Toulouse cedex 4, France\goodbreak
\and
California Institute of Technology, Pasadena, California, U.S.A.\goodbreak
\and
Computational Cosmology Center, Lawrence Berkeley National Laboratory, Berkeley, California, U.S.A.\goodbreak
\and
DTU Space, National Space Institute, Technical University of Denmark, Elektrovej 327, DK-2800 Kgs. Lyngby, Denmark\goodbreak
\and
D\'{e}partement de Physique Th\'{e}orique, Universit\'{e} de Gen\`{e}ve, 24, Quai E. Ansermet,1211 Gen\`{e}ve 4, Switzerland\goodbreak
\and
Departamento de Astrof\'{i}sica, Universidad de La Laguna (ULL), E-38206 La Laguna, Tenerife, Spain\goodbreak
\and
Departamento de F\'{\i}sica, Universidad de Oviedo, Avda. Calvo Sotelo s/n, Oviedo, Spain\goodbreak
\and
Department of Astrophysics/IMAPP, Radboud University Nijmegen, P.O. Box 9010, 6500 GL Nijmegen, The Netherlands\goodbreak
\and
Department of Physics \& Astronomy, University of British Columbia, 6224 Agricultural Road, Vancouver, British Columbia, Canada\goodbreak
\and
Department of Physics and Astronomy, Dana and David Dornsife College of Letter, Arts and Sciences, University of Southern California, Los Angeles, CA 90089, U.S.A.\goodbreak
\and
Department of Physics and Astronomy, University College London, London WC1E 6BT, U.K.\goodbreak
\and
Department of Physics and Astronomy, University of Sussex, Brighton BN1 9QH, U.K.\goodbreak
\and
Department of Physics, Gustaf H\"{a}llstr\"{o}min katu 2a, University of Helsinki, Helsinki, Finland\goodbreak
\and
Department of Physics, Princeton University, Princeton, New Jersey, U.S.A.\goodbreak
\and
Department of Physics, University of California, Santa Barbara, California, U.S.A.\goodbreak
\and
Department of Physics, University of Illinois at Urbana-Champaign, 1110 West Green Street, Urbana, Illinois, U.S.A.\goodbreak
\and
Dipartimento di Fisica e Astronomia G. Galilei, Universit\`{a} degli Studi di Padova, via Marzolo 8, 35131 Padova, Italy\goodbreak
\and
Dipartimento di Fisica e Astronomia, Alma Mater Studiorum, Universit\`{a} degli Studi di Bologna, Viale Berti Pichat 6/2, I-40127, Bologna, Italy\goodbreak
\and
Dipartimento di Fisica e Scienze della Terra, Universit\`{a} di Ferrara, Via Saragat 1, 44122 Ferrara, Italy\goodbreak
\and
Dipartimento di Fisica, Universit\`{a} La Sapienza, P. le A. Moro 2, Roma, Italy\goodbreak
\and
Dipartimento di Fisica, Universit\`{a} degli Studi di Milano, Via Celoria, 16, Milano, Italy\goodbreak
\and
Dipartimento di Fisica, Universit\`{a} degli Studi di Trieste, via A. Valerio 2, Trieste, Italy\goodbreak
\and
Dipartimento di Fisica, Universit\`{a} di Roma Tor Vergata, Via della Ricerca Scientifica, 1, Roma, Italy\goodbreak
\and
Dipartimento di Matematica, Universit\`{a} di Roma Tor Vergata, Via della Ricerca Scientifica, 1, Roma, Italy\goodbreak
\and
European Space Agency, ESAC, Planck Science Office, Camino bajo del Castillo, s/n, Urbanizaci\'{o}n Villafranca del Castillo, Villanueva de la Ca\~{n}ada, Madrid, Spain\goodbreak
\and
European Space Agency, ESTEC, Keplerlaan 1, 2201 AZ Noordwijk, The Netherlands\goodbreak
\and
Gran Sasso Science Institute, INFN, viale F. Crispi 7, 67100 L'Aquila, Italy\goodbreak
\and
HGSFP and University of Heidelberg, Theoretical Physics Department, Philosophenweg 16, 69120, Heidelberg, Germany\goodbreak
\and
Helsinki Institute of Physics, Gustaf H\"{a}llstr\"{o}min katu 2, University of Helsinki, Helsinki, Finland\goodbreak
\and
INAF - Osservatorio Astronomico di Padova, Vicolo dell'Osservatorio 5, Padova, Italy\goodbreak
\and
INAF - Osservatorio Astronomico di Roma, via di Frascati 33, Monte Porzio Catone, Italy\goodbreak
\and
INAF - Osservatorio Astronomico di Trieste, Via G.B. Tiepolo 11, Trieste, Italy\goodbreak
\and
INAF/IASF Bologna, Via Gobetti 101, Bologna, Italy\goodbreak
\and
INAF/IASF Milano, Via E. Bassini 15, Milano, Italy\goodbreak
\and
INFN - CNAF, viale Berti Pichat 6/2, 40127 Bologna, Italy\goodbreak
\and
INFN, Sezione di Bologna, viale Berti Pichat 6/2, 40127 Bologna, Italy\goodbreak
\and
INFN, Sezione di Ferrara, Via Saragat 1, 44122 Ferrara, Italy\goodbreak
\and
INFN, Sezione di Roma 1, Universit\`{a} di Roma Sapienza, Piazzale Aldo Moro 2, 00185, Roma, Italy\goodbreak
\and
INFN, Sezione di Roma 2, Universit\`{a} di Roma Tor Vergata, Via della Ricerca Scientifica, 1, Roma, Italy\goodbreak
\and
Imperial College London, Astrophysics group, Blackett Laboratory, Prince Consort Road, London, SW7 2AZ, U.K.\goodbreak
\and
Institut d'Astrophysique Spatiale, CNRS, Univ. Paris-Sud, Universit\'{e} Paris-Saclay, B\^{a}t. 121, 91405 Orsay cedex, France\goodbreak
\and
Institut d'Astrophysique de Paris, CNRS (UMR7095), 98 bis Boulevard Arago, F-75014, Paris, France\goodbreak
\and
Institute of Astronomy, University of Cambridge, Madingley Road, Cambridge CB3 0HA, U.K.\goodbreak
\and
Institute of Theoretical Astrophysics, University of Oslo, Blindern, Oslo, Norway\goodbreak
\and
Instituto de Astrof\'{\i}sica de Canarias, C/V\'{\i}a L\'{a}ctea s/n, La Laguna, Tenerife, Spain\goodbreak
\and
Instituto de F\'{\i}sica de Cantabria (CSIC-Universidad de Cantabria), Avda. de los Castros s/n, Santander, Spain\goodbreak
\and
Istituto Nazionale di Fisica Nucleare, Sezione di Padova, via Marzolo 8, I-35131 Padova, Italy\goodbreak
\and
Jet Propulsion Laboratory, California Institute of Technology, 4800 Oak Grove Drive, Pasadena, California, U.S.A.\goodbreak
\and
Jodrell Bank Centre for Astrophysics, Alan Turing Building, School of Physics and Astronomy, The University of Manchester, Oxford Road, Manchester, M13 9PL, U.K.\goodbreak
\and
Kavli Institute for Cosmological Physics, University of Chicago, Chicago, IL 60637, USA\goodbreak
\and
Kavli Institute for Cosmology Cambridge, Madingley Road, Cambridge, CB3 0HA, U.K.\goodbreak
\and
LAL, Universit\'{e} Paris-Sud, CNRS/IN2P3, Orsay, France\goodbreak
\and
LERMA, CNRS, Observatoire de Paris, 61 Avenue de l'Observatoire, Paris, France\goodbreak
\and
Laboratoire Traitement et Communication de l'Information, CNRS (UMR 5141) and T\'{e}l\'{e}com ParisTech, 46 rue Barrault F-75634 Paris Cedex 13, France\goodbreak
\and
Laboratoire de Physique Subatomique et Cosmologie, Universit\'{e} Grenoble-Alpes, CNRS/IN2P3, 53, rue des Martyrs, 38026 Grenoble Cedex, France\goodbreak
\and
Laboratoire de Physique Th\'{e}orique, Universit\'{e} Paris-Sud 11 \& CNRS, B\^{a}timent 210, 91405 Orsay, France\goodbreak
\and
Lawrence Berkeley National Laboratory, Berkeley, California, U.S.A.\goodbreak
\and
Max-Planck-Institut f\"{u}r Astrophysik, Karl-Schwarzschild-Str. 1, 85741 Garching, Germany\goodbreak
\and
Mullard Space Science Laboratory, University College London, Surrey RH5 6NT, U.K.\goodbreak
\and
Nicolaus Copernicus Astronomical Center, Bartycka 18, 00-716 Warsaw, Poland\goodbreak
\and
Nordita (Nordic Institute for Theoretical Physics), Roslagstullsbacken 23, SE-106 91 Stockholm, Sweden\goodbreak
\and
SISSA, Astrophysics Sector, via Bonomea 265, 34136, Trieste, Italy\goodbreak
\and
School of Chemistry and Physics, University of KwaZulu-Natal, Westville Campus, Private Bag X54001, Durban, 4000, South Africa\goodbreak
\and
School of Physics and Astronomy, Cardiff University, Queens Buildings, The Parade, Cardiff, CF24 3AA, U.K.\goodbreak
\and
School of Physics and Astronomy, University of Nottingham, Nottingham NG7 2RD, U.K.\goodbreak
\and
Simon Fraser University, Department of Physics, 8888 University Drive, Burnaby BC, Canada\goodbreak
\and
Sorbonne Universit\'{e}-UPMC, UMR7095, Institut d'Astrophysique de Paris, 98 bis Boulevard Arago, F-75014, Paris, France\goodbreak
\and
Space Sciences Laboratory, University of California, Berkeley, California, U.S.A.\goodbreak
\and
Sub-Department of Astrophysics, University of Oxford, Keble Road, Oxford OX1 3RH, U.K.\goodbreak
\and
The Oskar Klein Centre for Cosmoparticle Physics, Department of Physics,Stockholm University, AlbaNova, SE-106 91 Stockholm, Sweden\goodbreak
\and
UPMC Univ Paris 06, UMR7095, 98 bis Boulevard Arago, F-75014, Paris, France\goodbreak
\and
Universit\'{e} de Toulouse, UPS-OMP, IRAP, F-31028 Toulouse cedex 4, France\goodbreak
% ...
}


\abstract{Using the \Planck\ 2015 data release (PR2) temperature maps, we separate Galactic thermal dust emission from cosmic infrared background (CIB) anisotropies. For this purpose, we implement a specifically tailored component-separation method, the so-called generalized needlet internal linear combination ({\tt GNILC}) method, which uses spatial information (the angular power spectra) to disentangle the Galactic dust emission and CIB anisotropies. We produce significantly improved all-sky maps of \Planck\ thermal dust emission, with reduced CIB contamination, at $353$, $545$, and $857\GHz$. By reducing the CIB contamination of the thermal dust maps, we provide more accurate estimates of the local dust temperature and dust spectral index over the sky with reduced dispersion, especially at high Galactic latitudes above $b=\pm 20^\circ$. We find that the dust temperature is $T=(19.4\pm 1.3)$\,K and the dust spectral index is $\beta=1.6\pm 0.1$ averaged over the whole sky, while $T=(19.4\pm 1.5)$\,K and $\beta=1.6\pm 0.2$ on $21$\,\% of the sky at high latitudes. Moreover, subtracting the new CIB-removed thermal dust maps from the CMB-removed \Planck\ maps gives access to the CIB anisotropies over 60\% of the sky at Galactic latitudes $\vert b\vert > 20^\circ$. Because they are a significant improvement over previous \Planck\ products, the {\tt GNILC} maps are recommended for thermal dust science. The new CIB maps can be regarded as indirect tracers of the dark matter and they are recommended for exploring cross-correlations with lensing and large-scale structure optical surveys. The reconstructed {\tt GNILC} thermal dust and CIB maps are delivered as \Planck\ products.}

\maketitle


\section {Introduction }
\label{sec:introduction}

The various populations of dust grains in the Galaxy are heated by absorbing the ultraviolet emission from stars. By re-emitting the light at infrared frequencies, the heated dust grains are responsible for the thermal dust radiation of the Galaxy.
For this reason, the dust emission is a tracer of the gas and particle density in the interstellar medium \citep{planck2011-7.0,planck2013-p06b,planck2013-XVII,planck2014-XXVIII,planck2014-XXXII} and of the star formation activity in the Galaxy \citep{Draine2007}. 
The Galactic thermal dust emission is also one of the major astrophysical foregrounds for observations of the cosmic microwave background (CMB) \citep{planck2014-XXX,pb2015}. Incorrect modelling of thermal dust might be responsible for a significant bias on the cosmological parameters \citep[e.g.,][]{Remazeilles2016}. The characterization of Galactic thermal dust emission over the whole sky is therefore essential for the accurate subtraction of this foreground from the CMB observations. Accurate characterization of the Galactic dust is also useful for the analysis of supernovae observations, where the Galactic dust causes extinction \citep{Riess1996}.

An unresolved background of dusty star-forming early galaxies also generates diffuse emission, known as the cosmic infrared background radiation \citep{Puget1996,Gispert2000,Lagache2005}. The cosmic infrared background (CIB) anisotropies are a probe of star-formation history in the Universe and also an indirect tracer of the dark matter \citep{planck2011-6.6,planck2013-pip56}. At high frequencies ($\gtrsim$ 300\,GHz), the Galactic thermal dust emission and the CIB radiation both scale approximately as modified blackbodies. This makes it challenging to separate the dust and CIB components solely on the basis of their spectral properties \citep{planck2013-p06b,planck2014-a12}

The previously released \Planck\footnote{\Planck\ (\url{http://www.esa.int/Planck}) is a project of the European Space Agency (ESA) with instruments provided by two scientific consortia funded by ESA member states and led by Principal Investigators from France and Italy, telescope reflectors provided through a collaboration between ESA and a scientific consortium led and funded by Denmark, and additional contributions from NASA (USA).} dust maps -- the \Planck\ 2013 (P13) dust model \citep{planck2013-p06b} and the \Planck\ 2015 (P15) dust model \citep{planck2014-a12} -- have been produced by fitting a modified blackbody (MBB) spectrum to the \Planck\ data. For the P13 dust map, a standard $\chi^2$ fitting of the MBB spectrum was applied pixel by pixel to four maps, namely the CMB-removed \Planck\ temperature maps at $353$, $545$, and $857\GHz$ from the \Planck\ 2013 data release (hereafter, PR1), and a $100\micron$ map obtained from a combination of the IRIS map from \citet{mamd2005} and the map from \citet{SFD1998}.
The CMB removal in the \Planck\ frequency channels was performed by subtracting the \Planck\ {\tt SMICA} CMB map \citep{planck2013-p06} from the \Planck\ frequency maps. For the P15 dust map, a Bayesian fitting of the MBB spectrum was implemented on the \Planck\ 2015 data release (hereafter, PR2) temperature maps by using the full set of \Planck\ frequency channels. 

However, the \Planck\ dust models P13 and P15 still suffer from contamination by the CIB anisotropies. In particular, at high Galactic latitudes the contamination by CIB anisotropies adds significant uncertainty to the measured dust spectral index and dust temperature \citep{planck2013-p06b}.  

By definition, the spectral fitting employed in \citet{planck2013-p06b} and \citet{planck2014-a12} relied solely on the frequency information to reconstruct the Galactic thermal dust model from observations of the sky emission. Because the Galactic dust emission and the extragalactic CIB emission have such similar spectral indices in the \Planck\ bands, the result of these frequency-based fits is that the CIB anisotropies inevitably leak into the \Planck\ dust model maps. In order to disentangle the Galactic thermal dust emission and the extragalactic CIB emission, additional discriminating statistical information is required. 

The CIB temperature fluctuations have been successfully measured by \Planck\ in relatively small regions of the sky, where the Galactic dust contamination is low \citep{planck2013-pip56}, and  the angular power spectra of the CIB anisotropies have been computed at frequencies from $143\GHz$ to $3000\GHz$. 

In this work we perform the separation of the Galactic thermal dust and CIB components over a large area of the sky by exploiting not only the frequency spectral information but also the spatial information through the use of the \Planck\ CIB best-fit angular power spectra computed in \citet[][hereafter, CIB 2013]{planck2013-pip56}. The CIB power spectrum scales approximately as $\ell^{-1}$ \citep{planck2011-6.6}, while the dust power spectrum scales approximately as $\ell^{-2.7}$ \citep{planck2013-pip56}. This distinct spatial behaviour provides the necessary extra statistical information that enables robust separation of thermal dust emission and CIB radiation. 
Although the CIB 2013 angular power spectra have been estimated in only small areas of the sky, we assume that the statistics of the CIB anisotropies are the same in a larger area of the sky, because of the homogeneity and isotropy of the CIB emission.

The component-separation method employed in this work is the generalized needlet internal linear combination ({\tt GNILC}) method, first developed in \citet{Remazeilles2011b}. It is worth noting that the {\tt GNILC} method has also been applied in a different context to simulations of a radio intensity mapping experiment for separating the cosmological \hi\ $21$-cm temperature fluctuations and the Galactic synchrotron radiation in \citet{Olivari2016}, where the component-separation problem was similar.


This paper is organized as follows. In Sect. \ref{sec:data} we present the data used in the analysis. In Sect. \ref{sec:method} we give a summary of the component-separation method that we implement on the data to disentangle the Galactic dust emission and the CIB anisotropies; 
the full description of the method and validation on simulations are presented in Appendix \ref{sec:details}.
In Sect. \ref{sec:results1} we discuss the results for the Galactic thermal dust emission and the estimated spectral parameters. In Sect. \ref{sec:results2} we discuss the results for the CIB emission. In Sect. \ref{subsec:HI} we explore the correlations of the new dust and CIB maps with the \hi\ gas tracer. We conclude in Sect. \ref{sec:conclusions}.

\section {Data and preprocessing}
\label{sec:data}

\subsection{\Planck\ data }
\label{subsec:pr2}

The data used in this paper are the temperature full-mission sky maps \citep{planck2014-a07,planck2014-a09} of the \Planck\ 2015 data release (PR2) that have been made publicly available on the Planck Legacy Archive. We make use of the nine single-frequency maps from $30$ to $857$\,GHz from both LFI and HFI instruments.
As discussed in \citet{planck2013-pip88}, the zodiacal light emission is removed from the \Planck\ HFI temperature maps ($100$ to $857$\,GHz) by fitting different \Planck\ surveys of the sky with the COBE/DIRBE\footnote{Cosmic Background Explorer Diffuse Infrared Brightness Experiment.} zodiacal model \citep{Kelsall1998}. Because different \Planck\ surveys are taken at different times, the sky is observed through different column depths of interplanetary dust. Differencing two surveys removes all distant structure in the maps, such as Galactic and extra-galactic emission, but leaves a detectable Zodiacal signal. This difference signal is fit to extend the COBE zodiacal model to \Planck\ frequencies. The entire, un-differenced signal is then reconstructed from the model and removed from the data of each \Planck\ HFI bolometer prior to mapmaking \citep{planck2014-a09}.

We also make use of the \Planck\ temperature half-mission sky maps (hereafter, HM1 and HM2, as defined in \citet{planck2014-a09}) in the nine frequency channels, in order to estimate by their half-difference (see Eq.~\ref{eq:half-diff}) the local rms of the instrumental noise in the \Planck\ full-mission maps.

\subsection{The IRAS 100\,\micron\ map}
\label{subsec:iris_sfd}

Following \citet{planck2013-p06b}, in addition to the \Planck\ PR2 data we also use in this work the full-sky temperature map at 100\,\micron\ based on the combination of the IRIS map \citep{mamd2005} and the map of \citet[][hereafter SFD map]{SFD1998} both projected on the {\tt HEALPix} grid \citep{Gorski05} at $N_{\rm side}=2048$. The combined 100\,\micron\ map is compatible with the SFD map at angular scales larger than $30'$ and compatible with the IRIS map at smaller angular scales. The effective beam resolution of the combined 100\,\micron\ map is $4.3'$ and the noise rms level is $0.06$\,MJy\,sr$^{-1}$.

Note that residual low-level zodiacal light emission is present in the combined 100\,\micron\ map, because the zodiacal emission has not been corrected in the same way in the SFD map and in the IRIS map. We refer to the appendix of \citet{planck2013-p06b} for further discussion of this point.

\subsection{Preprocessing of the point sources }
\label{subsec:inpainting}

We make use of nine point source masks, one for each \Planck\ frequency channel, in order to remove the point sources detected in each frequency at a signal-to-noise ratio S/N\,$ > $5 in the second Planck Catalogue of Compact Sources, PCCS2 \citep{planck2014-a35}.

The masked pixels in each \Planck\ frequency map are filled in through a minimum curvature spline surface inpainting technique, implemented in the Planck Sky Model ({\tt PSM}) software package \citep{Delabrouille2013} and described in \citet{Remazeilles2015} and \citet{planck2014-a14}. For consistency, we also consider the source-subtracted version of the combined 100\,\micron\ map that is described in the appendix of \citet{planck2013-p06b}. The inpainted \Planck\ 2015 maps and the inpainted combined 100\,\micron\ map are the inputs to the component-separation algorithm described in the next section.

\section{Summary of the component-separation method }
\label{sec:method}

The component-separation technique that we follow in this work is based on \citet{Remazeilles2011b} and called {\tt GNILC}. 

In order to simplify the reading of the paper, we give a brief summary of the method employed in this work to separate the thermal dust and CIB anisotropies. A complete description of the formalism and technical details of {\tt GNILC} are presented in Appendix \ref{sec:details}.

Each frequency map is first decomposed on a needlet (spherical wavelet) frame \citep{Narcowich2006,Guilloux2007}. The localization properties of the needlets allow us to adapt the component separation to the local conditions of contamination in both harmonic space and real space \citep{Delabrouille2009,Remazeilles2011b,Basak2012,Remazeilles2013,Basak2013}. We define ten needlet windows, $\{h^{(j)}(\ell)\}_{1\leq j \leq 10}$, having a Gaussian shape and acting as bandpass filters in harmonic space, each of them selecting a specific subrange of angular scales (see Fig.~\ref{Fig:bands}). The spherical harmonic transform, $a_{\ell m}$, of each frequency map is bandpass filtered in harmonic space by the ten needlet windows. The inverse transform of the bandpass-filtered coefficient, $h^{(j)}(\ell)a_{\ell m}$, provides a needlet map at scale $j$, conserving only statistical information from the range of $\ell$ considered. Therefore, we have 100 input maps (10 frequencies times 10 needlet scales). The component separation is performed on each needlet scale independently. The main steps of the {\tt GNILC} algorithm are the following.
For each needlet scale, $j$, considered we perform seven steps:
\begin{enumerate}
\item Compute the frequency-frequency data covariance matrix, at pixel $p$, and scale $j$, 
\bea
\widehat{\tens{R}}^{j}_{ab}(p) = \sum_{p'\in \mathcal{D}(p)} \bdx^{j}_a(p)(\bdx^{j}_b(p))^{\sf{T}}, 
\eea
where $\mathcal{D}(p)$ is a domain of pixels centred at pixel $p$ and $\bdx^{j}_a(p)$ and $\bdx^{j}_b(p)$ are the needlet maps at scale $j$ of the observations for the pair of frequencies $a,b$. In practice, the domain of pixels, $\mathcal{D}(p)$, is defined by the convolution in real space of the product of the needlet maps with a Gaussian kernel. The width of the Gaussian kernel is a function of the needlet scale considered. 
\item Similarly, compute the frequency-frequency covariance matrix of the instrumental noise, at pixel $p$, and scale $j$, 
\bea
\widehat{\tens{R}}^{j}_{{\rm noise} ~ab}(p) = \sum_{p'\in \mathcal{D}(p)} \bdn^{j}_a(p)(\bdn^{j}_b(p))^{\sf{T}}, 
\eea
where the instrumental noise maps, $\bdn(p)$, are estimated from the half-difference of the half-mission HM1 and HM2 \Planck\ maps. 
\item Similarly, compute the frequency-frequency covariance matrix of the CMB, $\widehat{\tens{R}}^{j}_{\rm CMB}(p)$, and the frequency-frequency covariance matrix of the CIB, $\widehat{\tens{R}}^{j}_{\rm CIB}(p)$, but this time using CMB maps and CIB maps that are simulated from the \Planck\ CMB best-fit $C_\ell$ \citep{planck2013-p08} and the \Planck\ CIB best-fit $C^{a\times b}_\ell$ \citep{planck2013-pip56} respectively. The simulated maps were analyzed with the same needlet decomposition as was applied to the real data before computing the CMB and CIB covariance matrices.
\item Compute the ``nuisance'' covariance matrix, $\widehat{\tens{R}}_{\rm N}$, by co-adding the noise covariance matrix, the CMB covariance matrix, and the CIB covariance matrix:
\bea\label{eq:nuisance}
\widehat{\tens{R}}_{\rm N} = \widehat{\tens{R}}_{\rm CIB} + \widehat{\tens{R}}_{\rm CMB} + \widehat{\tens{R}}_{\rm noise}.
\eea  
\item Diagonalize the transformed data covariance matrix 
\bea
\widehat{\tens{R}}_{\rm N}^{-1/2}\widehat{\tens{R}}\,\widehat{\tens{R}}_{\rm N}^{-1/2} = \widehat{\tens{U}}\left[\begin{array}{ccc} \mu_1& &\\ & ... & \\ &  & \mu_{N_{\rm ch}}\end{array}\right]\widehat{\tens{U}}^{\sf{T}} \approx \widehat{\tens{U}}_{\rm S}\widehat{\tens{D}}_{\rm S}\widehat{\tens{U}}_{\rm S}^{\sf{T}} + \widehat{\tens{U}}_{\rm N}\widehat{\tens{U}}_{\rm N}^{\sf{T}},
\eea 
where $N_{\rm ch}$ is the number of frequency channels. In this representation, the eigenvalues that are close to unity correspond to the nuisance power (CIB plus CMB plus noise), while the $m$ eigenvalues larger than unity that are collected in the diagonal matrix $\widehat{\tens{D}}_{\rm S}$ correspond to the power of the Galactic signal. The matrix $\widehat{\tens{U}}_{\rm S}$ collects the $m$ eigenvectors spanning the Galactic signal subspace.
\item Compute the effective dimension, $m$, of the foreground signal subspace (number of Galactic degrees of freedom, or principal components) by minimizing the Akaike Information Criterion \citep[AIC,][]{Akaike1974}:
\bea
\min_{m \in [1,N_{\rm ch}]}  \left(2\,m\,+\,\sum_{i=m+1}^{N_{\rm ch}}\, \left( \mu_i - \log\mu_i -1 \right)\right),
\eea
where $\mu_i$ are the eigenvalues of $\widehat{\tens{R}}_{\rm N}^{-1/2}\widehat{\tens{R}}\,\widehat{\tens{R}}_{\rm N}^{-1/2}$.

\item Apply the $m$--dimensional ILC filter \citep{Remazeilles2011b} to the data in order to reconstruct the total Galactic signal at scale $j$:
\bea
\widehat{\bdf}^{j} &=& \widehat{\tens{F}}\left(\widehat{\tens{F}}^{\sf{T}} \widehat{\tens{R}}^{-1}\widehat{\tens{F}}\right)^{-1}\widehat{\tens{F}}^{\sf{T}}  \widehat{\tens{R}}^{-1}\bdx^{j},
\eea
where the estimated mixing matrix is given by
\bea
\widehat{\tens{F}} = \widehat{\tens{R}}_{\rm N}^{1/2} \widehat{\tens{U}}_{\rm S}
\eea
with $\widehat{\tens{U}}_{\rm S}$ collecting the $m$ eigenvectors selected by the AIC criterion at scale $j$.
\end{enumerate}

The reconstructed Galactic signal maps are finally \mbox{synthesized} as follows. We transform the estimated maps, $\widehat{\bdf}^{j}$, to spherical harmonic coefficients, then bandpass filter the harmonic coefficients by the respective needlet window, $h^{j}_\ell$, and transform back to maps in real space. This operation provides one reconstructed Galactic signal map per needlet scale. We co-add these maps to obtain, for each frequency channel, the complete reconstructed Galactic signal map on the whole range of angular scales. The needlet windows are chosen so that ${\sum_{j=1}^{10} \left(h^{j}_\ell\right)^2 = 1}$, therefore conserving the total power in the synthesis.

The reconstruction of the CIB maps is performed as follows. In step 4, we replace Eq.~(\ref{eq:nuisance}) by $\widehat{\tens{R}}_{\rm N} = \widehat{\tens{R}}_{\rm CMB} + \widehat{\tens{R}}_{\rm noise}$ so that the reconstructed signal is the sum of the Galactic dust plus the CIB. We then subtract the reconstructed Galactic dust (only), $\widehat{\bdf}$, from the Galactic dust plus CIB reconstruction.

Note that the priors on the CMB and CIB angular power spectra are only used for estimating the dimension, $m$, of the Galactic signal subspace (step 5), not for the ILC filtering (step 7) in the reconstruction of the components of the emission.

We have validated the {\tt GNILC} method on the \Planck\ full focal plane simulations \citep{planck2014-a14} before applying it to the \Planck\ data. The results on simulations are presented in 
Sect. \ref{subsec:simu} of Appendix \ref{sec:details}.

\section{{\tt GNILC} results on the thermal dust}
\label{sec:results1}

\begin{figure}
\begin{center}
\includegraphics[width=0.5\columnwidth]{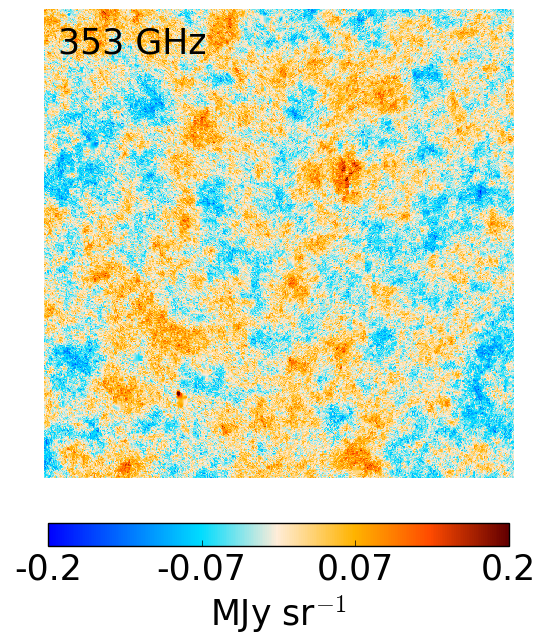}~
\includegraphics[width=0.5\columnwidth]{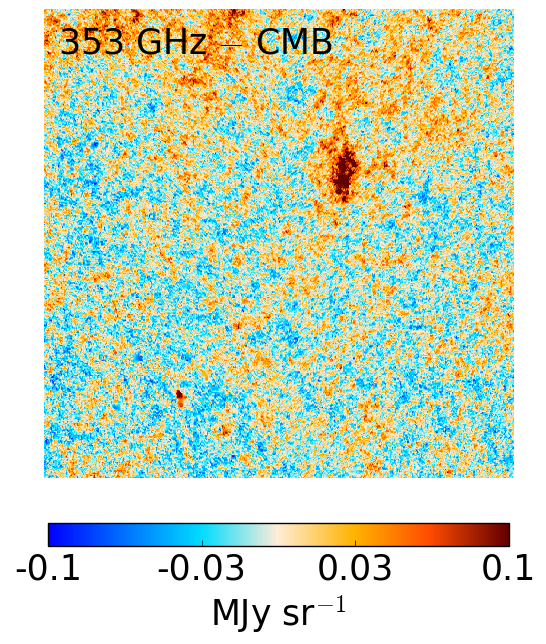}~\\
\includegraphics[width=0.5\columnwidth]{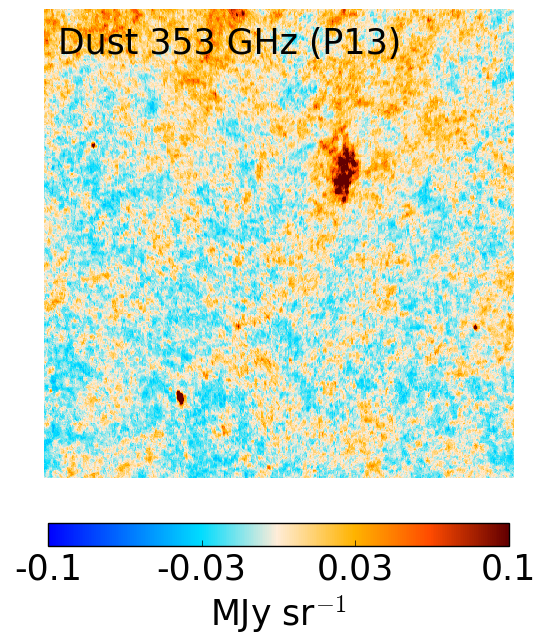}~
\includegraphics[width=0.5\columnwidth]{figures/fig03a}~\\
\includegraphics[width=0.5\columnwidth]{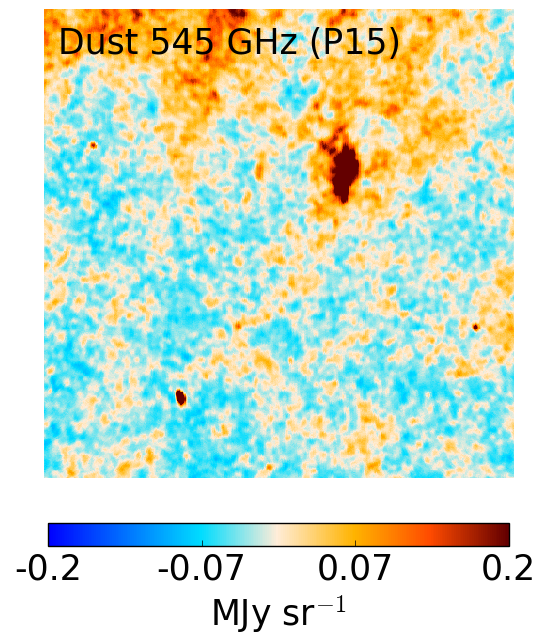}~
\includegraphics[width=0.5\columnwidth]{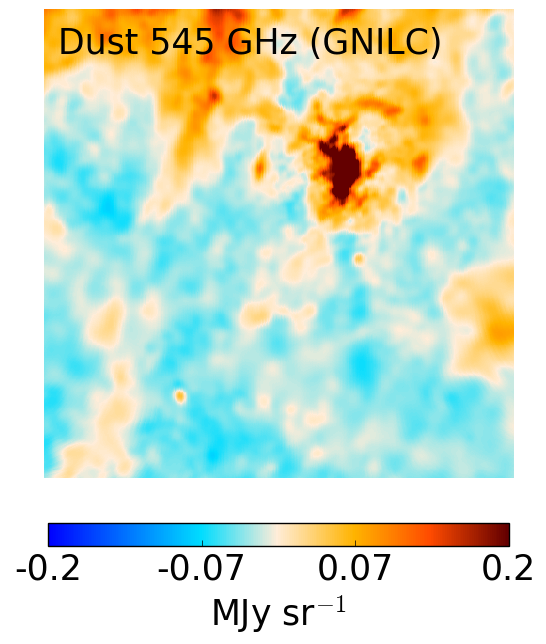}\\
\end{center}
  \caption{A $12^\circ.5\times 12^\circ.5$ gnomonic projection of the sky centred at high latitude $(l,b)=(90^\circ,-80^\circ)$. {\it Top left}: {\it Planck} $353$-GHz map. {\it Top right}: CMB-removed {\it Planck} $353$-GHz map. {\it Middle left}: dust model P13 at $353$\,GHz (MBB fit on CMB-removed {\it Planck} maps). {\it Middle right}: {\tt GNILC} dust map at $353$\,GHz. {\it Bottom left}: dust model P15 at $545$\,GHz ({\tt Commander} Bayesian fitting). {\it Bottom right}: {\tt GNILC} dust map at $545$\,GHz. Maps at $353$\,GHz are shown at $5'$ resolution, while maps at $545$\,GHz are smoothed to $7.5'$ resolution. The {\tt GNILC} dust maps have a non-uniform resolution (see Fig.~\ref{Fig:beaming}) with $5'$ resolution kept in regions of bright dust emission. In each image the local mean intensity has been subtracted for this comparison.}
  \label{Fig:patches}
\end{figure}

\begin{figure}
  \begin{center}
    \includegraphics[width=0.5\textwidth]{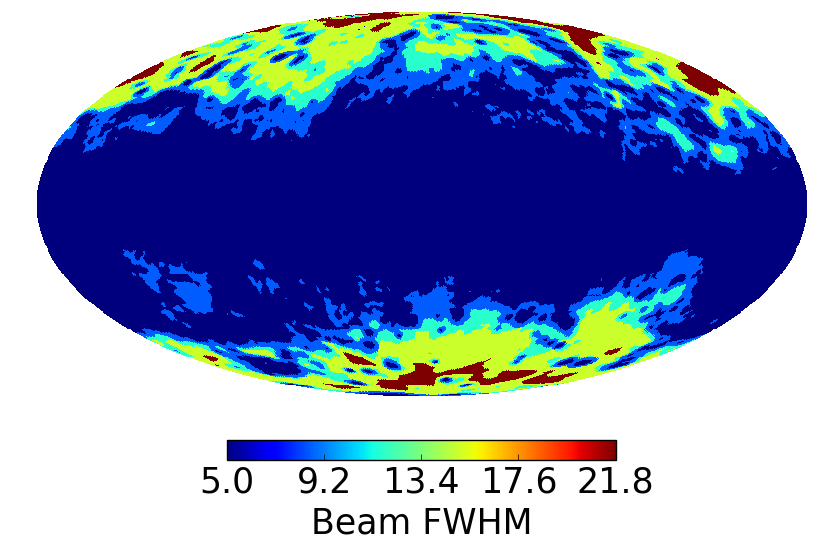}~\\
    \includegraphics[width=0.5\columnwidth]{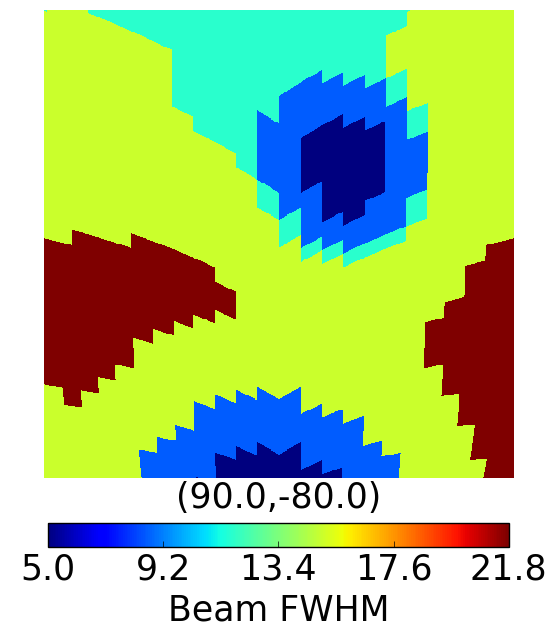}~
  \end{center}
  \caption{Effective beam FWHM of the {\tt GNILC} dust maps on the whole sky (\emph{top panel}) and on a $12^\circ.5\times 12^\circ.5$ area of the sky centred at high latitude $(l,b)=(90^\circ,-80^\circ)$ (\emph{bottom panel}). The spatially varying beam FWHM is the same for all frequencies. {\tt GNILC} preserves the $5'$-scale power of the thermal dust in the high signal-to-noise regions of the sky.}
\label{Fig:beaming}
\end{figure}

\begin{figure*}
  \begin{center}
    \includegraphics[width=\textwidth]{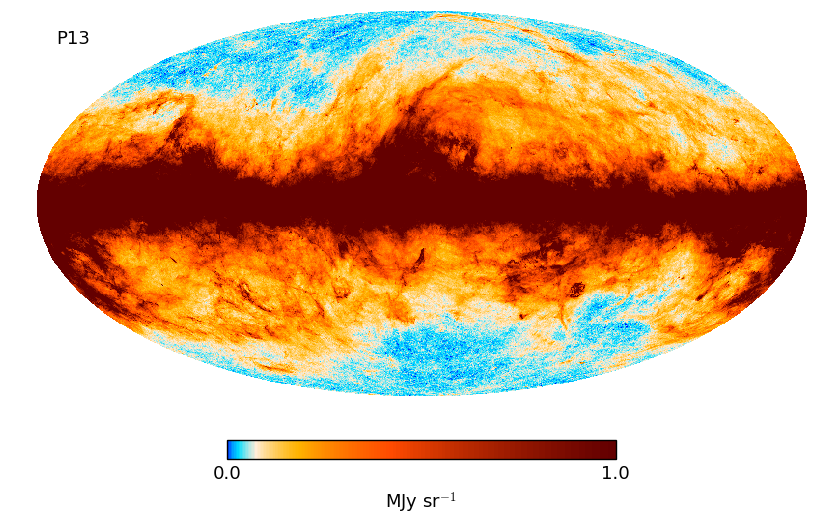}~\\
    \includegraphics[width=\textwidth]{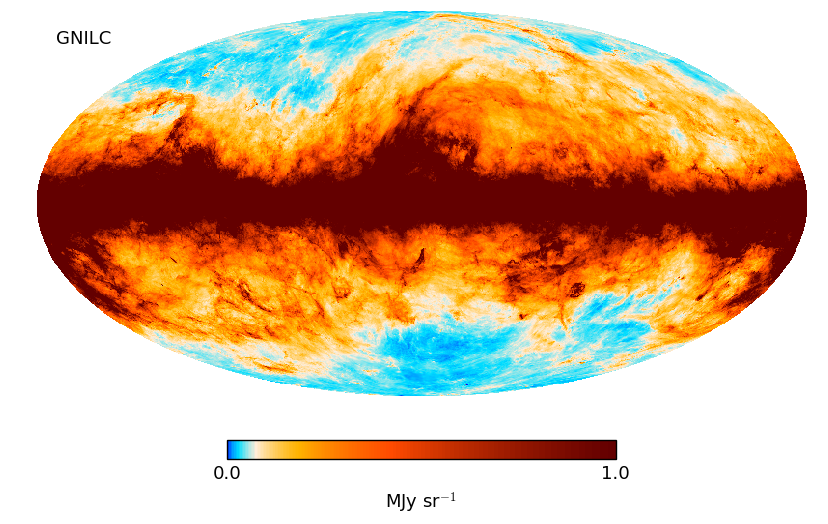}~
  \end{center}
  \caption{Full-sky map of the Galactic thermal dust emission: \Planck\ 2013 (P13) thermal dust model at $353$\,GHz and $5'$ resolution ({\it top panel}), suffering from CIB contamination at high latitudes, and the {\tt GNILC} dust map (this work) at $353$\,GHz and $5'$ resolution ({\it bottom panel}), for which the CIB is clearly filtered out at high-latitudes. 
A logarithmic colour scale is used here to highlight the low-intensity emission at high latitudes.
The effective local beam of the {\tt GNILC} dust maps is shown in Fig.~\ref{Fig:beaming}.}
  \label{Fig:full-sky}
\end{figure*}

\begin{figure*}
\begin{center}
\includegraphics[width=0.33\textwidth]{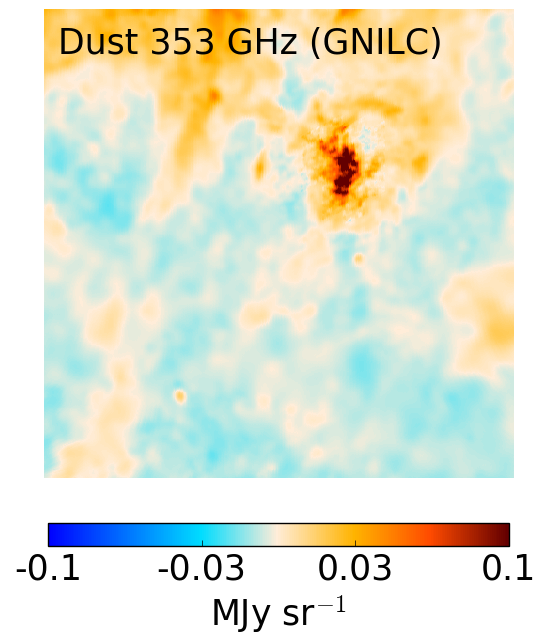}~
\includegraphics[width=0.33\textwidth]{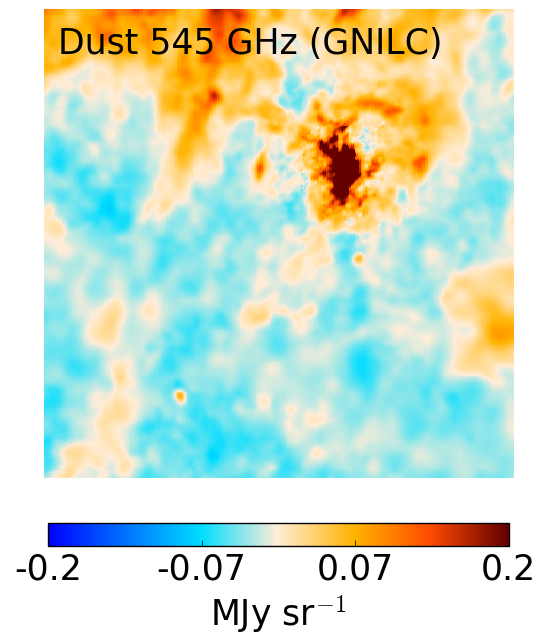}~
\includegraphics[width=0.33\textwidth]{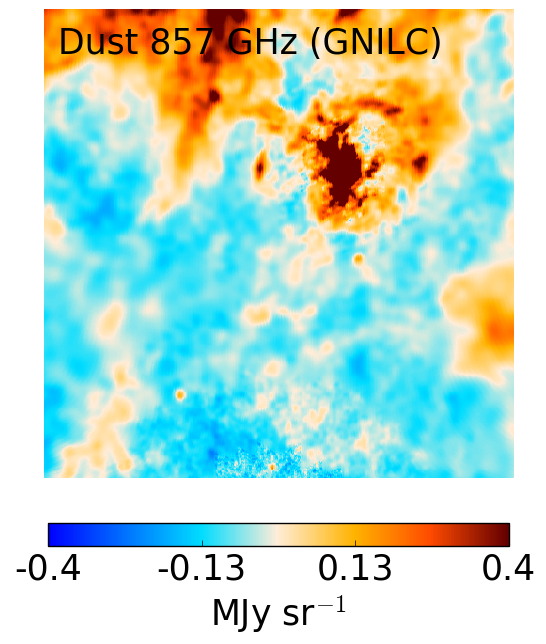}
\end{center}
  \caption{{\tt GNILC} dust maps at $353$\,GHz ({\it left panel}), $545$\,GHz ({\it middle panel}), and $857$\,GHz ({\it right panel}) on a $12^\circ.5\times 12^\circ.5$ gnomonic projection of the sky centred at high latitude, $(l,b)=(90^\circ,-80^\circ)$. {\tt GNILC} filters out the CIB anisotropies while preserving the small-scale dust signal (see bottom panel of Fig.~\ref{Fig:beaming}). In these images, the local mean intensity of each map has been subtracted.}
  \label{Fig:dust}
\end{figure*}

\begin{figure*}
\begin{center}
\includegraphics[width=0.33\textwidth]{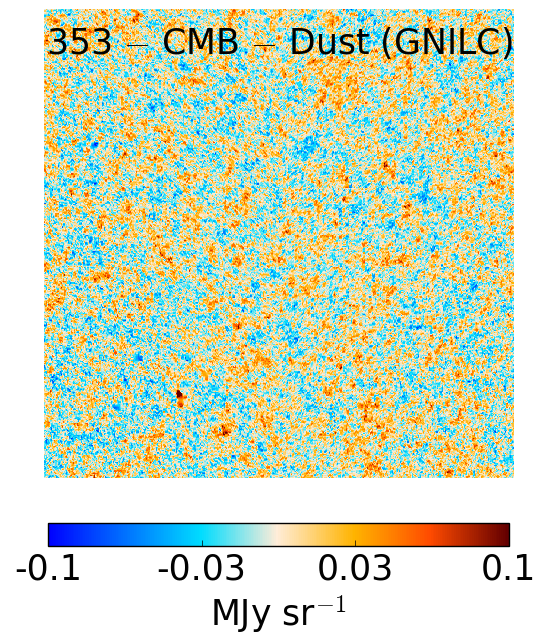}~
\includegraphics[width=0.33\textwidth]{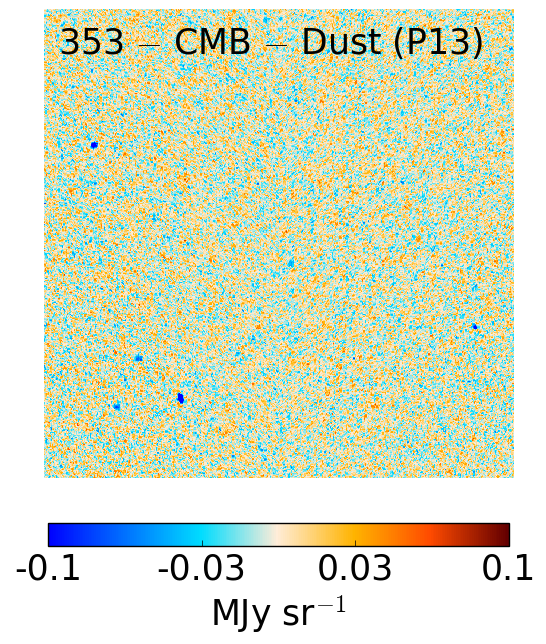}~
\includegraphics[width=0.33\textwidth]{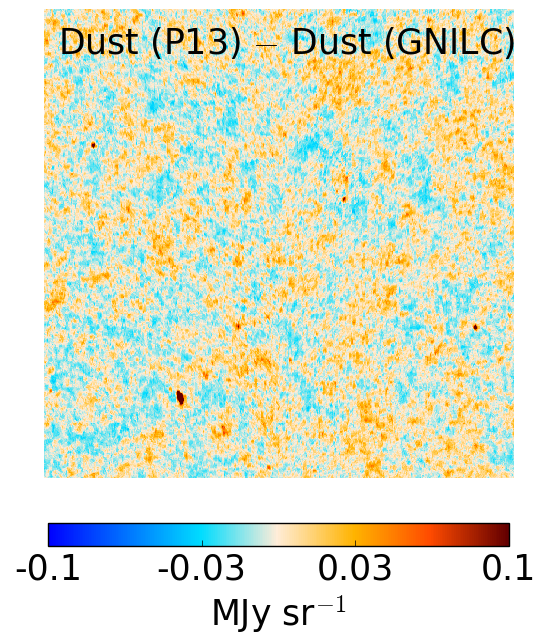}~
\end{center}
  \caption{A $12^\circ.5\times 12^\circ.5$ gnomonic projection of the sky centred at high latitude, $(l,b)=(90^\circ,-80^\circ)$. {\it Left}: difference map ({\it Planck} $353$\,GHz $-$ {\it Planck} CMB $-$ {\tt GNILC} dust) reveals the CIB anisotropies at $353$\,GHz. {\it Middle}: difference map ({\it Planck} $353$\,GHz - {\it Planck} CMB - dust model P13) revealing only the instrumental noise, because the dust model P13, like the {\it Planck} observations at $353$\,GHz, still contains the CIB signal. {\it Right}: difference (dust model P13 - {\tt GNILC} dust) revealing the amount of CIB contamination in the dust model P13 with respect to the {\tt GNILC} dust map. In these images, the local mean intensity of each map has been subtracted.}
  \label{Fig:diff}
\end{figure*}
\begin{figure}[!h]
\begin{center}
\includegraphics[width=\columnwidth]{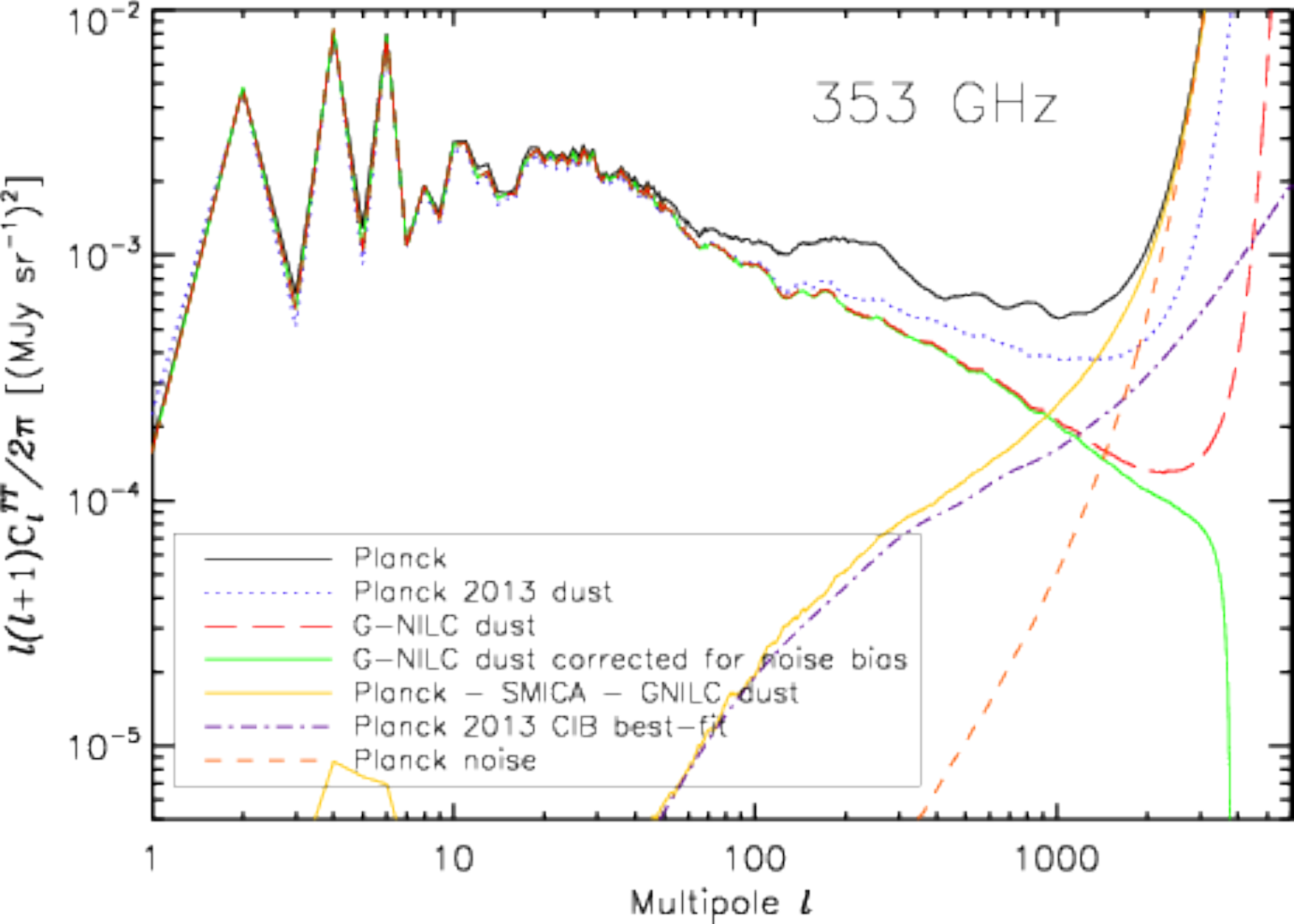}~\\
\includegraphics[width=\columnwidth]{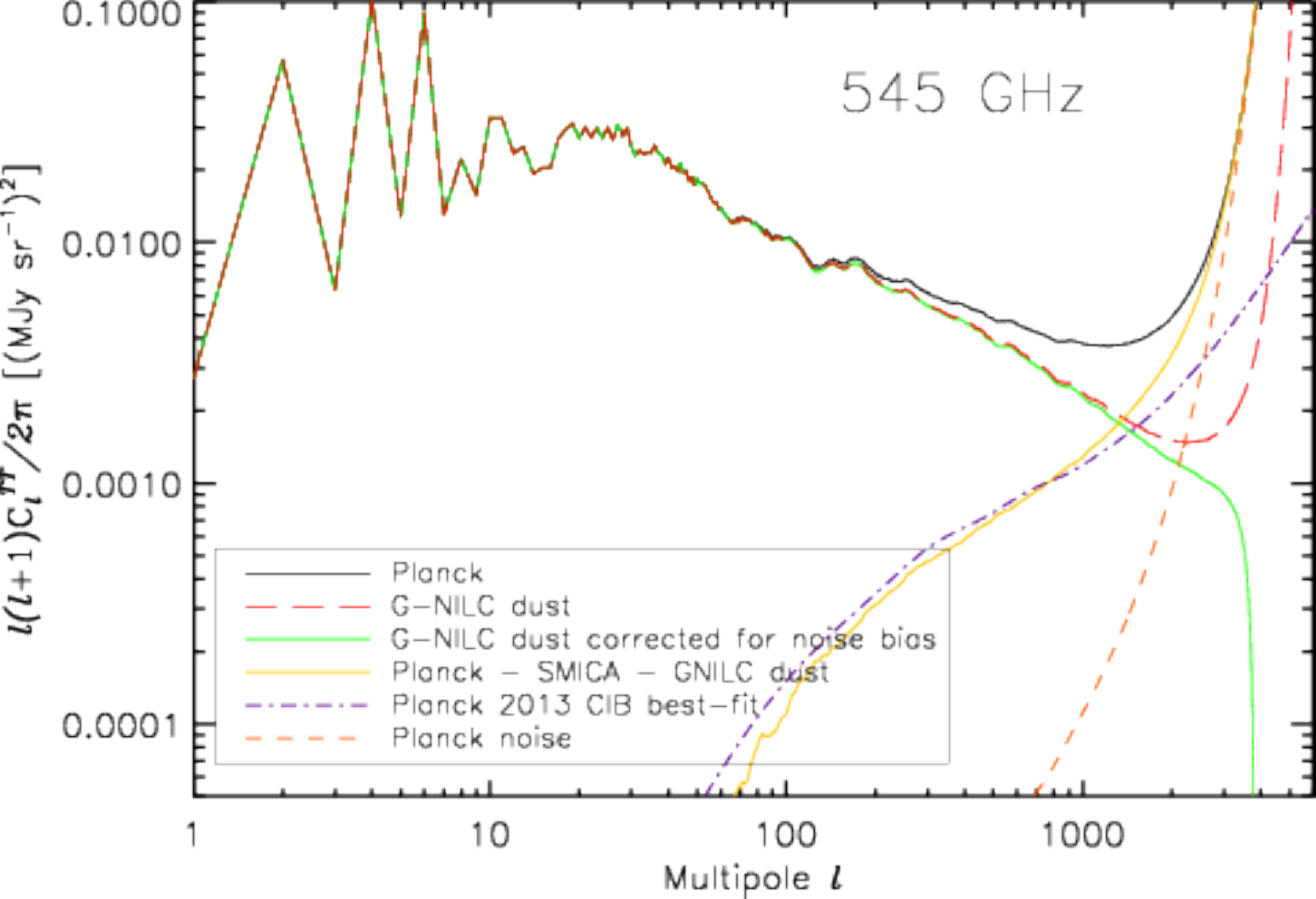}~\\
\includegraphics[width=\columnwidth]{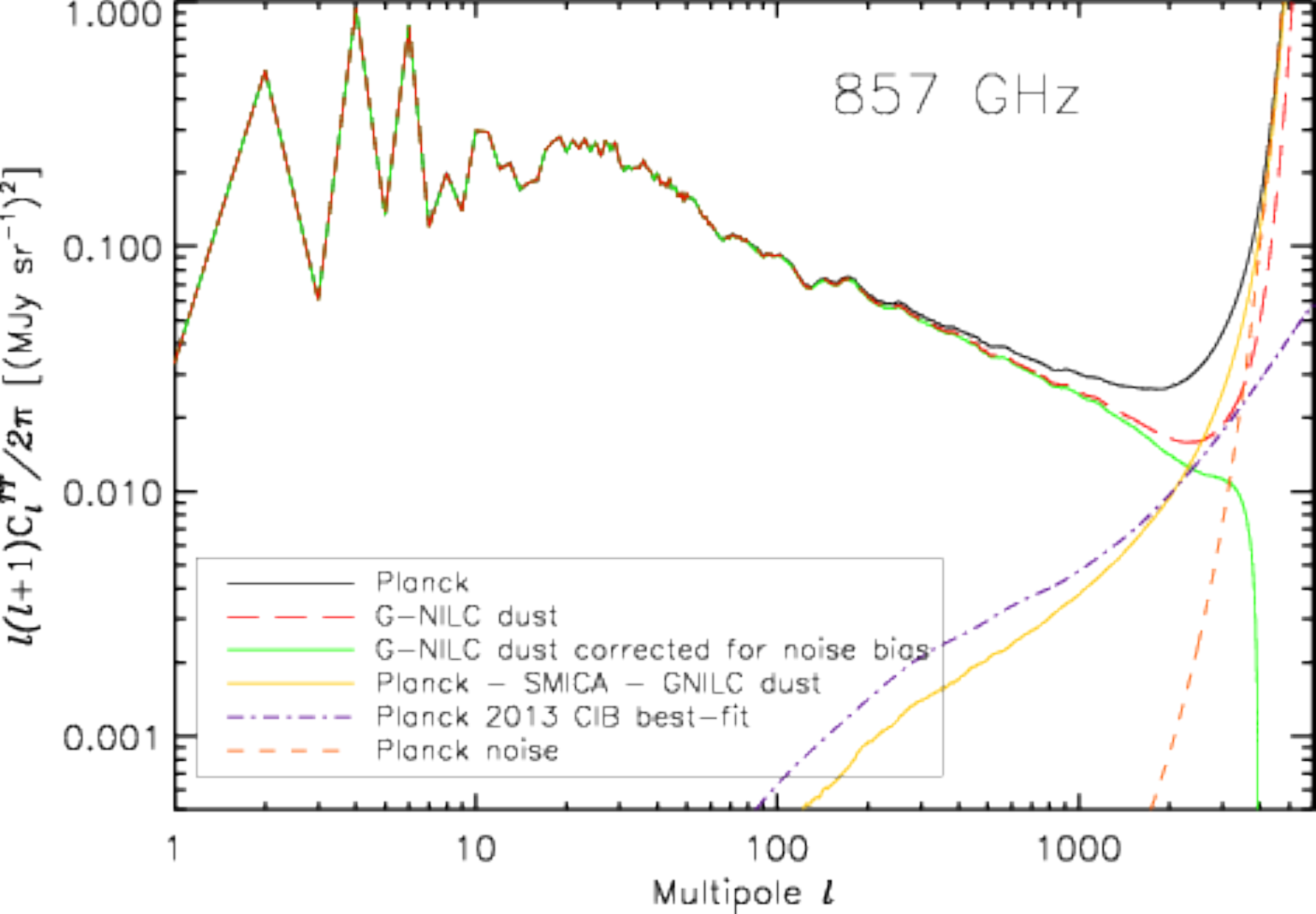}~
\end{center}
  \caption{Angular power spectra of the various maps at $353$\,GHz (\emph{top panel}), $545$\,GHz (\emph{middle panel}), and $857$\,GHz (\emph{bottom panel}), on a fraction of the sky, $f_{\rm sky}=57$\,\%\,: \Planck\ map (solid black line), dust model P13 \citep[dotted blue line,][]{planck2013-p06b}, {\tt GNILC} dust map (long dashed red line), {\tt GNILC} dust map corrected for the residual noise (solid green line), and {\tt GNILC} residual map (\Planck\ map $-$ \Planck\ CMB $-$ {\tt GNILC} dust, solid yellow line), which is compared to the \Planck\ CIB best-fit power spectrum \citep[dash-dot purple line,][]{planck2013-pip56} and the \Planck\ noise power spectrum (dashed orange line).}
  \label{Fig:ps}
\end{figure}

\subsection{Dust maps and power spectra}
\label{subsec:dustmaps}

In Fig.~\ref{Fig:patches} we compare various maps projected onto a high Galactic latitude $12^\circ.5\times 12^\circ.5$ area centred at ${(l,b)=(90^\circ,-80^\circ)}$. In the top left panel of Fig.~\ref{Fig:patches}, the \Planck\ $353$-GHz channel map is shown. At 353\,GHz the CMB radiation is clearly visible in the \Planck\ observation map at high Galactic latitude, through degree-scale temperature fluctuations which are typical in size of the CMB anisotropies.
In the top right panel of Fig.~\ref{Fig:patches}, the \Planck\ $353$-GHz map is shown after subtraction of the {\it Planck} CMB map \citep[i.e., the {\tt SMICA} map from][]{planck2013-p06}. The CMB-removed {\it Planck} $353$-GHz map reveals the thermal dust emission, but is still quite noisy and contaminated by the CIB temperature anisotropies. The dust model P13 at $353$\,GHz, which has been computed by fitting an MBB spectrum to the CMB-removed \Planck\ maps \citep{planck2013-p06b}, is plotted in the middle left panel of Fig.~\ref{Fig:patches}. Because of the similar spectral signatures of the thermal dust and the CIB, the dust model P13 resulting from the spectral fitting can not avoid the leakage of the CIB fluctuations into the dust map. Conversely, in the {\tt GNILC} dust map at $353$\,GHz produced in this work, the CIB anisotropies have been successfully filtered out, while the $5'$-scale dust emission has been conserved in the map. Note that all the maps are shown at $5'$ resolution, but the {\tt GNILC} dust map has a local effective beam resolution that is shown in Fig.~\ref{Fig:beaming}. The local beam resolution of the {\tt GNILC} dust maps is not the result of a local smoothing of the maps, but the result of the thresholding of the needlet coefficients that depends on local signal-to-noise ratio. In some high-latitude regions of the sky, beyond a certain angular scale, the power of the dust is found to be consistent with zero, i.e., the dimension of the Galactic signal subspace selected by the AIC criterion is $m=0$ (see Fig.~\ref{Fig:dof}), because the sky observations in this needlet domain become compatible with the CIB-plus-CMB-plus-noise model.

In the bottom panels of Fig.~\ref{Fig:patches}, we compare the dust model P15 at $545$\,GHz from \citet{planck2014-a12} and the {\tt GNILC} dust map at $545$\,GHz produced in this work. 
The dust model P15 has been obtained by using a Bayesian fitting method, {\tt Commander} \citep{Eriksen2008}, instead of the $\chi^2$ fitting method used for the dust model P13 in \citet{planck2013-p06b}. {\tt Commander} makes an overall fit of many foreground parameters, including those of the thermal dust component (intensity, spectral index, and temperature). However, the {\tt Commander} fitting again does not make any distinction between the CIB and the Galactic thermal dust, both sharing a similar MBB spectrum. In summary, thermal dust and CIB are still fitted as a single component in constructing the dust models P13 and P15. 
 The dust model P15 at $545$\,GHz still shows CIB anisotropies at high latitude, whereas those CIB anisotropies have been successfully filtered out in the {\tt GNILC} dust map at $545$\,GHz (see Fig.~\ref{Fig:patches}). Unlike the dust models P13 and P15, the {\tt GNILC} dust maps are not the result of any fit of a dust model, but the result of a component-separation procedure solely based on prior assumptions on the CIB, CMB, and noise angular power spectra.

Figure \ref{Fig:full-sky} shows the {\tt GNILC} all-sky map of the thermal dust at $353$\,GHz in the bottom panel. This map can be compared to the dust model P13 at $353$\,GHz from \citet{planck2013-p06b}, shown in the top panel. While the dust model P13 still shows visible small-scale contamination by CIB anisotropies at high latitude, in the {\tt GNILC} dust map the CIB fluctuations are clearly filtered out at high latitude. The $12^\circ.5\times 12^\circ.5$ gnomonic projections, centred at ${(l,b)=(90^\circ,-80^\circ)}$, of the various {\tt GNILC} maps of the dust at $353$, $545$, and $857$\,GHz, are shown in Fig.~\ref{Fig:dust}.

At high Galactic latitude and small angular scales, the AIC criterion can select a dimension zero for the Galactic signal subspace, considering that in this region the Galactic signal is completely buried under the CIB and noise signals, and therefore the dust is compatible with zero. This aspect of the {\tt GNILC} filtering is visible in the bottom panel of Fig.~\ref{Fig:dof}, where in the high-latitude region at $5'$ scale there are no Galactic degrees of freedom selected by the AIC criterion. Therefore, in practice the {\tt GNILC} filtering is equivalent to a local smoothing of the sky map, depending on the relative power of the Galactic dust with respect to the local contamination by the CIB, the CMB, and the instrumental noise.  The effective local beam FWHM over the sky of the {\tt GNILC} dust maps is plotted in Fig.~\ref{Fig:beaming}. Over $65$\,\% of the sky, where the dust signal is significant, the $5'$ beam resolution is preserved by the {\tt GNILC} filtering. 

It is interesting to look at the residual map given by the difference between the CMB-removed \Planck\ map and the dust map, i.e., the difference map ({\it Planck} map $-$ {\it Planck} CMB map $-$ dust map). In the case where the {\tt GNILC} dust map is used in the subtraction, the residual map clearly shows the CIB anisotropies plus the instrumental noise (left panel of Fig.~\ref{Fig:diff}), as expected. Conversely, if the \Planck\ 2103 dust model is used for the subtraction in place of the {\tt GNILC} dust map then the residual map shows the instrumental noise only (middle panel of Fig.~\ref{Fig:diff}). This, again, indicates that the CIB anisotropies have leaked into the dust model P13, and therefore that the CIB anisotropies can not be recovered in the residual map. In the right panel of Fig.~\ref{Fig:diff}, we plot the difference between the \Planck\ 2103 dust model and the {\tt GNILC} dust map at $353$\,GHz, highlighting the amount of CIB leaking into the dust model P13.

The resulting angular power spectrum of the various dust maps and the residual maps at $353$, $545$, and $857$\,GHz are plotted in Fig.~\ref{Fig:ps}. We have used the {\tt HEALPix} routine {\tt anafast} \citep{Gorski05} for computing the angular power spectrum of the maps. The amplitude of the power spectrum of the {\tt GNILC} dust map at $353$\,GHz (long dashed red line) is reduced by a factor of $2$ at $\ell\approx 1000$ with respect to the dust model P13 (dotted blue line) because of the removal of the CIB contamination. The power spectrum of the {\tt GNILC} dust map is steeper than the power spectrum of the dust model P13. The {\tt GNILC} dust power spectrum scales as a power-law $C_\ell\approx \ell^{-2.7}$. For completeness, the {\tt GNILC} dust power spectrum corrected for residual noise is overplotted as a solid green line.

The angular power spectrum of the {\tt GNILC} residual map, i.e. of the difference map (\Planck\ map $-$ \Planck\ CMB $-$ {\tt GNILC} dust), typically shows the power of the CIB plus noise that has been filtered out in the {\tt GNILC} dust map (solid yellow line). As illustrated in the top panel of Fig.~\ref{Fig:ps}, the angular power spectrum of the {\tt GNILC} residual map successfully matches the sum of the {\it Planck} CIB best-fit power spectrum at $353$\,GHz (dash-dot purple line) and the {\it Planck} $353$-GHz instrument noise power spectrum (dashed orange line). At $545$ and $857$\,GHz the angular power spectrum of the {\tt GNILC} residual map is below the {\it Planck} CIB best-fit power spectrum, which means that some amount of residual CIB contamination is still left in the {\tt GNILC} dust maps at $545$ and $857$\,GHz.

\subsection{Thermal dust temperature and spectral index}
\label{subsec:fitting}

\begin{figure*}
\begin{center}
\includegraphics[width=0.5\textwidth]{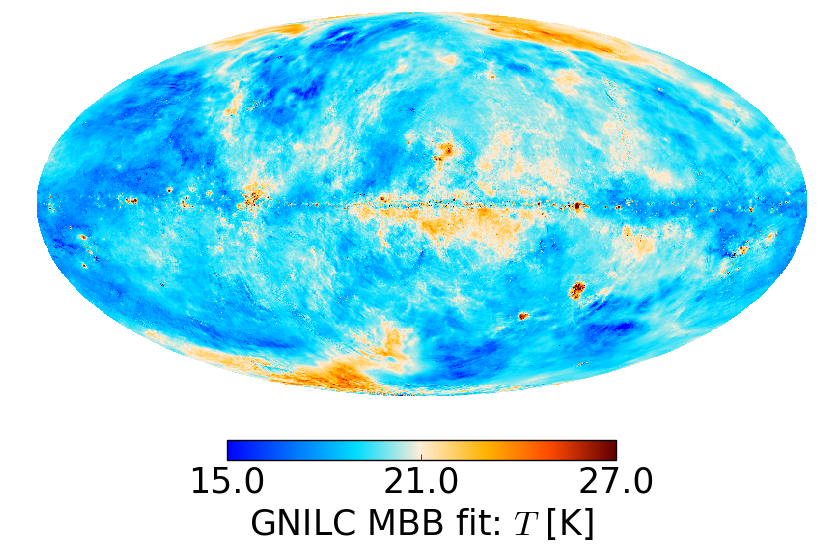}~
\includegraphics[width=0.5\textwidth]{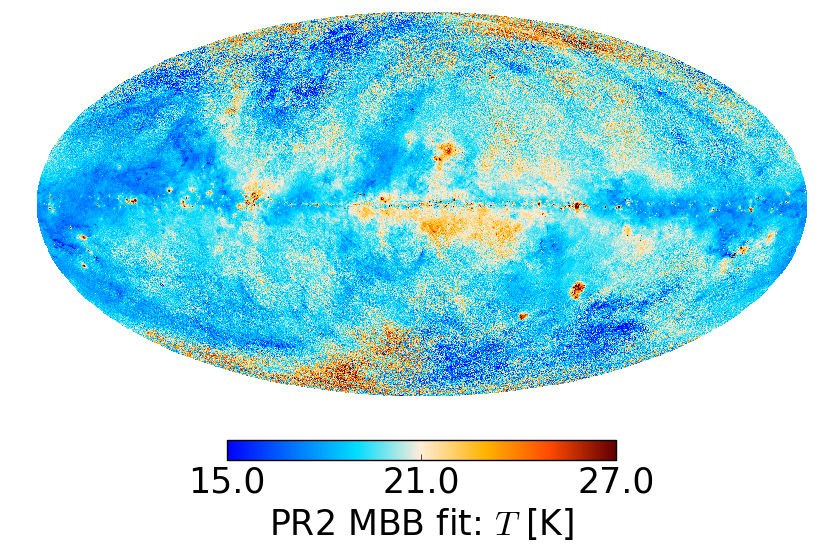}~\\
\includegraphics[width=0.5\textwidth]{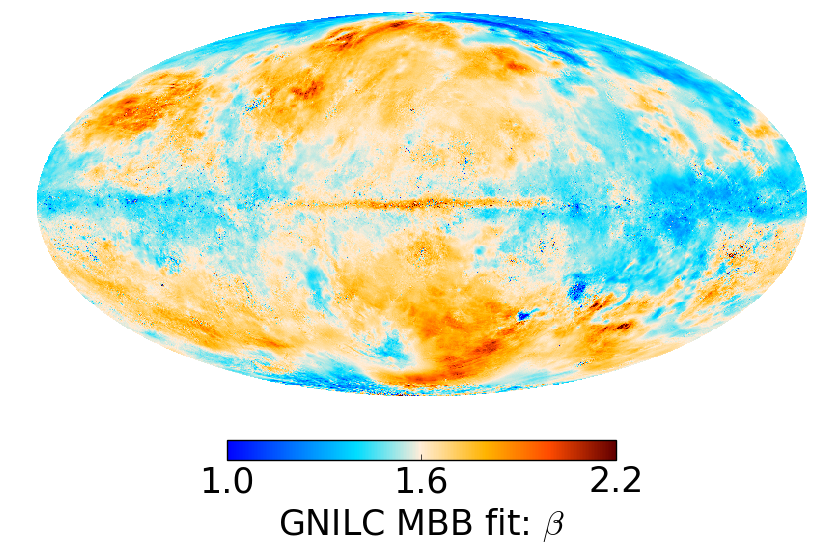}~
\includegraphics[width=0.5\textwidth]{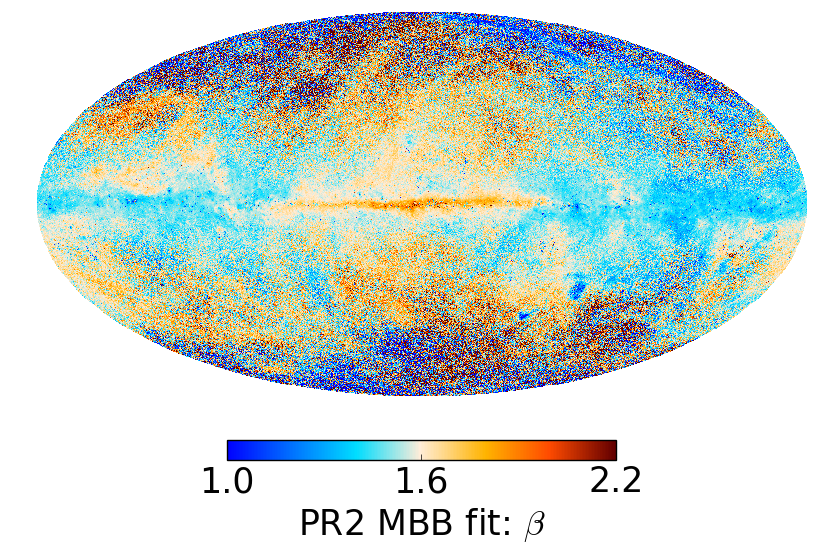}~\\
\includegraphics[width=0.5\textwidth]{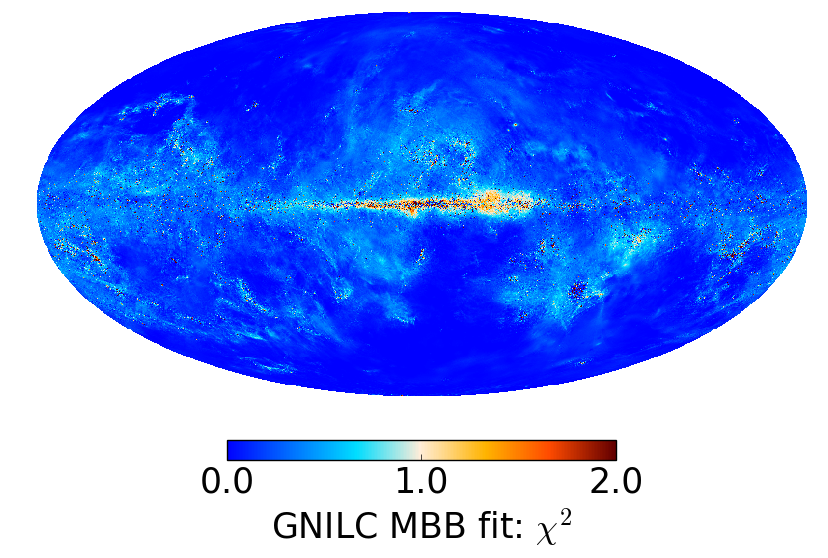}~
\includegraphics[width=0.5\textwidth]{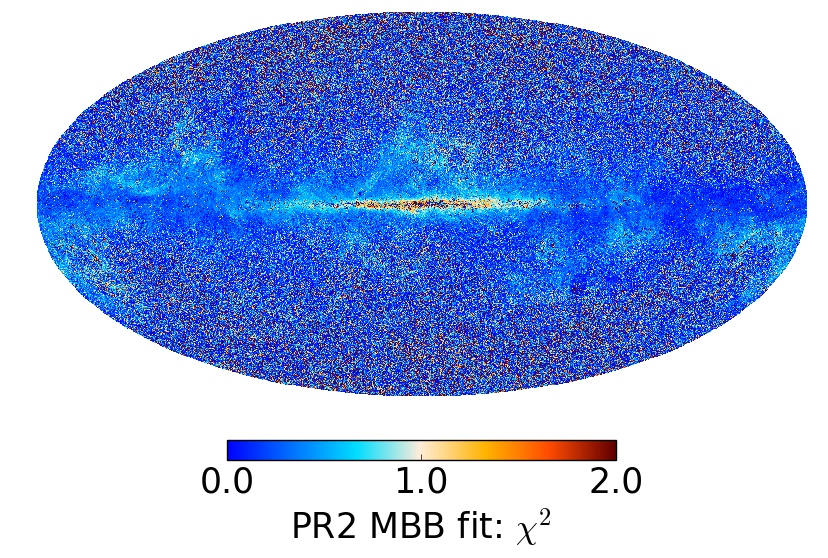}~\\
\end{center}
  \caption{Full-sky thermal dust parameter maps: temperature (\emph{top row}), spectral index (\emph{middle row}), and map of the $\chi^2$ statistic of the fit (\emph{bottom row}). \emph{Left panels}: {\tt GNILC} modified blackbody (MBB) fit. \emph{Right panels}: PR2 modified blackbody (MBB) fit a la model P13.}
  \label{Fig:fitting}
\end{figure*}

 Following \citet{planck2013-p06b}, we fit in each pixel a modified blackbody (MBB) spectral model to the {\tt GNILC} dust maps at $353$, $545$, $857$, and $3000$\,GHz in order to estimate the dust temperature, spectral index, and optical depth over the sky. We also performed an analysis similar to that of \citet{planck2013-p06b}, which used the PR1 data, by fitting the same MBB model to the CMB-removed PR2 maps, in place of the {\tt GNILC} dust maps. We refer to this as the \mbox{PR2 MBB fit}. This allows us to highlight the improvement in estimating the dust temperature and spectral index after filtering out the CIB anisotropies with {\tt GNILC}.

\begin{figure}
\begin{center}
\includegraphics[width=0.25\textwidth]{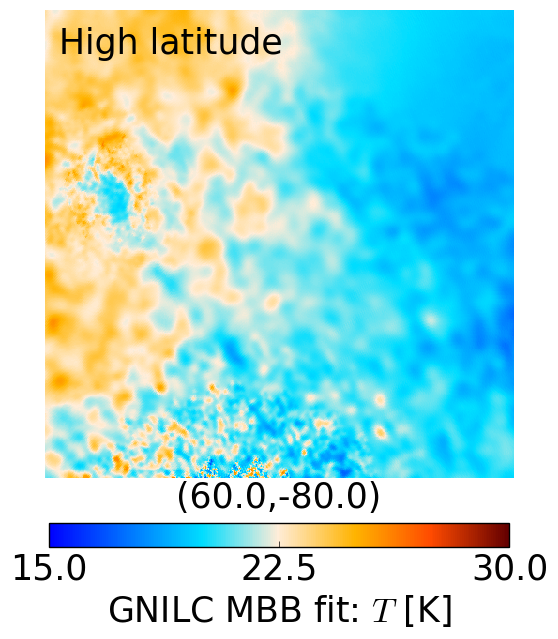}~
\includegraphics[width=0.25\textwidth]{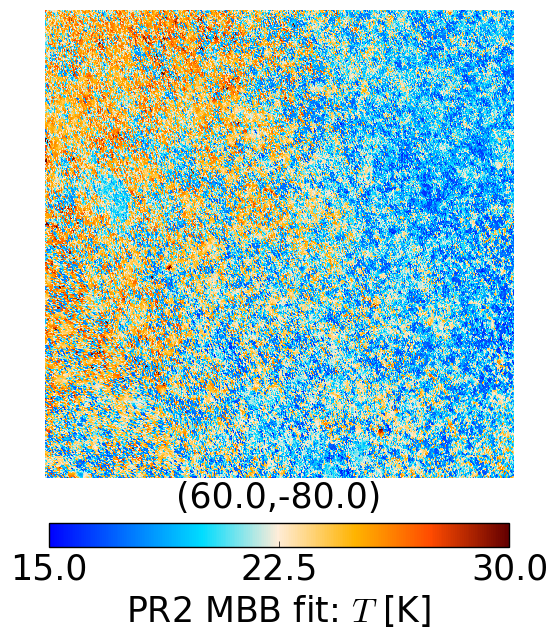}~\\
\includegraphics[width=0.25\textwidth]{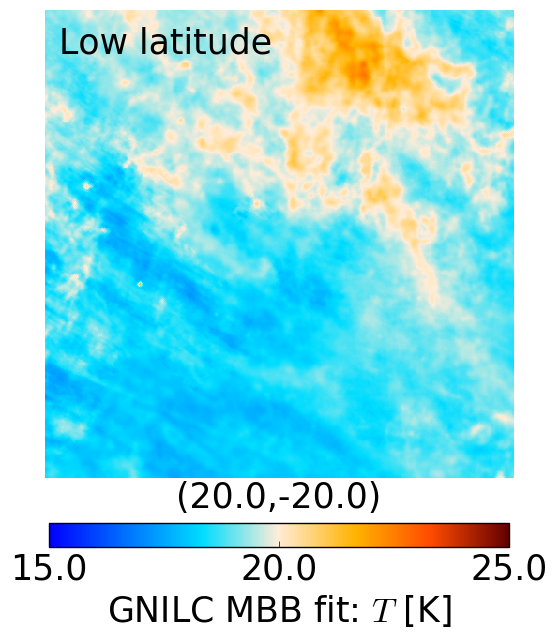}~
\includegraphics[width=0.25\textwidth]{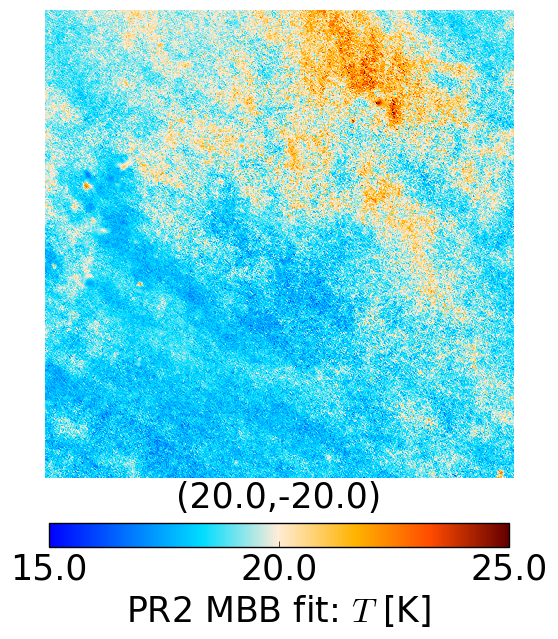}~
\end{center}
  \caption{$12^\circ.5\times 12^\circ.5$ gnomonic projections of the dust temperature maps at high latitude, $b=-80^\circ$ (\emph{top panels}) and low latitude, $b=-20^\circ$ (\emph{bottom panels}). \emph{Left}: {\tt GNILC} MBB fit. \emph{Right}: PR2 MBB fit according to model P13.}
  \label{Fig:fitting2}
\end{figure}
\begin{figure}
\begin{center}
\includegraphics[width=0.25\textwidth]{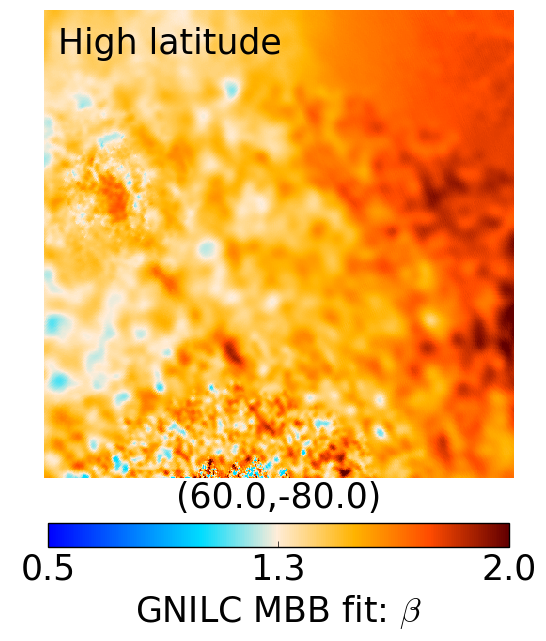}~
\includegraphics[width=0.25\textwidth]{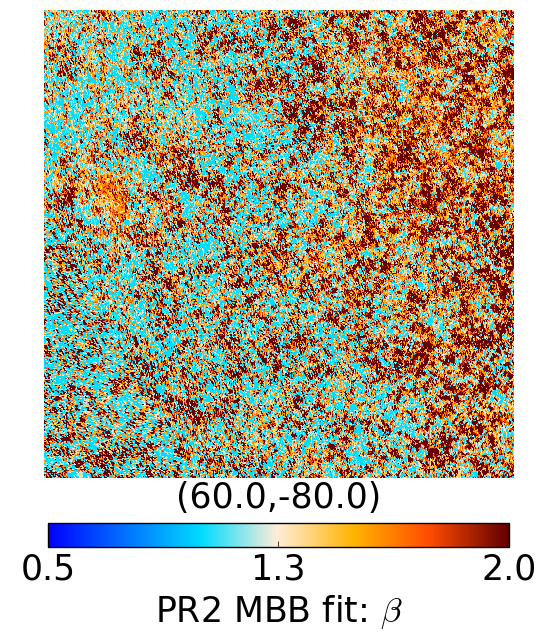}~\\
\includegraphics[width=0.25\textwidth]{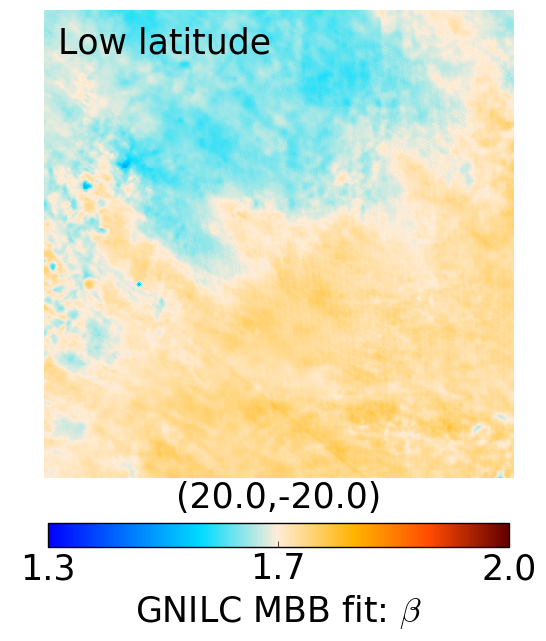}~
\includegraphics[width=0.25\textwidth]{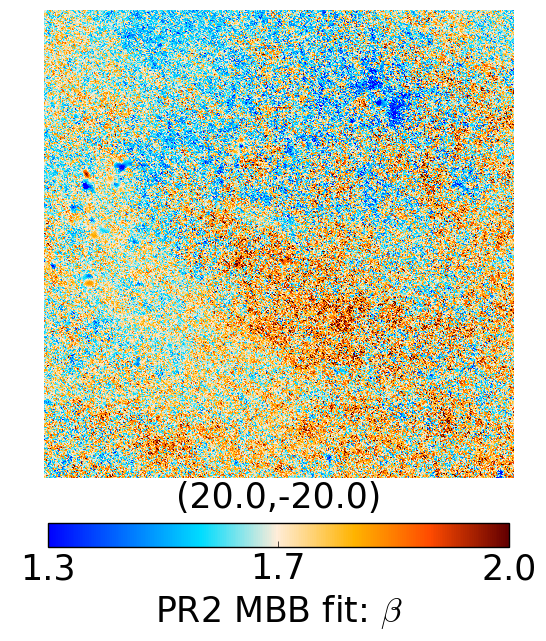}~
\end{center}
  \caption{$12^\circ.5\times 12^\circ.5$ gnomonic projections of the dust spectral index maps at high latitude, $b=-80^\circ$ (\emph{top panels}) and low latitude, $b=-20^\circ$ (\emph{bottom panels}). \emph{Left}: {\tt GNILC} MBB fit. \emph{Right}: PR2 MBB fit according to model P13.}
  \label{Fig:fitting2beta}
\end{figure}

\begin{figure}
\begin{center}
\includegraphics[width=\columnwidth]{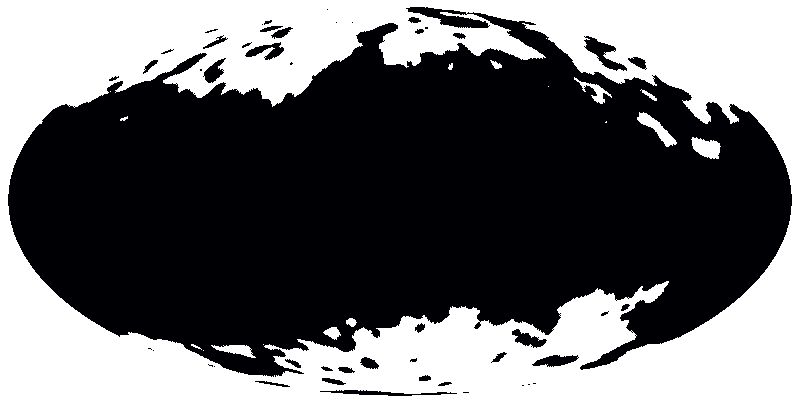}~\\
\includegraphics[width=\columnwidth]{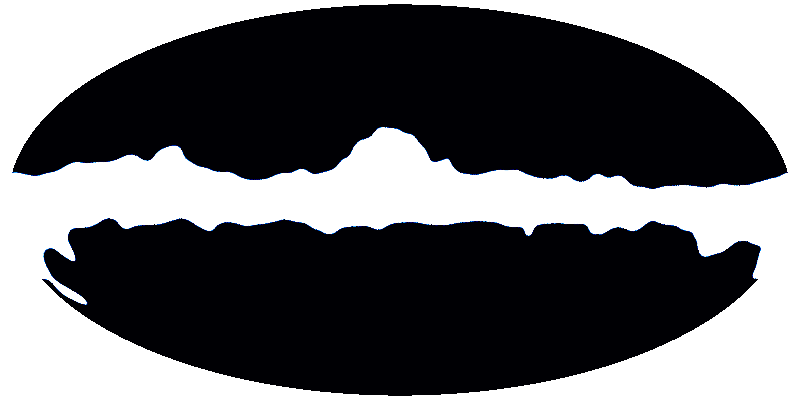}~
\end{center}
  \caption{The high-latitude area of the sky with $f_{\rm sky}=21\%$ (\emph{top}) and the low-latitude area of the sky with $f_{\rm sky}=20\%$ (\emph{bottom}) that are considered in Fig.~\ref{Fig:hist} and Table~\ref{tab:fit}.}
  \label{Fig:regions}
\end{figure}

\begin{table}
\begingroup
\newdimen\tblskip \tblskip=5pt
\caption{Mean and dispersion of the dust temperature and spectral index in different areas of the sky (full sky, high latitude, low latitude). \emph{Top}: {\tt GNILC} MBB fit.  \emph{Middle}: PR2 MBB fit a la dust model P13.  \emph{Bottom}: dust model P15 ({\tt Commander} $60'$).}                          
\label{tab:fit}                            
\nointerlineskip
\vskip -3mm
\footnotesize
\setbox\tablebox=\vbox{
   \newdimen\digitwidth 
   \setbox0=\hbox{\rm 0} 
   \digitwidth=\wd0 
   \catcode`*=\active 
   \def*{\kern\digitwidth}
   \newdimen\signwidth 
   \setbox0=\hbox{+} 
   \signwidth=\wd0 
   \catcode`!=\active 
   \def!{\kern\signwidth}
\halign{\tabskip 0em\hbox to 2.0cm{#\leaderfil}\tabskip 1em&
     \hfil#\hfil \tabskip 1em&
     \hfil#\hfil \tabskip 1em&
     \hfil#\hfil \tabskip 1em&
     \hfil#\hfil \tabskip 1em&
     \hfil#\hfil \tabskip 0em\cr                        
\noalign{\doubleline}
 Area   & $f_{\rm sky}$ & $\langle T_{\tt GNILC} \rangle$  &   $\sigma\left(T_{\tt GNILC}\right)$   &    $\langle \beta_{\tt GNILC} \rangle$  &   $\sigma\left(\beta_{\tt GNILC}\right)$  \cr                                 
\omit& [\%] & [K]  &   [K]  &  \omit   & \omit \cr 
\noalign{\vskip 3pt\hrule\vskip 5pt}
Full sky    &   100      &   19.40       &    1.26     &   1.60      &    0.13\cr
High latitude &   21      &   19.41       &    1.54     &   1.63      &    0.17\cr
Low latitude    &   20      &   19.19       &    1.49     &   1.54      &    0.11\cr     
\noalign{\vskip 5pt\hrule\vskip 3pt}
\noalign{\doubleline}
Area   & $f_{\rm sky}$ & $\langle T_{\tt PR2} \rangle$  &   $\sigma\left(T_{\tt PR2}\right)$   &    $\langle \beta_{\tt PR2} \rangle$  &   $\sigma\left(\beta_{\tt PR2}\right)$    \cr
\omit  & [\%] & [K]  &   [K]  &  \omit   &  \omit  \cr
\noalign{\vskip 3pt\hrule\vskip 5pt}
Full sky    &   100      &   19.50       &    1.70     &   1.59      &    0.27\cr
High latitude &   21      &   19.56       &    2.46     &   1.64      &    0.45\cr
Low latitude    &   20      &   19.18      &    1.50     &   1.55      &    0.10\cr
\noalign{\vskip 5pt\hrule\vskip 3pt}
\noalign{\doubleline}
Area   & $f_{\rm sky}$ & $\langle T_{\tt P15} \rangle$  &   $\sigma\left(T_{\tt P15}\right)$   &    $\langle \beta_{\tt P15} \rangle$  &   $\sigma\left(\beta_{\tt P15}\right)$    \cr
\omit  & [\%] & [K]  &   [K]  &   \omit  & \omit   \cr
\noalign{\vskip 3pt\hrule\vskip 5pt}
Full sky    &   100      &   20.93       &    2.25     &   1.54      &    0.05\cr
High latitude &   21      &   23.25       &    1.67     &   1.55      &    0.05\cr
Low latitude    &   20      &   18.63       &    1.96     &   1.57      &    0.05\cr
\noalign{\vskip 5pt\hrule\vskip 3pt}}}
\endPlancktable                    
\endgroup
\end{table}                        

\begin{figure*}
\begin{center}
\includegraphics[width=\columnwidth]{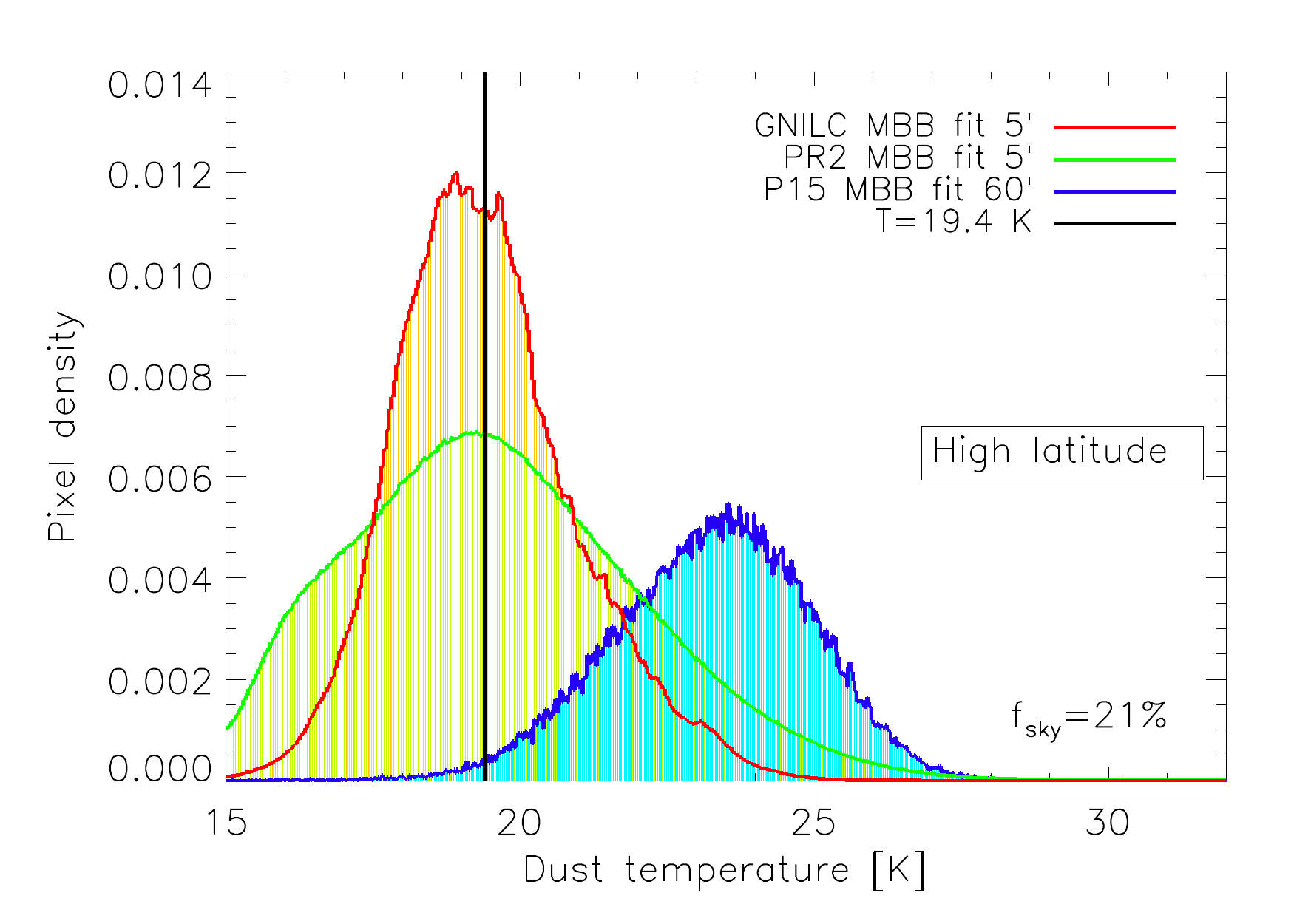}~
\includegraphics[width=\columnwidth]{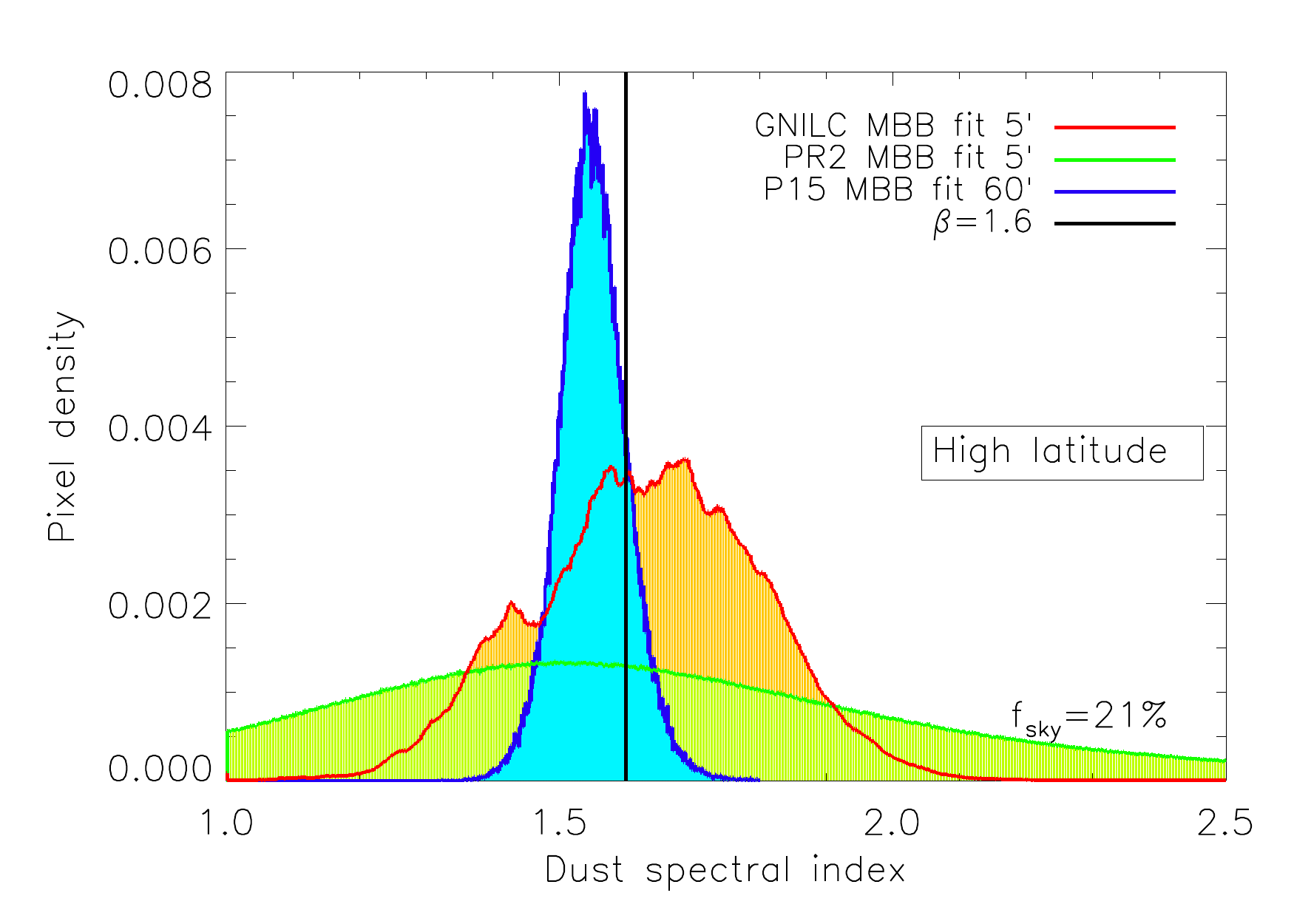}~\\
\includegraphics[width=\columnwidth]{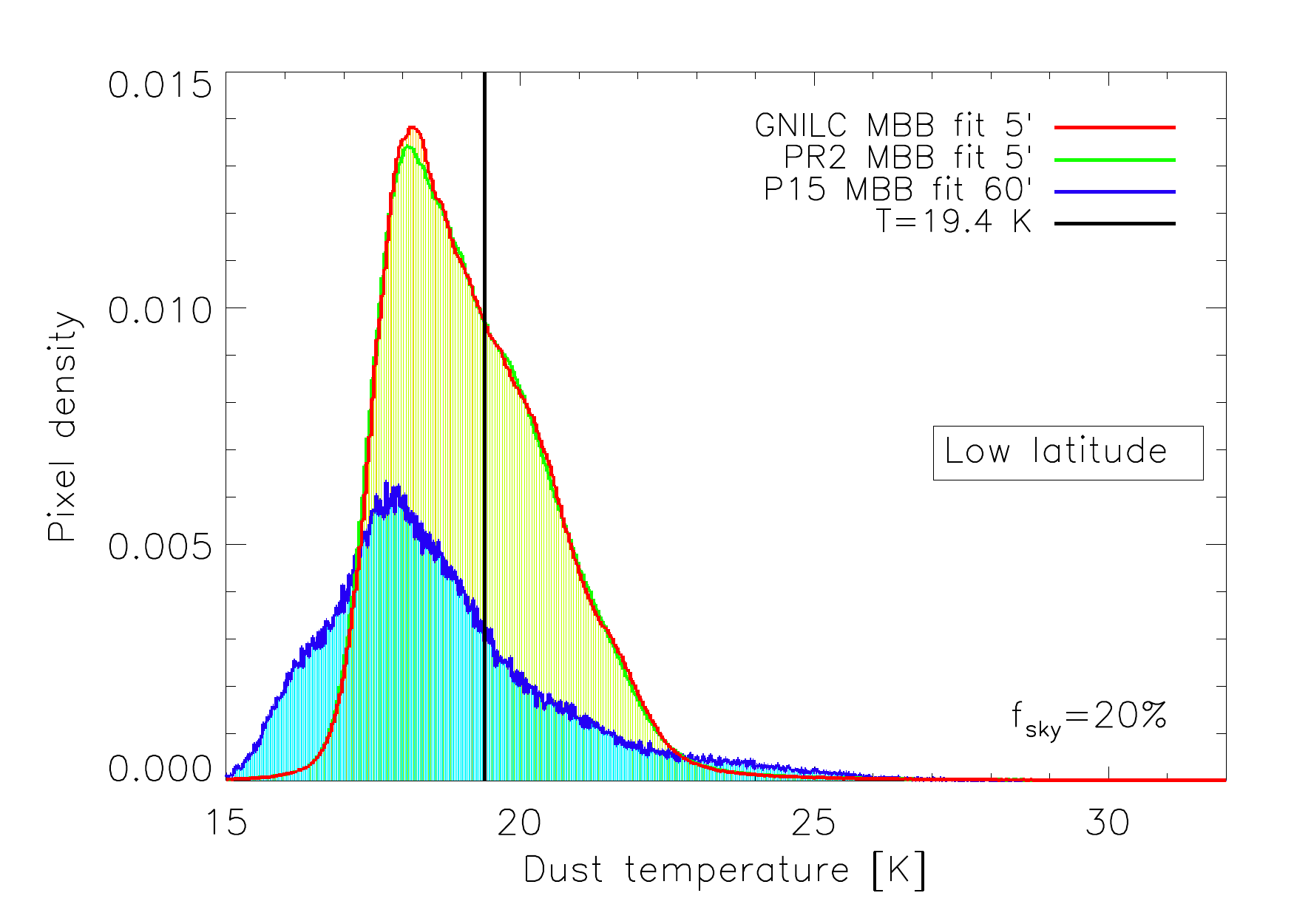}~
\includegraphics[width=\columnwidth]{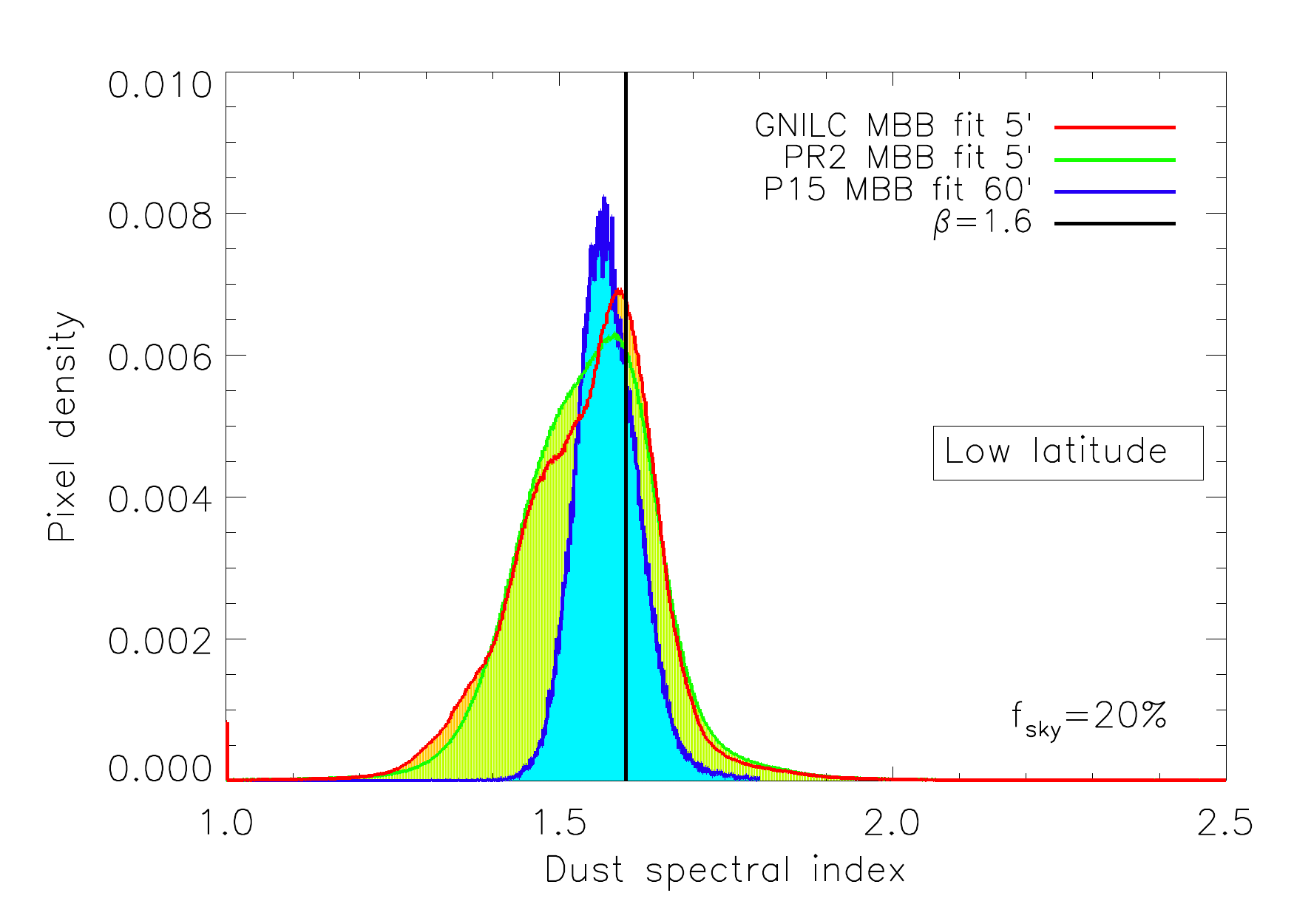}~\\
\includegraphics[width=\columnwidth]{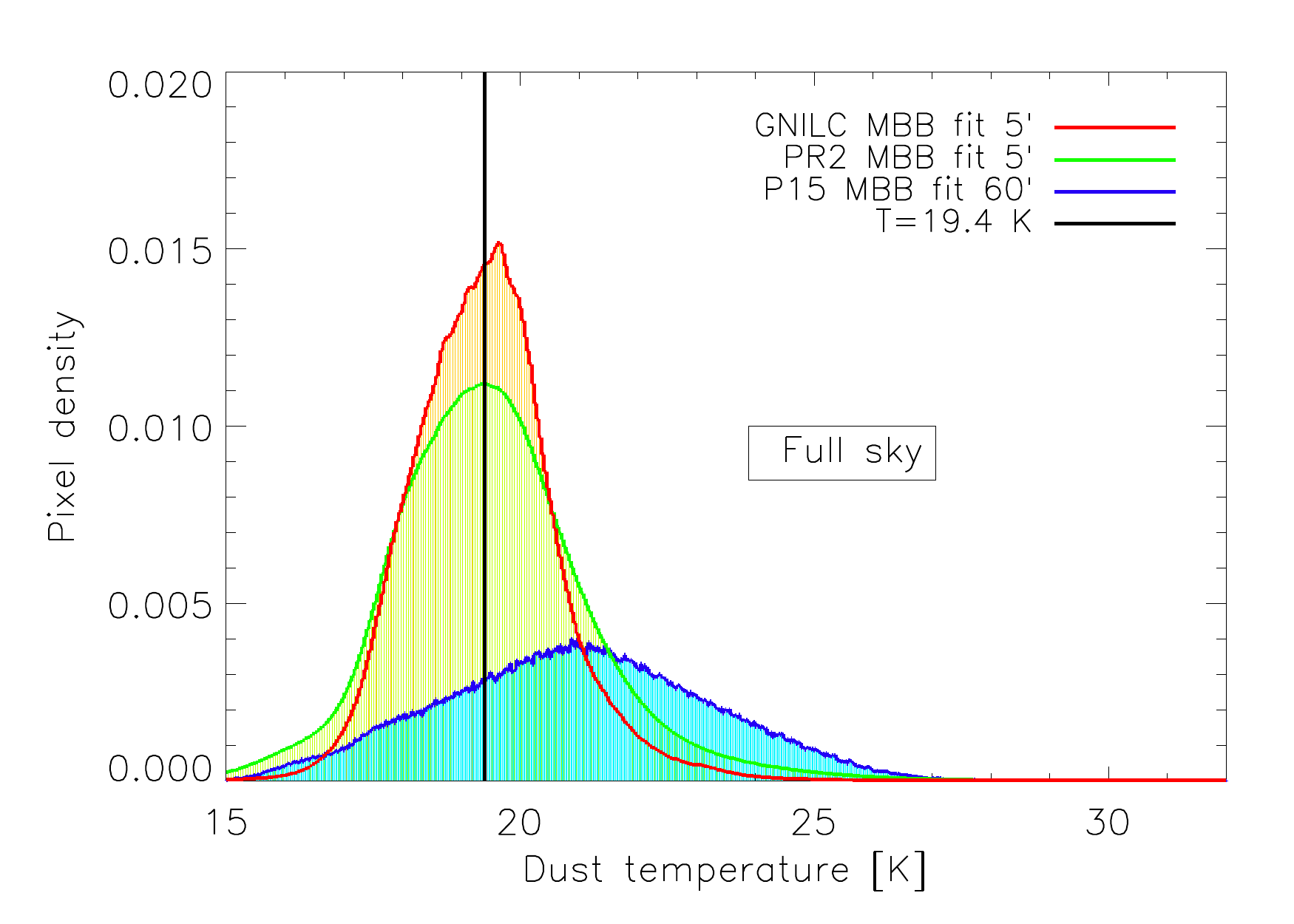}~
\includegraphics[width=\columnwidth]{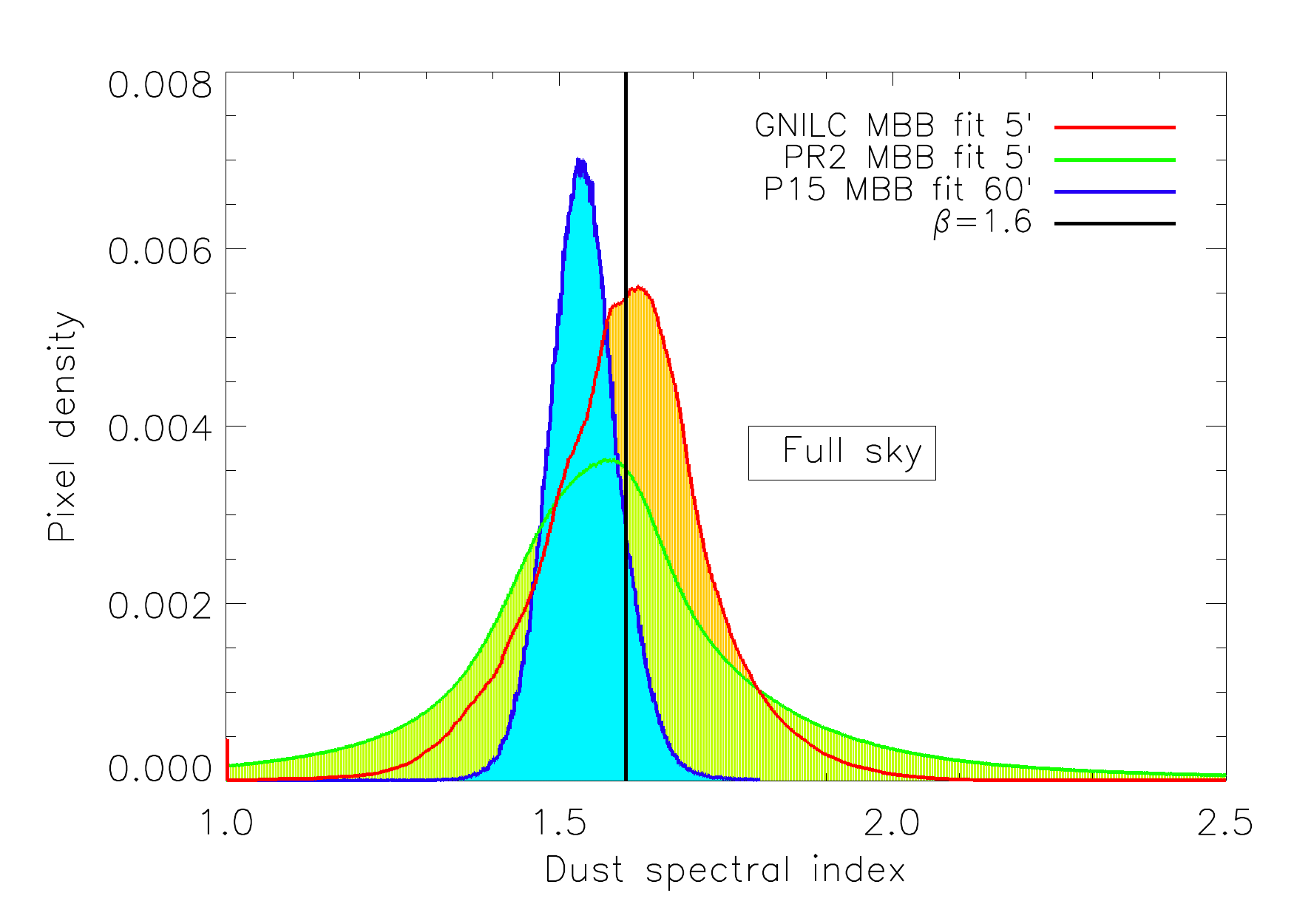}~
\end{center}
  \caption{Normalized histograms of $T_{\rm dust}$ and $\beta_{\rm dust}$ at $5'$ resolution for the {\tt GNILC} MBB fit (red contours) and the PR2 MBB fit a la model P13 (green contours). The normalized histograms for the dust model P15 ({\tt Commander} fit at $60'$ resolution) are overplotted (blue contours). The histograms are computed from the subset of pixels corresponding to either the high-latitude area in the sky with $f_{\rm sky}=21\%$ (\emph{upper panels}), the low-latitude area in the sky with $f_{\rm sky}=20\%$ (\emph{middle panels}), or the whole sky (\emph{lower panels}). Due to CIB contamination at high-latitude, the PR2 MBB fits show larger dispersion than the {\tt GNILC} MBB fits in the distributions of $T_{\rm dust}$ and $\beta_{\rm dust}$.}
  \label{Fig:hist}
\end{figure*}

\begin{figure*}
  \begin{center}
    \includegraphics[width=0.5\textwidth]{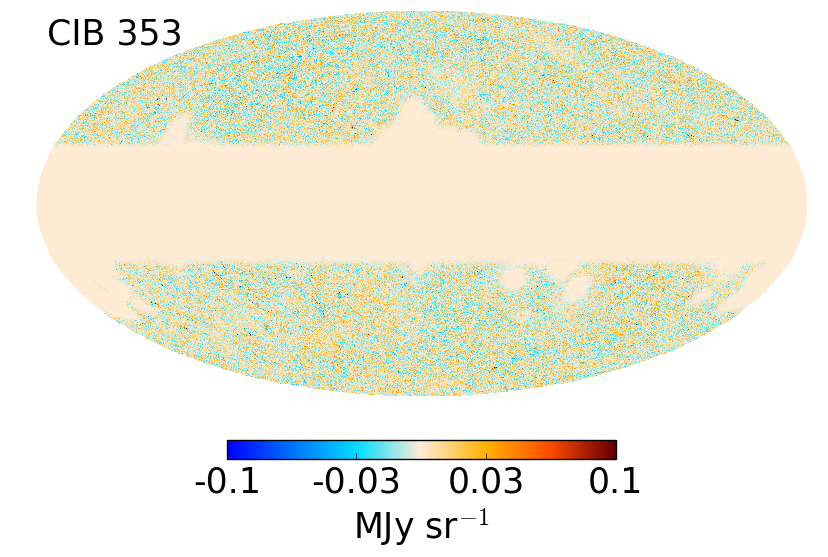}~
    \includegraphics[width=0.5\textwidth]{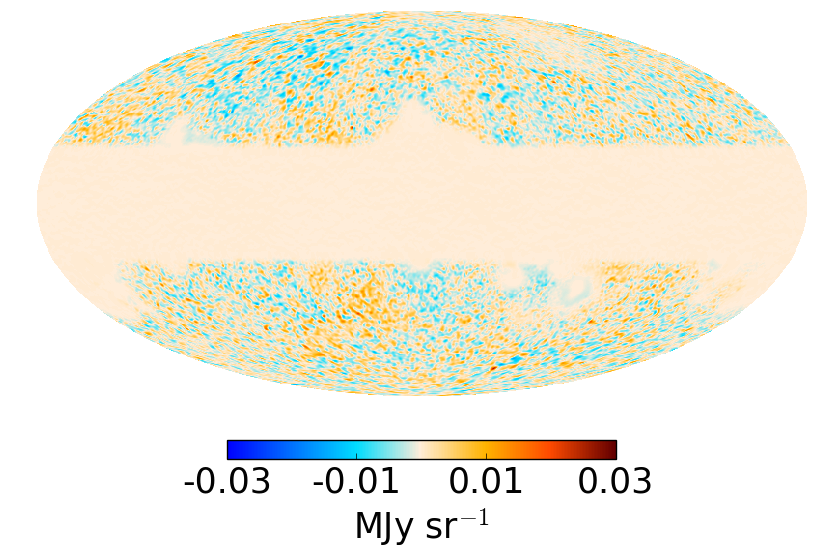}~\\
    \includegraphics[width=0.5\textwidth]{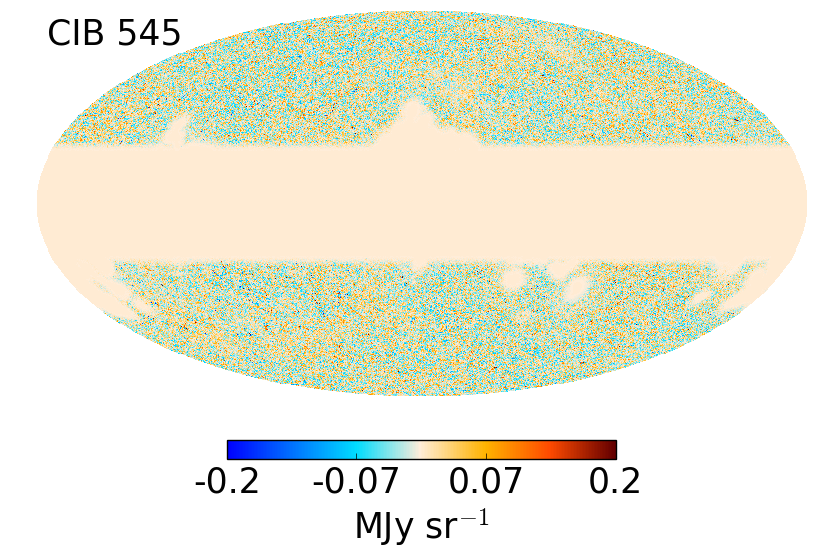}~
    \includegraphics[width=0.5\textwidth]{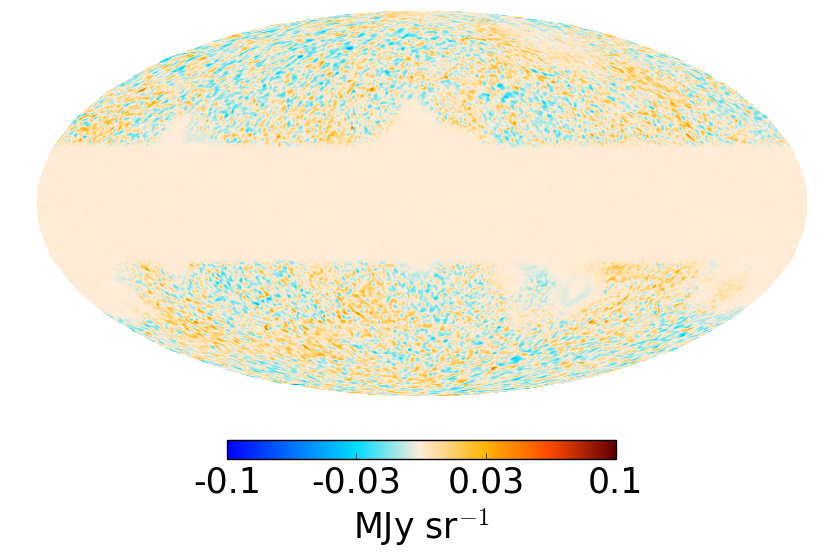}~\\
    \includegraphics[width=0.5\textwidth]{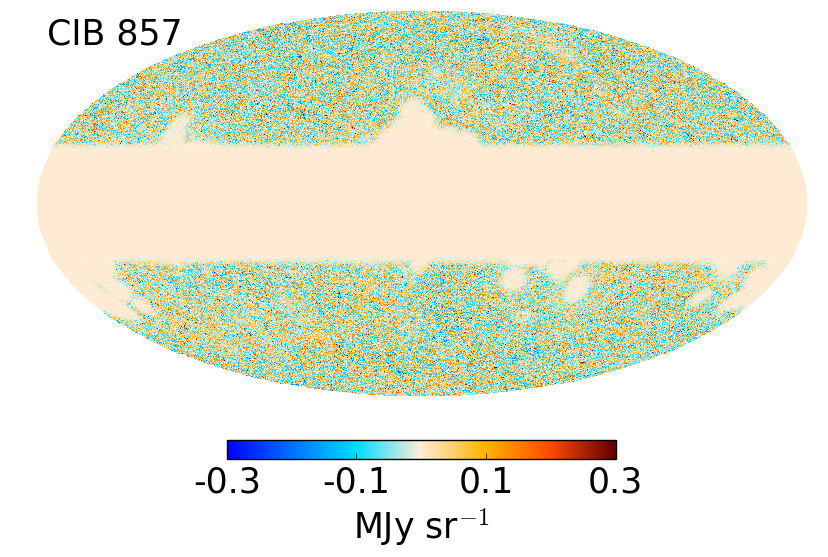}~
    \includegraphics[width=0.5\textwidth]{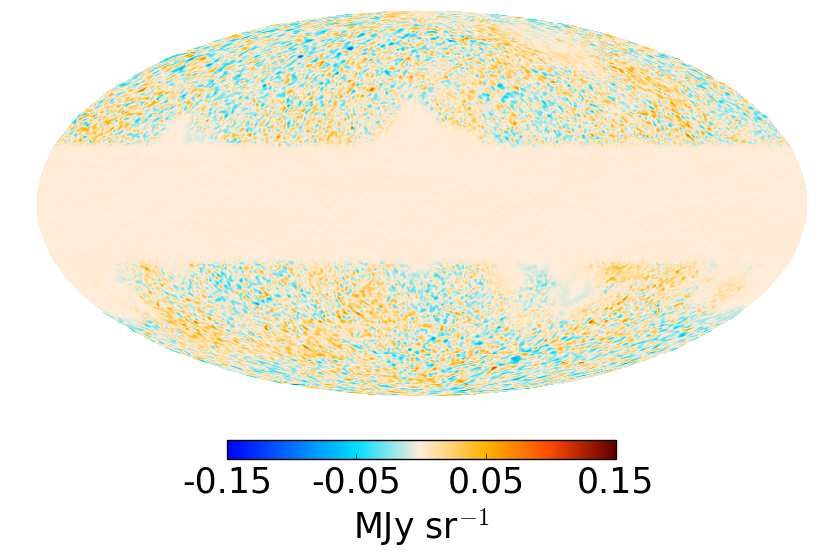}
  \end{center}
  \caption{{\tt GNILC} CIB maps for a large fraction of the sky at $353$\,GHz (\emph{top}), $545$\,GHz (\emph{middle}), and $857$\,GHz (\emph{bottom}). Apart from thermal dust reconstruction, the {\tt GNILC} component-separation method gives access to CIB anisotropies over $57$\,\% of the sky. The left panels show the {\tt GNILC} CIB maps at full resolution while the right panels show the CIB maps smoothed to one degree resolution.}
  \label{Fig:full-sky-cib}
\end{figure*}

\subsubsection{$\chi^2$ fitting  }
\label{subsec:algorithm}

The model of dust emission that we fit to the data is a modified blackbody (MBB) spectrum with three parameters:
\bea
I_\nu(p) = \tau_0(p) \left(\nu/{\nu}_0\right)^{\,\beta(p)} B_\nu\left(T(p)\right),
\eea
where $\nu_0=353$\,GHz is the reference frequency, $\tau_0(p)$ the dust optical depth at $353$\,GHz in pixel $p$, $T(p)$ the dust temperature in pixel $p$, and $\beta(p)$ the dust spectral index in pixel $p$. The function $B_\nu\left(T\right)$ is the Planck law for blackbody radiation.

We use a standard $\chi^2$ fitting method as in \citet{planck2013-p06b}. However, there a two-step approach was adopted for the fit; in order to limit the fluctuations in the estimated parameters induced by the noise and the CIB contamination, the spectral index parameter was estimated at $30'$ resolution in a first step, then the temperature and the optical depth were fit at $5'$ resolution. 
Given that we already have cleaned the thermal dust from CIB contamination at $353$, $545$, $857$, and $3000$ GHz by using the {\tt GNILC} component-separation method, there is no reason to perform a low-resolution MBB fit on the cleaned {\tt GNILC} maps. Therefore, we will fit the three parameters $\tau_0$, $\beta$, and $T$ simultaneously at full resolution ($5'$), instead of following the two-step approach adopted in \citet{planck2013-p06b}. For the fit, we use the frequency data at $353$, $545$, $857$, and $3000$\,GHz, either from the unfiltered PR2 data (i.e., inputs similar to those used to produce the dust model P13) or from the CIB-removed {\tt GNILC} dust maps. In this way, we will highlight the improvement in the estimated dust parameters resulting from the removal of the CIB fluctuations with {\tt GNILC}.

In most of the images presented in this paper the local average of the dust maps has been subtracted to facilitate side-by-side comparisons of the different versions of the \Planck\ dust map in terms of contamination by CIB fluctuations. There is no subtraction of any offset in the released {\tt GNILC} products themselves. In order to fit for the dust spectral parameters, $\tau_0$, $T$, and $\beta$, the offsets of the {\tt GNILC} dust maps have been estimated by correlation with the \hi\ map at high latitude in the exact same way as described in \citet{planck2013-p06b}. The offsets of the {\tt GNILC} dust maps are found to be $0.1248$, $0.3356$, $0.5561$, and $0.1128$\,MJy\,sr$^{-1}$ at $353$, $545$, $857$, and $3000$\,GHz respectively.
The uncertainties on the absolute calibration of the \Planck\ channels have been estimated from the observation of planets \citep{planck2014-a09}; they are $1.2$\% at $353$\,GHz, $6.3$\% at $545$\,GHz, and $6.1$\% at $857$\,GHz. The calibration incertainty at $3000$\,GHz is $13.5$\% \citep{mamd2005}.
Calibration uncertainties and offset uncertainties have been taken into account in the $\chi^2$ fitting, following the procedure detailed in the appendix of \citet{planck2013-p06b}.

\begin{figure*}
\begin{center}
\includegraphics[width=0.33\textwidth]{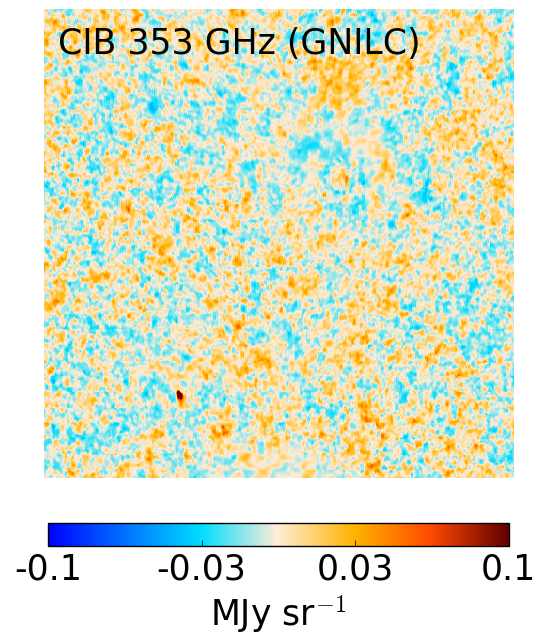}~
\includegraphics[width=0.33\textwidth]{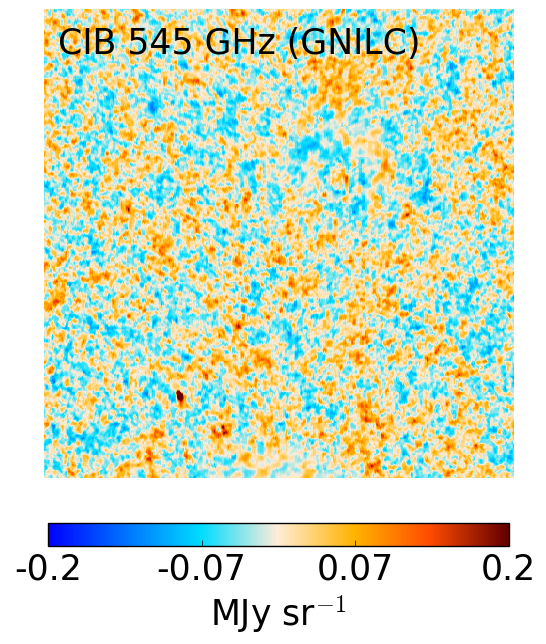}~
\includegraphics[width=0.33\textwidth]{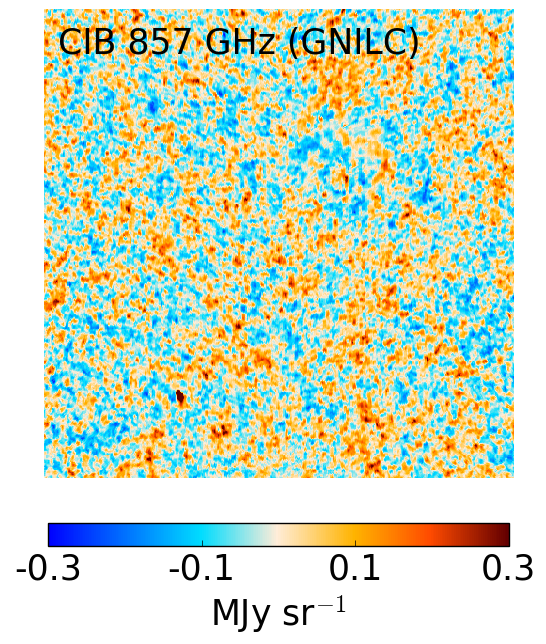}
\end{center}
  \caption{{\tt GNILC} CIB maps at $353$\,GHz ({\it left}), $545$\,GHz ({\it middle}), and $857$\,GHz ({\it right}) on a $12^\circ.5\times 12^\circ.5$ gnomonic projection of the sky centred at high latitude, $(l,b)=(90^\circ,-80^\circ)$. The partial spatial correlation of the CIB anisotropies between pairs of frequencies is clearly visible. In this figure, the local mean intensity of each map has been subtracted.}
  \label{Fig:cib}
\end{figure*}

\subsubsection{Parameter maps  }
\label{subsec:params}

The results of the MBB fit to the {\tt GNILC} dust maps are shown on the left panels of Fig.~\ref{Fig:fitting}. The estimated {\tt GNILC} temperature map and {\tt GNILC} spectral index map are compared to the PR2 MBB fit temperature map and the PR2 MBB fit spectral index map at $5'$ resolution. The PR2 MBB fit is similar to the dust model P13 of \citet{planck2013-p06b}, i.e., the CIB anisotropies have not been filtered out, except that the model fitting is applied to the PR2 data instead of the PR1 data and not performed in two steps, but carried out simultaneously for the three dust parameters at $5'$ resolution. The impact of the CIB contamination on the measurement of the dust temperature and dust spectral index is particularly significant at high latitude in the PR2 MBB fit.

 In the bottom panels of Fig.~\ref{Fig:fitting}, we plot the resulting $\chi^2$ map of both the {\tt GNILC} MBB fit and the PR2 MBB fit. This provides a direct measurement of the goodness-of-fit. The reason for some reduced $\chi^2$ values being smaller than unity is mostly due to the fact that calibration uncertainties are included per pixel in the fit, to give less weight to data points with larger uncertainty \mbox{\citep{planck2013-p06b}}. However, the exact scale of $\chi^2$ is not relevant here, what is important is that the pixel-to-pixel differences in the goodness-of-fit are strongly reduced for the {\tt GNILC} MBB fit because of the removal of the CIB contamination in the {\tt GNILC} dust maps. Clearly, the CIB-filtered {\tt GNILC} maps lead to a better fitting of the MBB model over the sky than the unfiltered PR2 maps. Near the Galactic plane, the $\chi^2$ values between the {\tt GNILC} MBB fit and the unfiltered PR2 MBB fit are consistent because the CIB contamination plays a negligible role where the dust emission is bright. 
For a given spectral model of thermal dust, here a single MBB model, because of the removal of CIB fluctuations prior to fitting, the {\tt GNILC} maps provide higher precision than the unfiltered PR2 maps. However, the MBB model might not be the best parametrization of the thermal dust emission in the inner Galactic plane region, which shows high values of the $\chi^2$ statistic for both PR2 and {\tt GNILC} fits. It is likely that multiple MBB components of dust might contribute to the emission along the line of sight, in which case the exact parametrization of the thermal dust spectral energy distribution in the inner Galactic plane region is not trivial and might need more than three effective parameters.

 In Figs.~\ref{Fig:fitting2} and \ref{Fig:fitting2beta} we compare the {\tt GNILC} and PR2 MBB fit for temperature and spectral index respectively, at low and high latitudes in the sky.
  The improvement from {\tt GNILC} in terms of the reduction of the CIB contamination is particularly visible at high latitude for both temperature and spectral index.

In Fig.~\ref{Fig:regions} we define low- and high-latitude areas of the sky to look at the evolution of the distribution of the dust temperature and spectral index with respect to latitude.  \mbox{Figure \ref{Fig:hist}} shows the normalized histograms of the temperature map, $T$,  and the spectral index map, $\beta$, for three different fits: {\tt GNILC} (red contours); PR2 MBB fit a la model P13 (green contours); and dust model P15 (blue contours). Note that the dust model P15 is a low-resolution Bayesian fit at $60'$ resolution with Gaussian priors on $T$ ($23 \pm 3$ K) and $\beta$ ($1.55 \pm 0.1$).
We distinguish three areas in the sky: the high-latitude area defined in  Fig.~\ref{Fig:regions}, covering $21$\,\% of the sky (\emph{top panels}); the low-latitude area defined in  Fig.~\ref{Fig:regions}, covering $20$\,\% of the sky (\emph{middle panels}); and the whole sky (\emph{bottom panels}). 
As a complement to Fig.~\ref{Fig:hist}, Table~\ref{tab:fit} summarizes the mean best-fit values of the dust parameters, along with their $1\sigma$ errors, for the three different products ({\tt GNILC} MBB fit, PR2 MBB fit similar to the dust model P13, and dust model P15) in the three different areas of the sky.  
  The histograms in Fig.~\ref{Fig:hist} highlight the impact of the CIB anisotropies on the dust spectral parameters: at high latitude the CIB contamination increases the scatter in the temperature and spectral index distributions. The removal of the CIB anisotropies with {\tt GNILC} reduces the dispersion in the dust temperature by $40$\,\% at high latitude ($30$\,\% on the whole sky) with respect to the PR2 MBB fit and by $10$\,\% with respect to the dust model P15 (Table~\ref{tab:fit}), even though the P15 temperature fit is smoothed to $60'$ resolution. The impact of the CIB removal with {\tt GNILC} is even more significant for the dust spectral index, with the dispersion reduced by $60$\,\% at high latitude ($50$\,\% on the whole sky) with respect to the PR2 MBB fit. The $1\sigma$ error on the P15 spectral index has a lowest value of 0.05 in any area of the sky for two reasons: first, the resolution of the P15 spectral index is much lower ($60'$); second, a tight prior has been imposed on $\beta$ in the {\tt Commander} fit \citep{planck2014-a12}. With {\tt GNILC} we find a dust temperature of $T=(19.4\pm 1.3)$\,K and a dust spectral index of $\beta=1.6\pm 0.1$ as the best-fit values on the whole sky, where the error bars show the dispersion over the sky of the parameter values.

\section{{\tt GNILC} results on the CIB}
\label{sec:results2}

\subsection{CIB maps}
\label{subsec:cibmaps}

\begin{figure*}
\begin{center}
\includegraphics[width=0.3\textwidth]{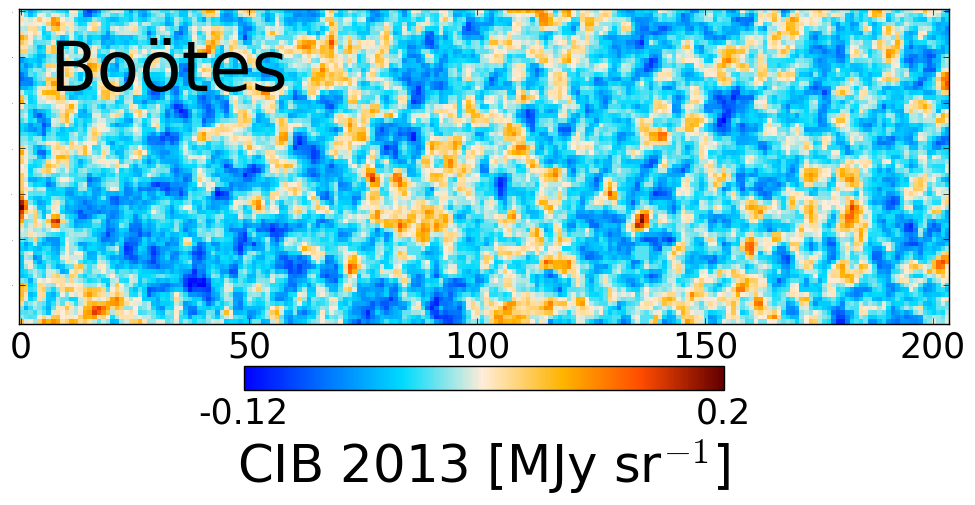}~
\includegraphics[width=0.3\textwidth]{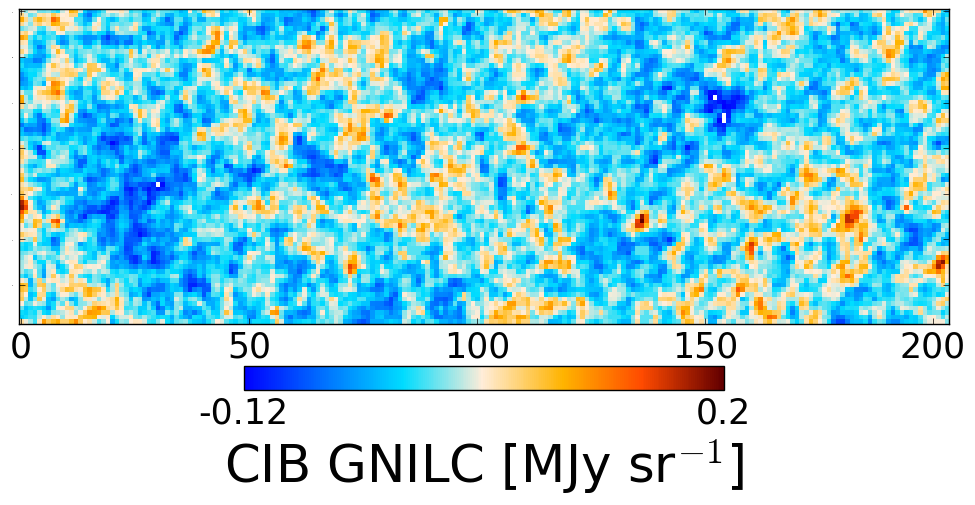}~
\includegraphics[width=0.3\textwidth]{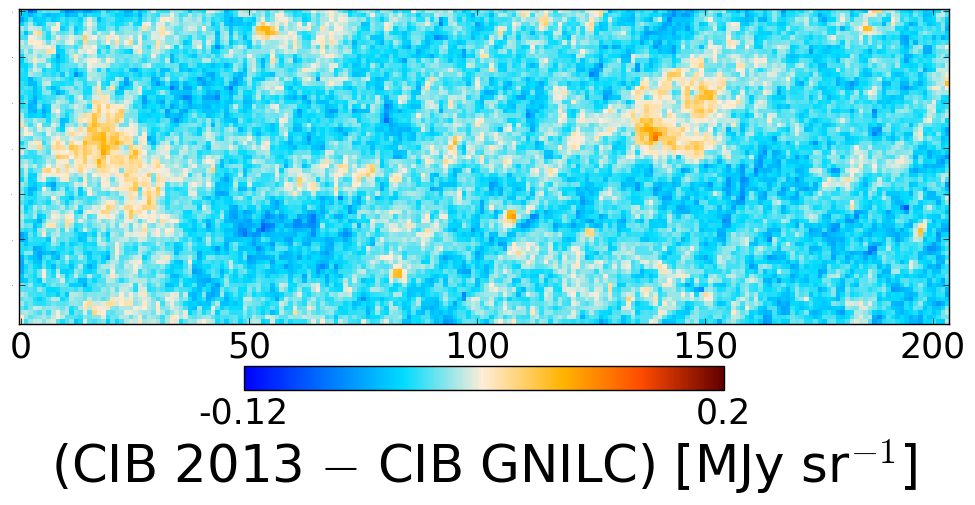}~\\
\includegraphics[width=0.3\textwidth]{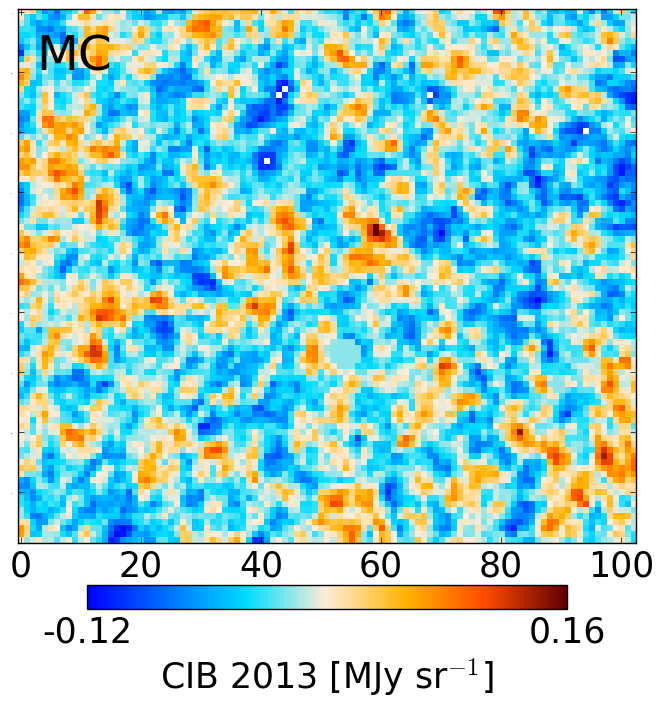}~
\includegraphics[width=0.3\textwidth]{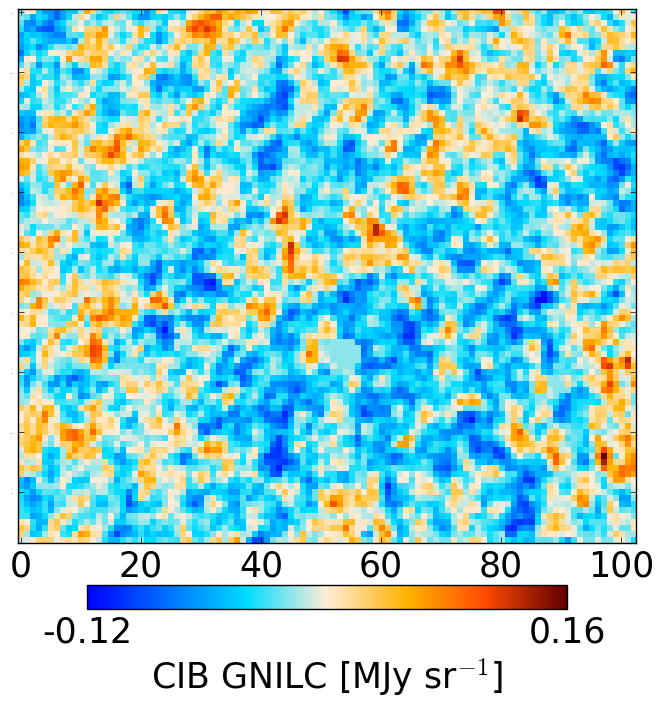}~
\includegraphics[width=0.3\textwidth]{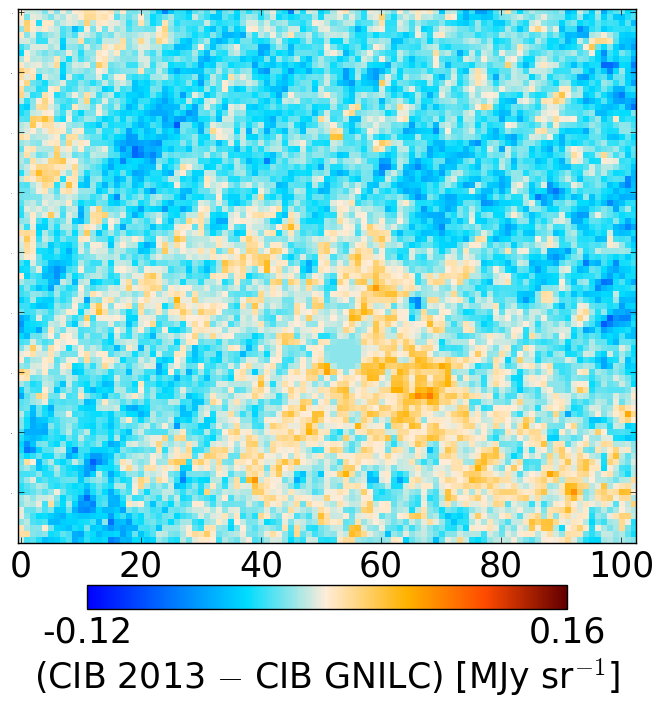}~\\
\includegraphics[width=0.3\textwidth]{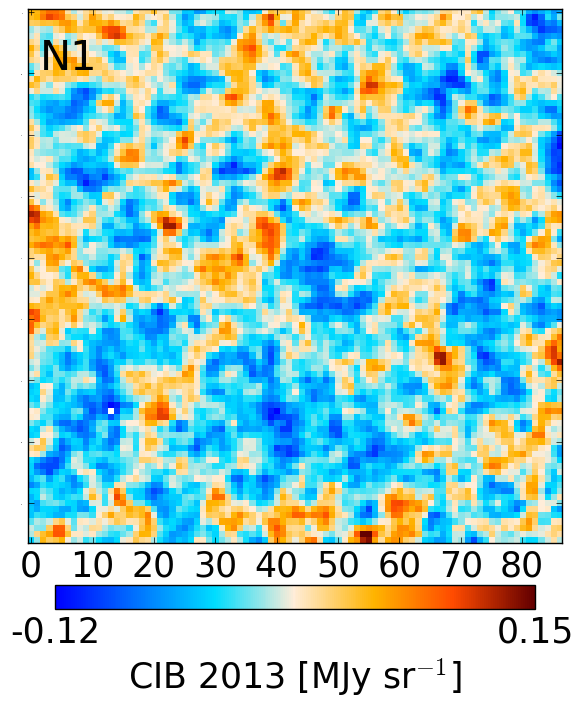}~
\includegraphics[width=0.3\textwidth]{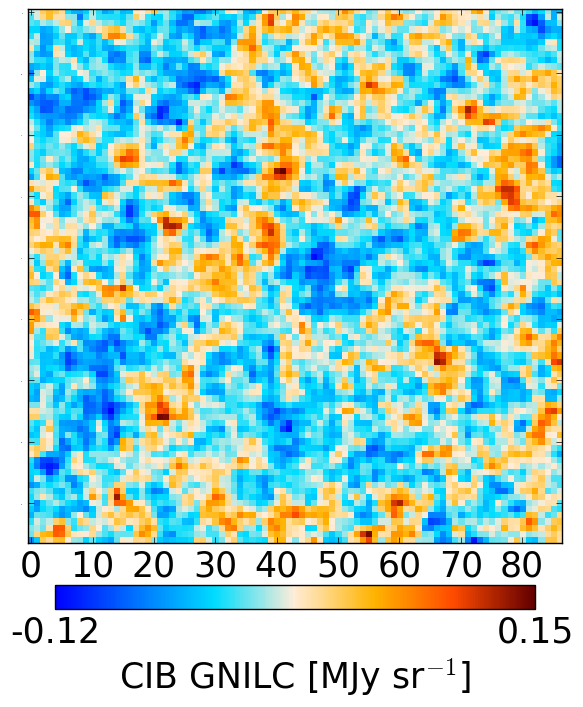}~
\includegraphics[width=0.3\textwidth]{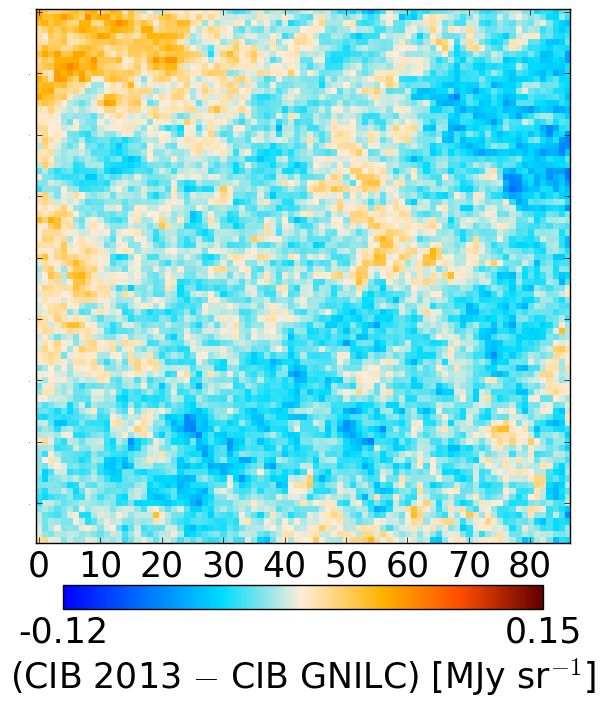}~\\
\includegraphics[width=0.3\textwidth]{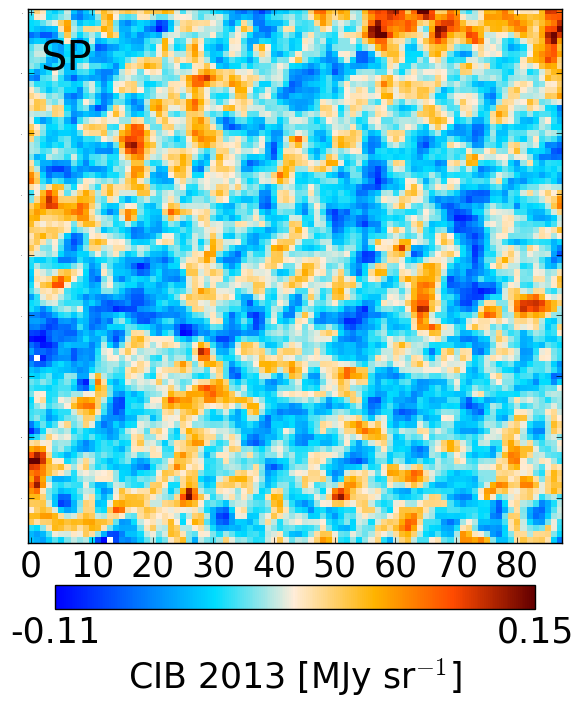}~
\includegraphics[width=0.3\textwidth]{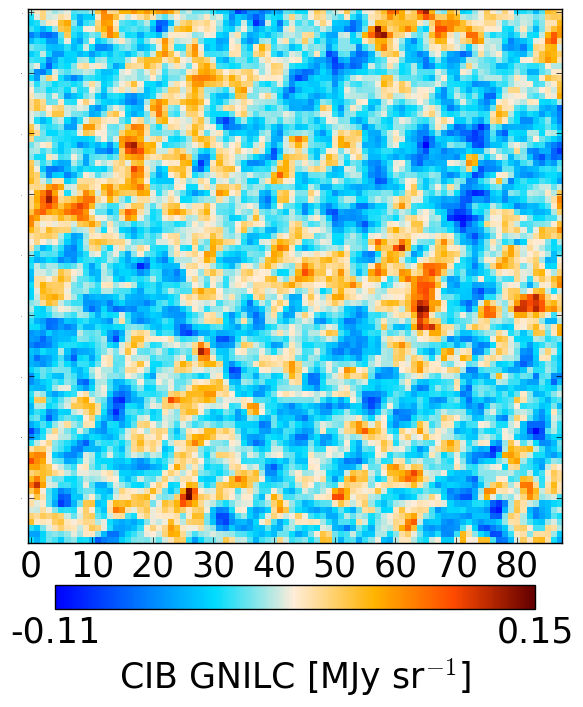}~
\includegraphics[width=0.3\textwidth]{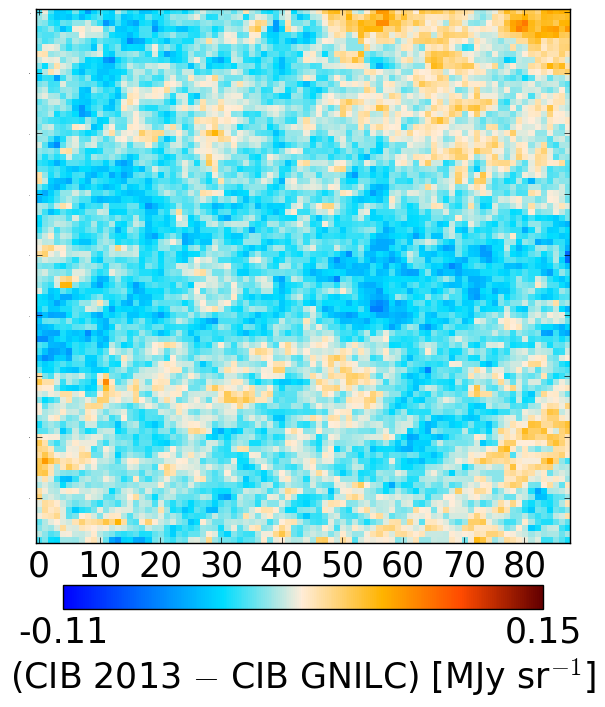}~
\end{center}
  \caption{Maps of cosmic infrared background (CIB) anisotropies in GHIGLS fields \citep{Martin2015}. \emph{Left column}: CIB 2013 \citep{planck2013-pip56}. \emph{Middle column}: CIB {\tt GNILC}. \emph{Right column}: difference (CIB 2013 $-$ GNILC). From top to bottom row, the GHIGLS fields are: Bo{\"o}tes; MC; N1; and SP. The size and the location in the sky of each field are defined in table 1 of \citet{Martin2015}.}
  \label{Fig:cibcomp}
\end{figure*}
\begin{figure*}
\begin{center}
\includegraphics[width=0.9\columnwidth]{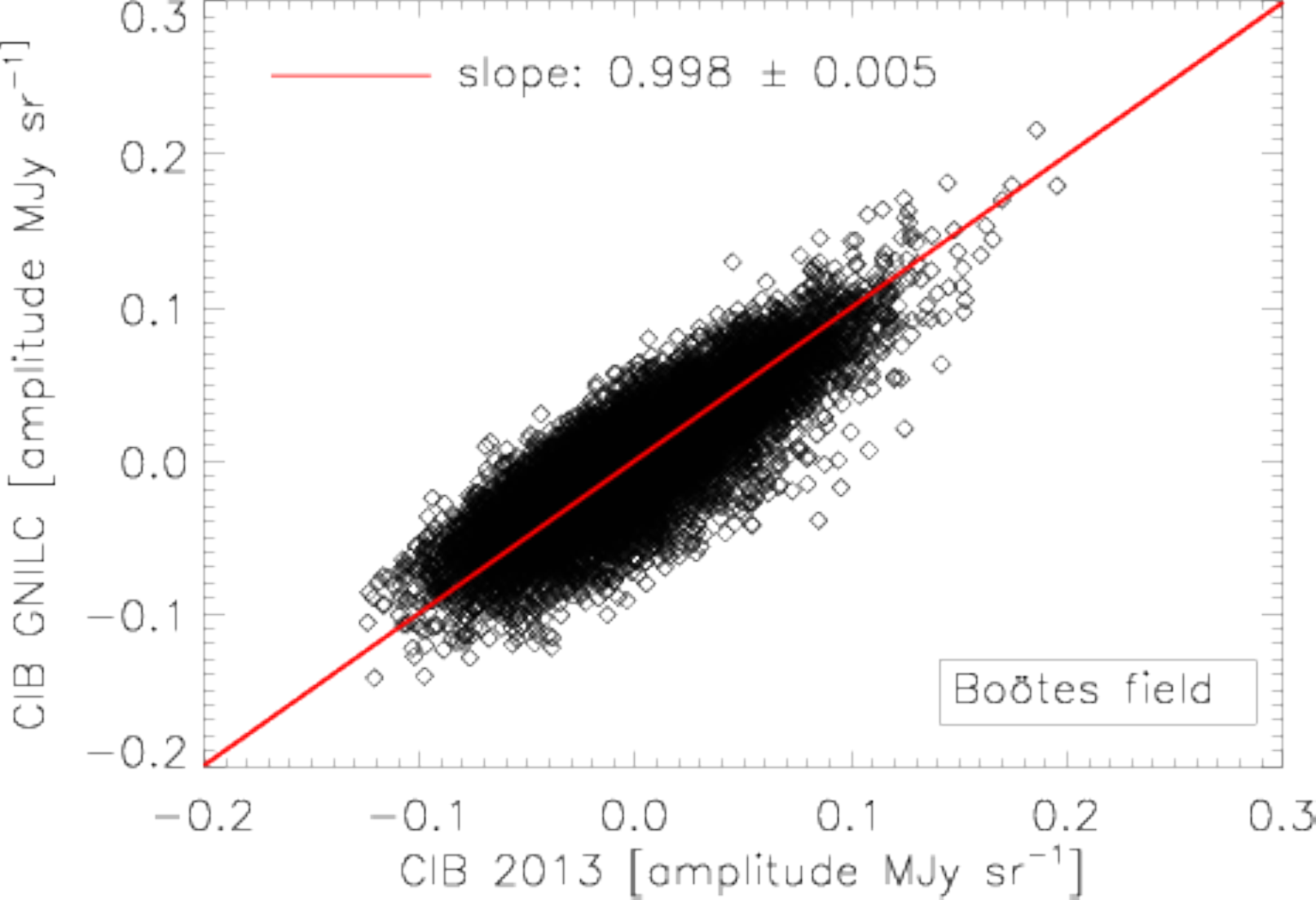}~
\includegraphics[width=0.9\columnwidth]{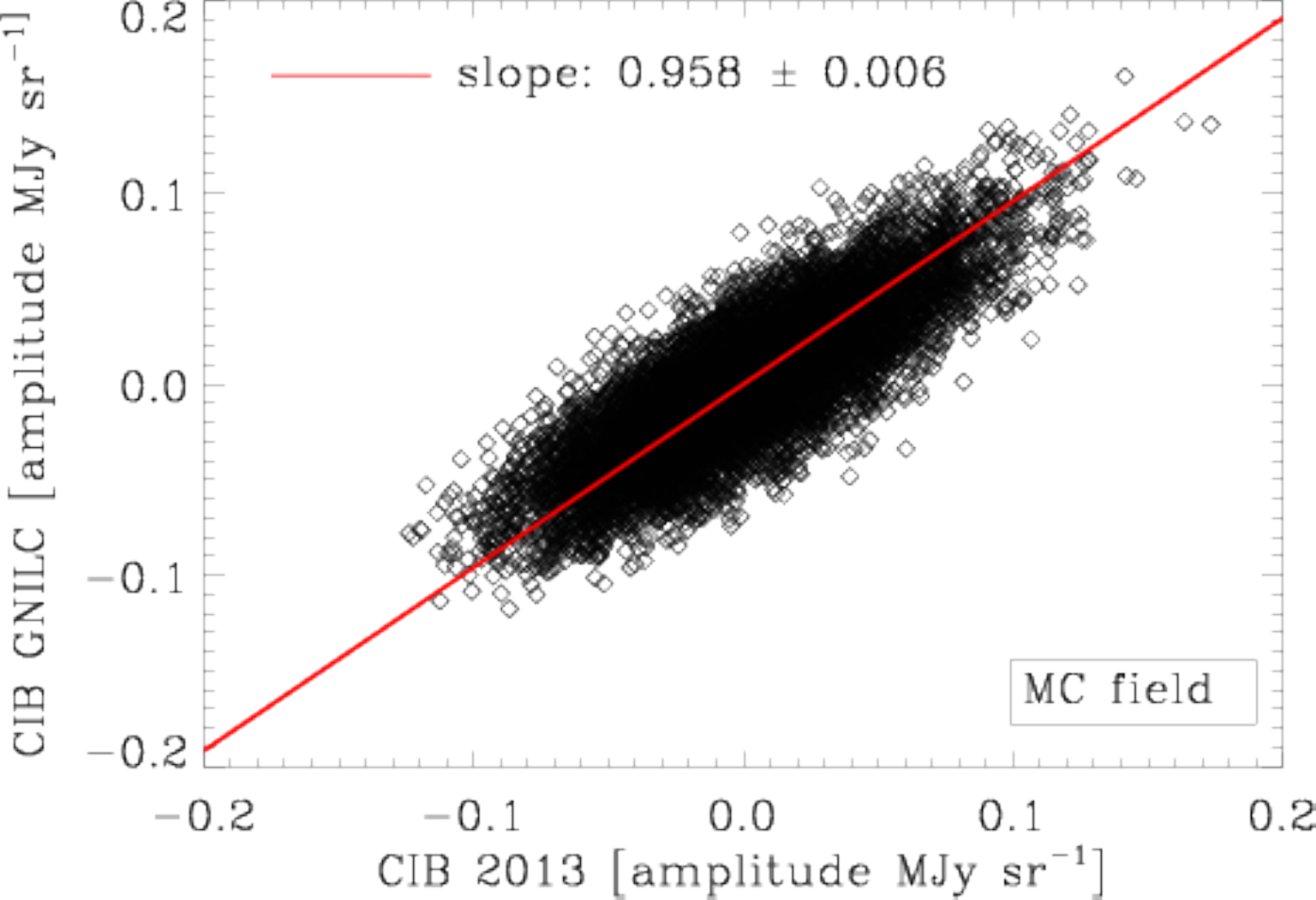}~\\
\includegraphics[width=0.9\columnwidth]{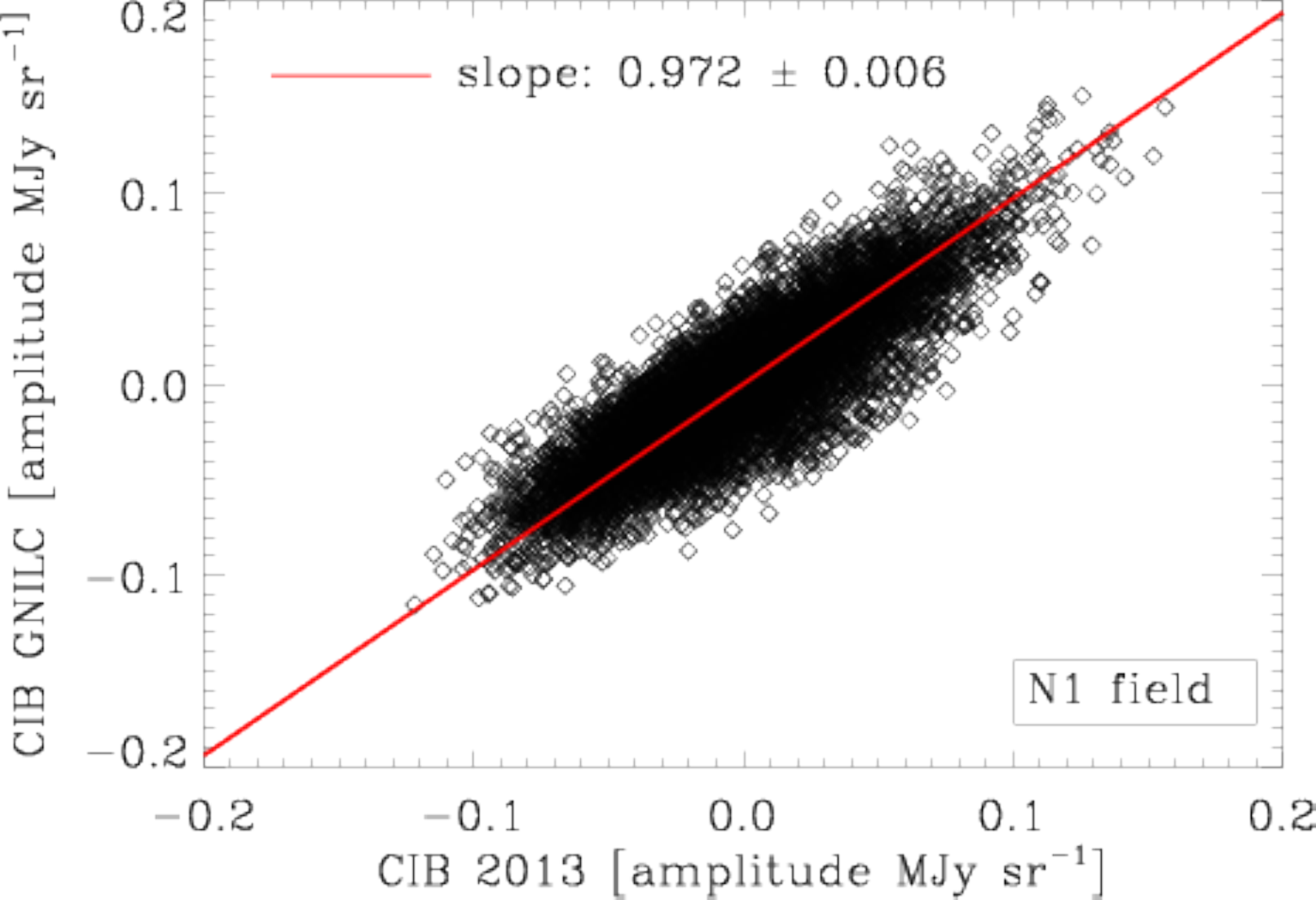}~
\includegraphics[width=0.9\columnwidth]{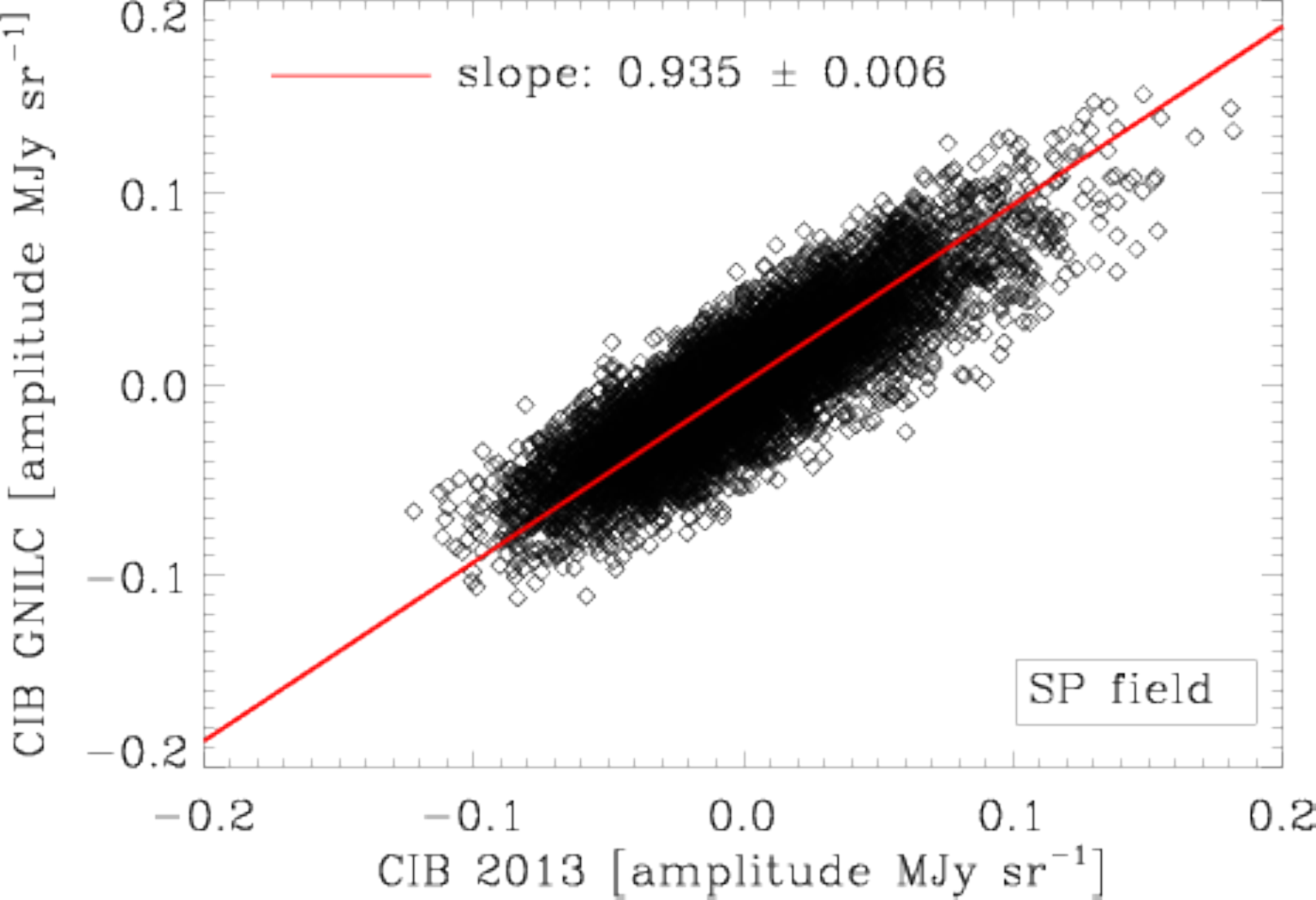}~
\end{center}
  \caption{$T$--$T$ scatter plot between CIB {\tt GNILC} and CIB 2013. The fields are Bo{\"o}tes, MC, N1, and SP.}
  \label{Fig:cibcomptt}
\end{figure*}

The {\tt GNILC} method is flexible by allowing either the recovery of the dust map with the CIB-plus-CMB-plus-noise filtered out (shown in this paper) or the recovery of the dust-plus-CIB map with the removal of the CMB-plus-noise only, depending on whether or not one uses the prior on the CIB power spectrum. Therefore, from the difference between the unfiltered {\tt GNILC} dust-plus-CIB map and the CIB-filtered {\tt GNILC} dust map we are able to reveal the CIB anisotropies at different frequencies over a large area of the sky. The resulting {\tt GNILC} CIB maps at $353$, $545$, and $857$\,GHz, reconstructed over a large fraction of the sky, are shown in Fig.~\ref{Fig:full-sky-cib}, both at full resolution and smoothed to $1^\circ$ resolution. We can see residual zodiacal light emission along the ecliptic plane in the low-resolution CIB maps. This residual comes from the combination by {\tt GNILC} of different data sets, namely the \Planck, IRAS, and SFD maps, in which the zodiacal light has been corrected differently. The zodiacal light emission has been reduced in \Planck\ data to a negligible level compared to the CMB and dust emissions but is still at a level comparable to the amplitude of the CIB emission.

The {\tt GNILC} products give us access to the CIB anisotropies on a much larger fraction of the sky (approximately 57\,\%) than the \Planck\ CIB maps produced  in \citet{planck2013-pip56}. The {\tt GNILC} CIB maps can be used as tracers of the dark matter distribution because they allow for an exploration over large areas of the sky of the cross-correlations between CIB and CMB lensing fields \citep{planck2013-p13} or between CIB and other tracers of large-scale structure \citep{Serra2014}.

In Fig.~\ref{Fig:cib} we show the {\tt GNILC} CIB maps at $353$, $545$, and $857$\,GHz in a high-latitude $12^\circ.5\times 12^\circ.5$ region of the sky centred at $(l,b)=(90^\circ,-80^\circ)$. The partial spatial correlation of the CIB anisotropies between pairs of frequencies decreases when the ratio between the two frequencies is further from unity, as expected from the redshift distribution of the CIB anisotropies \citep{planck2013-pip56}.

\subsection{{\tt GNILC} CIB versus CIB 2013 in small fields  }
\label{subsec:cibcomp}

We now check the consistency between the new {\tt GNILC} CIB maps and the CIB 2013 maps from \citet{planck2013-pip56}. In Fig.~\ref{Fig:cibcomp}, we compare the {\tt GNILC} CIB map at $545$\,GHz with the CIB 2013 map at $545$\,GHz in four different fields of the Green Bank Telescope \hi\ Intermediate Galactic Latitude Survey (GHIGLS) defined in \citet{Martin2015}: Bo{\"o}tes, MC, N1, and SP.
The difference map (CIB 2013 $-$ {\tt GNILC} CIB) is also shown within the same fields. In Fig.~\ref{Fig:cibcomptt} we plot the $T$--$T$ correlation between the {\tt GNILC} CIB map and the CIB 2013 map in the common fields. We have used a least-squares bisector linear regression \citep{Isobe1990} for computing the correlation coefficient between both products.

Figures \ref{Fig:cibcomp} and \ref{Fig:cibcomptt} show that the {\tt GNILC} CIB maps are consistent with the CIB 2013 maps within the fields considered. 
In particular the Pearson correlation coefficient between the {\tt GNILC} CIB maps and the CIB 2013 maps is larger than $0.8$ in all fields, with a $T$--$T$ slope of $0.998\pm 0.005$ in the Bo{\"o}tes field, $0.958\pm 0.006$ in the MC field, $0.972\pm 0.006$ in the N1 field, and $0.935\pm 0.006$ in the SP field. Despite the high correlation between the {\tt GNILC} CIB maps and the CIB 2013 maps within the GHIGLS fields, the correlation is not perfect because both sets of maps were produced from different data releases, respectively PR2 and PR1, for which there have been changes in the calibration coefficients. In addition, estimates of the optical beam resolution have slightly changed between both data releases, therefore not guaranteeing the exact same resolution of both CIB products.

\section{Correlations of the CIB and dust maps with the \hi\ map}
\label{subsec:HI}

In the diffuse interstellar medium the dust emission is tightly correlated with the line emission of neutral hydrogen (\hi). In this respect, as a tracer of the Galactic dust emission the \hi\ emission map can be used to detect any residual Galactic dust emission in the {\tt GNILC} CIB maps. 

\begin{figure}
\begin{center}
\includegraphics[width=1.05\columnwidth]{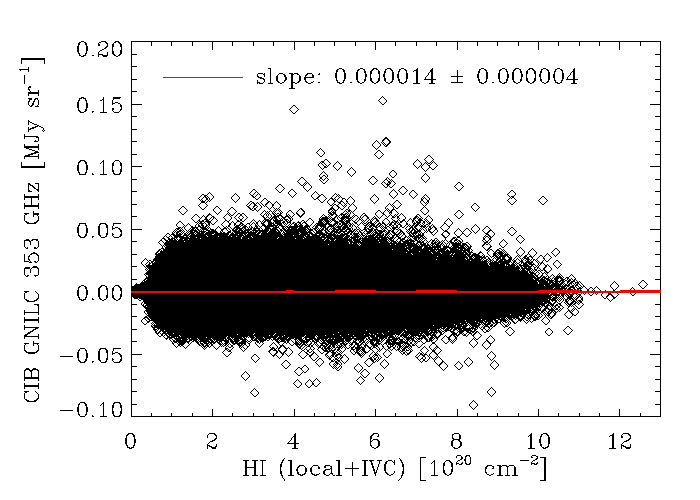}~
\end{center}
  \caption{$T$--$T$ scatter plot between the \hi\ map and the {\tt GNILC} CIB map at $353$\,GHz over 57\,\% of the sky (see Fig.~\ref{Fig:full-sky-cib}).}
  \label{Fig:cib-h1}
\end{figure}

We compute the $T$--$T$ correlation between the \hi\ map (local plus intermediate velocity clouds) of the LAB survey \citep{LAB2005,Land2007} and the {\tt GNILC} CIB maps in order to detect any Galactic residual in the reconstructed CIB maps. Figure \ref{Fig:cib-h1} shows that the correlation between the {\tt GNILC} CIB map at $353$\,GHz and the \hi\ map is consistent with zero (Pearson correlation coefficient of $0.004$), therefore showing no significant residual Galactic emission in the {\tt GNILC} CIB map.

We also compute the $T$--$T$ correlation between the dust map at $353$\,GHz and the \hi\ map at high Galactic latitude.
The high latitude region is defined as the area of the sky where the local beam FWHM of the {\tt GNILC} dust map is larger than $15'$ (Fig.~\ref{Fig:beaming}). The maps are degraded to {\tt HEALPix} $N_{\rm side}=256$.

\begin{figure}
\begin{center}
\includegraphics[width=1.025\columnwidth]{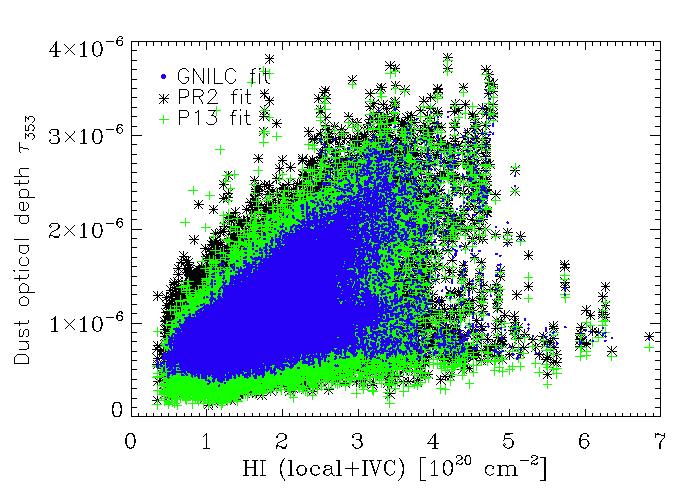}~
\end{center}
  \caption{$T$--$T$ scatter plot at high Galactic latitude between the \hi\ map and the dust optical depth maps at $353$\,GHz: dust model P13 (green), PR2 MBB fit (black), {\tt GNILC} MBB fit (blue). While the slope of the correlation with the \hi\ map is consistent for all the dust optical depth maps, the scatter is smallest for the {\tt GNILC} dust optical depth map because of the removal of the CIB temperature anisotropies.}
  \label{Fig:h1}
\end{figure}

Figure \ref{Fig:h1} shows the scatter plot between the \hi\ map and the dust optical depth map for the dust model P13 (green), the PR2 MBB fit (black), and the CIB-filtered {\tt GNILC} MBB fit (blue). 
The PR2 MBB fit shows larger scatter than the dust model P13 because in the former the fit is performed in one step at $5'$ resolution for all three dust parameters, while in the latter the fit was performed in two steps, with the spectral index first fitted at lower resolution ($30'$), which slightly reduces the scatter due to CIB contamination. Due to the large reduction of the CIB contamination in the {\tt GNILC} maps, the correlation with the \hi\ shows the most reduced scatter. 

\begin{figure}
\begin{center}
\includegraphics[width=\columnwidth]{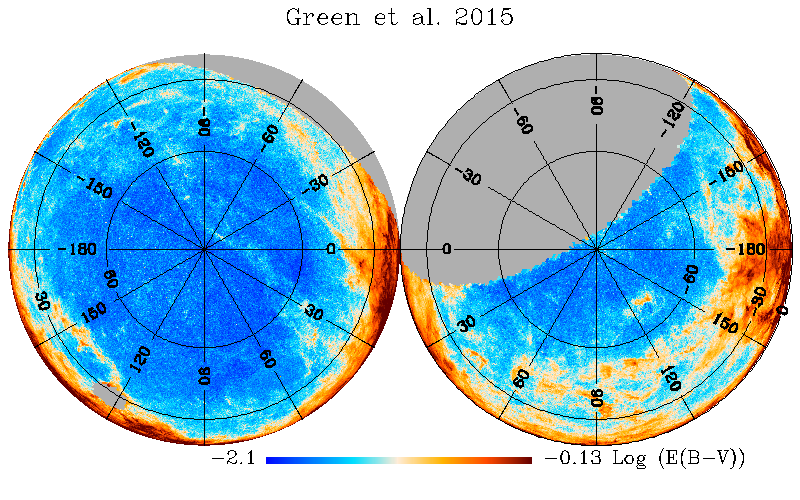}~\\
\includegraphics[width=\columnwidth]{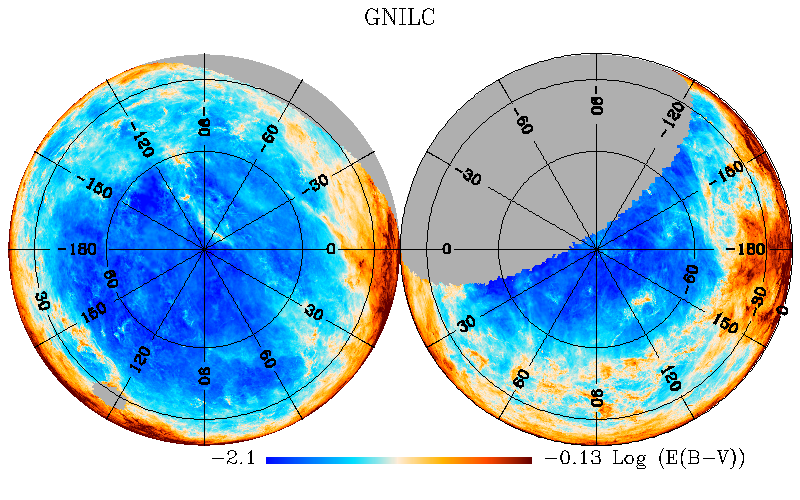}~
\end{center}
  \caption{Polar orthographic projections of the $E(B-V)$ maps at north (\emph{left}) and south (\emph{right}) Galactic poles in the Pan-STARRS 1 survey area \citep{Onaka2008}. \emph{Top}: $E(B-V)$ map from \cite{Green2015}. \emph{Bottom}: {\tt GNILC} $E(B-V)$ map. }
  \label{Fig:green_vs_gnilc}
\end{figure}
\begin{figure}
\begin{center}
\includegraphics[width=\columnwidth]{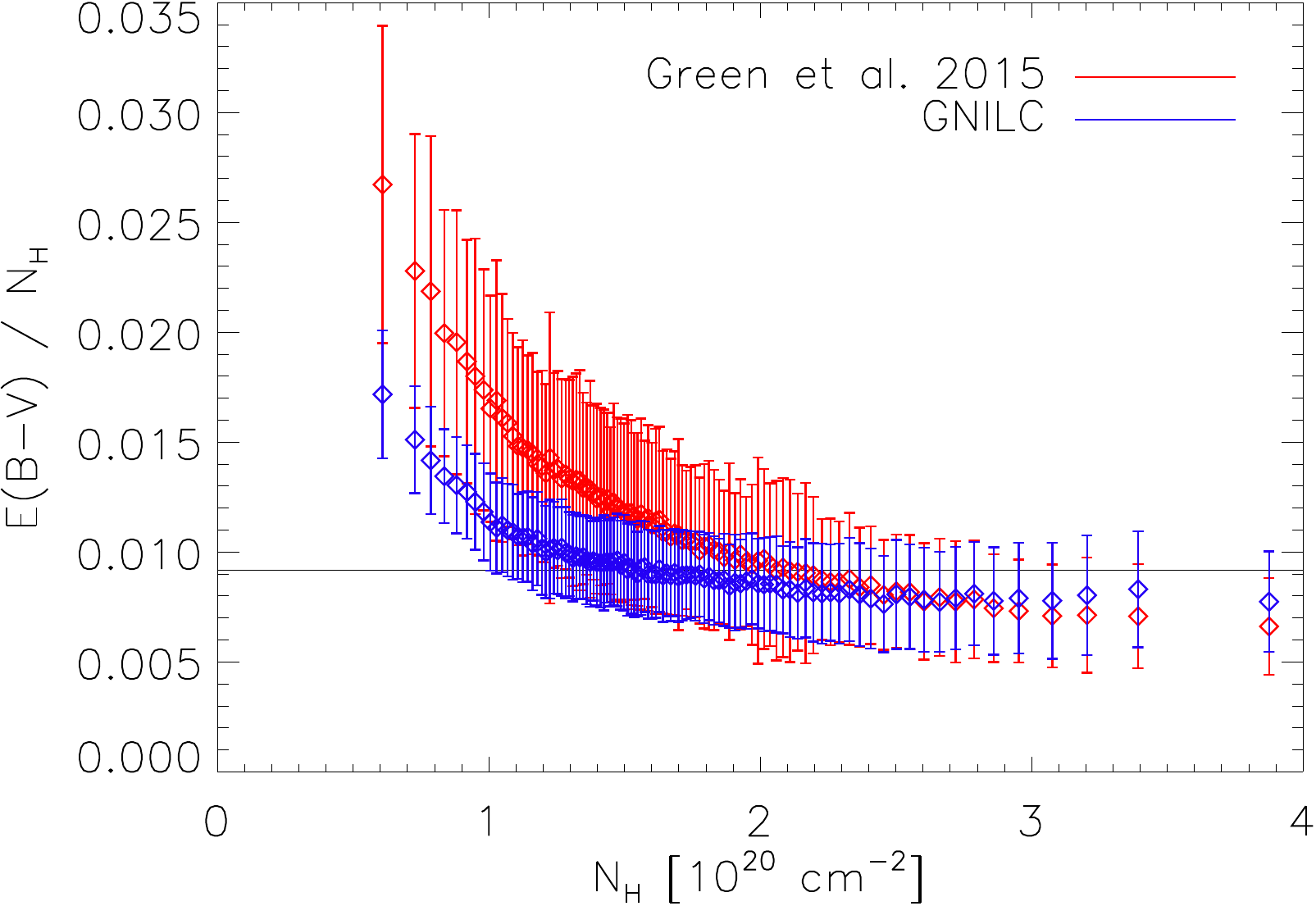}~
\end{center}
  \caption{$T$--$T$ scatter plot between the ratio $E(B-V)/N_H$ and the gas column density $N_H$ for the two-dimensional projection of the E(B-V) map of \cite{Green2015} (red diamonds) and the {\tt GNILC} E(B-V) map (blue diamonds) in the high-latitude region of the sky defined in Fig. \ref{Fig:regions}. Each point is the average of $E(B-V)/N_H$ values in a bin of $N_H$. The bin size varies such that there is always the same number of samples per bin.}
  \label{Fig:green_vs_gnilc2}
\end{figure}

The particles producing thermal dust emission also cause extinction of the light from stars and quasars, which is quantified by Galactic reddening, $E(B-V)$. 
We develop a {\tt GNILC} reddening $E(B-V)$ map by multiplying the {\tt GNILC} dust optical depth map, $\tau_{353}$, by the factor $1.49\times 10^{-4}$\,mag derived in \cite{planck2013-p06b} from the correlation between the reddening of quasars and dust optical depth along the same line of sight.
Recently, \cite{Green2015} have produced a three-dimensional dust reddening $E(B-V)$ map based on stars in the Pan-STARRS 1 survey. 
We projected their three-dimensional dust reddening into a two-dimensional map by computing the median reddening in the farthest distance bin.  
Figure \ref{Fig:green_vs_gnilc} compares this with the {\tt GNILC} $E(B-V)$ map, focusing on common area in the north and south Galactic caps where the gas column density is low.
Although it shows spatial structure similar to the {\tt GNILC} $E(B-V)$ map, the $E(B-V)$ map from \cite{Green2015} is noisier. 
Just as there is a good correlation of $\tau_{353}$ and the \hi\ gas column density, $N_H$, the correlation of reddening $E(B-V)$ and $N_H$ has been long established \citep[e.g., ][]{Savage1972}. 
To explore this further for the low column density area of the sky defined in Fig. \ref{Fig:regions}, in Fig. \ref{Fig:green_vs_gnilc2} we plot the ratio $E(B-V) / N_H$ binned with respect to $N_H$ for both {\tt GNILC} and the reddening map from \cite{Green2015}.  
For {\tt GNILC} the trend with $N_H$ is quite flat; the horizontal line plotted is compatible with the behaviour for the opacity $\tau_{353}/N_H$ found in \cite{planck2013-p06b} over the same range in $N_H$ (see their Figure 20).
On the other hand, for the $E(B-V)$ map from \cite{Green2015} we see a strong dependence of the ratio on $N_H$, i.e., a lack of linearity between this measure of $E(B-V)$ and $N_H$. 
The binned results from the map of \cite{Green2015} also show a larger dispersion. 
From these three perspectives, the {\tt GNILC} optical depth map appears to provide a better template for $E(B-V)$ studies. 
Note that both products show an excess ratio at the lowest column densities ($N_H < 1\times 10^{20}\,{\rm cm}^{-2}$); this could be the signature of dust mixed with ionized hydrogen, which is not traced by neutral \hi\ emission \citep[see discussion in][]{planck2013-p06b}.

\section{Conclusions}
\label{sec:conclusions}

We have produced a significantly improved all-sky map of the Galactic thermal dust emission from the \Planck\ data and the IRAS 100\,\micron\ map. We have fitted a modified blackbody model in each pixel to the {\tt GNILC} dust maps produced at 353, 545, 857, and 3000 GHz. The new \Planck\ {\tt GNILC} dust model has been compared with the dust models P13 and P15 and has been shown to be significantly less contaminated by CIB and noise.

By exploiting the distinct signature of Galactic dust and extragalactic CIB angular power spectra, the {\tt GNILC} method successfully separates the Galactic thermal dust emission from the CIB anisotropies in the {\it Planck} PR2 maps.

We have reduced the dispersion due to CIB contamination of the estimated dust temperature and spectral index in the {\tt GNILC} dust map with respect to the P13 dust map by a factor 1.3 for $T$ on the whole sky (1.6 at high latitude) and a factor 2.1 for $\beta$ on the whole sky (2.6 at high latitude).

The {\tt GNILC} dust map at $353$\,GHz has already been implemented as the thermal dust model in the released \Planck\ simulations of the sky \citep{planck2014-a14}.

 The {\tt GNILC} method, presented in this work, also gives access to the CIB anisotropies over a large fraction of the sky. Within the small fields considered in \citet{planck2013-pip56}, the {\tt GNILC} CIB maps and the CIB 2013 maps are found to be consistent with a Pearson correlation coefficient larger than $0.8$. The {\tt GNILC} CIB maps can be very useful as indirect tracers of the dark matter over a large area of the sky and they are recommended for the investigation of cross-correlations with galaxy weak lensing data and other tracers of large-scale structure.


The new \Planck\ {\tt GNILC} products are made publicly available on the Planck Legacy Archive.\footnote{\label{pla}\url{http://pla.esac.esa.int/pla}} They include:
\begin{itemize}
\item the CIB-removed {\tt GNILC} thermal dust maps at $353$, $545$, and $857$\,GHz;
\item the {\tt GNILC} CIB maps at $353$, $545$, and $857$\,GHz;
\item the {\tt GNILC} dust optical depth map;
\item the {\tt GNILC} dust spectral index map;
\item the {\tt GNILC} dust temperature map;
\item the {\tt GNILC} effective beam map for the thermal dust.
\end{itemize}


\begin{acknowledgements}
 The Planck Collaboration acknowledges the support of: ESA; CNES, and
CNRS/INSU-IN2P3-INP (France); ASI, CNR, and INAF (Italy); NASA and DoE
(USA); STFC and UKSA (UK); CSIC, MINECO, JA, and RES (Spain); Tekes, AoF,
and CSC (Finland); DLR and MPG (Germany); CSA (Canada); DTU Space
(Denmark); SER/SSO (Switzerland); RCN (Norway); SFI (Ireland);
FCT/MCTES (Portugal); ERC and PRACE (EU). A description of the Planck
Collaboration and a list of its members, indicating which technical
or scientific activities they have been involved in, can be found at
\href{http://www.cosmos.esa.int/web/planck/planck-collaboration}{http://www.cosmos.esa.int/web/planck/planck-collaboration}. Some of the results in this paper have been derived using the \healpix\ package. The research leading to these results has received funding from the ERC Grant no. 307209.
\end{acknowledgements}

\bibliographystyle{aat}
\bibliography{Planck_bib,Dust_CIB_separation_gnilc_RevisionAA}

\appendix
\section{Description of the {\tt GNILC} method }
\label{sec:details}

The generalized needlet internal linear combination, {\tt GNILC} \citep{Remazeilles2011b}, is a component-separation method designed to reconstruct the diffuse emission of a complex component originating from multiple correlated sources of emission, such as the Galactic foreground emission or the cosmic infrared background radiation. 

{\tt GNILC} is a multi-dimensional generalization (Sect. \ref{subsec:ilc}) of the standard internal linear combination (ILC) method, which has been extensively used to reconstruct one-dimensional components such as the CMB emission \citep{Bennett2003,planck2013-p06} or the Sunyaev-Zeldovich (SZ) signal \citep{Remazeilles2013,planck2014-a28}. 
A two-dimensional extension of the ILC, the so-called Constrained ILC, was first developed by \citet{Remazeilles2011a} to reconstruct both maps of the CMB and the SZ components, with vanishing contamination from one into the other. {\tt GNILC} is a further generalization in which the dimension of the ILC filter, which is related to the dimension of the signal subspace, is no longer fixed, but varies with both the direction in the sky and the angular scale, depending on the effective local signal-to-noise ratio, i.e., the local conditions of contamination in both real space and harmonic space. 

In this work, the signal is Galactic and the noise contributions consist of the CIB, the CMB, and the instrumental noise. The effective signal-to-noise ratio is determined locally both over the sky and over different angular scales by decomposing the data onto a wavelet (needlet) frame and by making use of a prior on the CIB power spectrum (Sect. \ref{subsec:covar}). The dimension of the Galactic signal is estimated locally in each wavelet domain through a modified principal component analysis (PCA) constrained by the power spectrum of the CIB (Sect. \ref{subsec:cpca}). The effective dimension of the Galactic subspace, given by the local number of principal components, is not determined ad hoc but through a statistical model selection by the Akaike Information Criterion (Sect. \ref{subsec:aic}). 
The prior of the CIB power spectrum is only used at the stage of determining the number of principal components, i.e. the dimension of the Galactic signal subspace. There is no prior assumption about the Galactic signal. 

\subsection{Multi-dimensional Internal Linear Combination}\label{subsec:ilc}

We model the sky observation, $x_i(p)$, at frequency channel $i$ and in the direction $p$ in the sky (pixel), as the combination of the Galactic foreground emission, the CIB emission, the CMB emission, and the instrumental noise: 
\bea\label{eq:model}
x_i(p) = f_i(p) + s_i^{\rm CIB}(p) + a_i s^{\rm CMB}(p)  + n_i(p),
\eea 
assuming no correlations between the different components of emission.
Equation (\ref{eq:model}) can be recast in the $N_{\rm ch}\times 1$ vector form, where $N_{\rm ch}$ is the number of frequency channels:
\bea\label{eq:modelv}
\bdx (p) = \bdf (p) + \bds ^{\rm CIB}(p) + \bda s^{\rm CMB}(p) + \bdn(p),
\eea
Here {$\bdx = (x_i)_{1\leq i \leq N_{\rm ch}}$} collects the $N_{\rm ch}$ observation maps, each of them being a mixture of the Galactic foreground emission, $\bdf$ (i.e., the thermal dust emission at high frequencies), the CIB emission, $\bds^{\rm CIB}(p)$, the CMB emission, $s^{\rm CMB}(p)$, scaling with a known spectral distribution, $\bda$, and the instrumental noise, $\bdn$.

The $N_{\rm ch}\times N_{\rm ch}$ frequency-frequency covariance matrix of the sky observations, ${\tens{R}(p) = (\tens{R}_{ij}(p))_{1\leq i,j \leq N_{\rm ch}}} = \langle \bdx(p)\bdx(p)^{\sf{T}}\rangle$, is
\bea
\tens{R} = \tens{R}_{\bdf} + \tens{R}_{\rm CIB} + \tens{R}_{\rm CMB} + \tens{R}_{\rm noise}
\eea
where ${\tens{R}_{\bdf} = \langle \bdf\bdf^{\sf{T}}\rangle}$ is the covariance matrix of the Galactic signal, ${\tens{R}_{\rm CIB} = \langle \bds ^{\rm CIB}\left(\bds ^{\rm CIB}\right)^{\sf{T}}\rangle}$ the covariance matrix of the CIB, ${\tens{R}_{\rm CMB} = \langle \left(s^{\rm CMB}\right)^2\rangle \bda\bda^{\sf{T}}}$ the covariance matrix of the CMB, and ${\tens{R}_{\rm noise} = \langle \bdn\bdn^{\sf{T}}\rangle}$ the covariance matrix of the noise. 

The Galactic foreground signal, $\bdf$, is a complex multi-component emission emanating from various physical processes (e.g., thermal dust emission, synchrotron emission, and free-free emission) with spectral properties varying over the sky. The number of degrees of freedom of the Galactic foreground signal would be infinite in the case of an infinitely narrow beam. However, in practice the beam is finite and the observations are limited by the number of frequency channels and the level of noise so that the effective number of Galactic degrees of freedom required to describe the Galactic emission is finite.
In addition, the various physical components of the diffuse Galactic emission are correlated, therefore the Galactic signal, $\bdf$, can be represented as the superposition of a relatively small number, $m$, of independent (not physical) templates, $\bdt$:
\bea
\bdf = \tens{F} \bdt,
\eea
where $\tens{F}$ is an $N_{\rm ch}\times m$ mixing matrix giving the contribution from the templates in each frequency channel. Therefore, the covariance matrix of the Galactic signal is an $N_{\rm ch}\times N_{\rm ch}$ matrix of rank $m$:
\bea
\tens{R}_{\bdf} = \tens{F} \tens{R}_{\bdt} \tens{F}^{\sf{T}}
\eea
where $\tens{R}_{\bdt}=\langle \bdt \bdt^{\sf{T}}\rangle$ is a full-rank $m\times m$ matrix.

 We now address the problem of estimating the set of maps, $\bdf(p)$,
i.e., determining a ``catch-all'' foreground component comprising the emission of the diffuse Galactic interstellar medium.  The  objective is to construct
estimated maps, $\widehat \bdf(p)$, that are good matches to what would be
observed by the instrument in the absence of CMB, CIB, and noise.

For extracting such an emission component from multi-frequency observations, we propose to generalize the internal linear combination (ILC) method to address the case of such a ``multi-dimensional component'' ($m$--dimensional, with ${m < N_{\rm ch}}$). 
We consider the estimation of $\bdf$ by a weighted linear operation
\bea
  \label{eq:2}
  \widehat{\bdf} =\tens{W} \bdx,
\eea
where the $N_{\rm ch}\times N_{\rm ch}$ weight matrix, $\tens{W}$, is designed to offer unit response to the Galactic foreground emission while minimizing the total variance of the vector estimate $\widehat{\bdf}$.
Stated otherwise, the matrix $\tens{W}$ is the minimizer of $E(\vert\vert\tens{W} \bdx\vert\vert^2)$ under the constraint $\tens{W}\tens{F}=\tens{F}$. The
weights matrix, $\tens{W}$, thus solves the following constrained variance minimization problem
\bea\label{eq:leminprob}
  \min_{\tens{W}\tens{F}=\tens{F}} {\rm Tr}\left(\tens{W} \widehat{\tens{R}} \tens{W}^{\sf{T}}\right),  
\eea
where $\widehat{\tens{R}}$ is the covariance matrix of the observations, $\bdx$. This problem can be solved by introducing a Lagrange
multiplier matrix, $\tens{\Lambda}$, and the Lagrangian
\bea
  \label{qqq:lagrange} 
  \mathcal{L}(\tens{W}, \tens{\Lambda})= 
  {\rm Tr}\left(\tens{W} \widehat{\tens{R}} \tens{W}^{\sf{T}}\right)
  -
  {\rm Tr}\left(\tens{\Lambda}^{\sf{T}}(\tens{W}\tens{F}-\tens{F})\right). 
\eea
By differentiating Eq.~(\ref{qqq:lagrange}) with respect to $\tens{W}$, one finds that $\partial \mathcal{L}(\tens{W},\tens{\Lambda})/\partial\tens{W} = 0$ is solved by
\bea
  \label{qqq:semisol}
  2 \tens{W} \widehat{\tens{R}} =  \tens{\Lambda} \tens{F}^{\sf{T}}.
\eea
By imposing the constraint $\tens{W}\tens{F}=\tens{F}$ on Eq.~(\ref{qqq:semisol}), one then finds that $\tens{\Lambda} = 2 \tens{F}(\tens{F}^{\sf{T}} \widehat{\tens{R}}^{-1} \tens{F})^{-1}$.  Hence, the solution of the Eq.~(\ref{eq:leminprob}) is given by the ILC weight matrix
\bea
  \label{qqq:forfilter} 
  \tens{W} = 
  \tens{F}\left(\tens{F}^{\sf{T}} \widehat{\tens{R}}^{-1}\tens{F}\right)^{-1}\tens{F}^{\sf{T}}  \widehat{\tens{R}}^{-1}.  
\eea
Multi-dimensional ILC appears as a direct generalization of the one-dimensional ILC of \citet{Bennett2003}.
The mixing matrix, $\tens{F}$, of the Galactic signal and its dimension, $m$, are the unknowns of the problem. However, it is important to notice that expression~(\ref{qqq:forfilter}) for $\tens{W}$ is invariant if $\tens{F}$ is changed into $\tens{F} \tens{T}$ for any invertible matrix $\tens{T}$.  Hence, implementing the multi-dimensional ILC filter~(\ref{qqq:forfilter}) only requires that the foreground mixing matrix, $\tens{F}$, be known up to right multiplication by an invertible factor \citep{Remazeilles2011b}.  In other words, the only meaningful and mandatory quantity for implementing a multi-dimensional ILC is not the ``true'' mixing matrix but the column space of $\tens{F}$, i.e. the dimension, $m$, of the Galactic signal subspace. This is the purpose of the Sects. \ref{subsec:cpca} and \ref{subsec:aic}.

The frequency-frequency covariance matrix of the observations, $\tens{R}$, can be estimated from the observation maps across the frequency channels. The coefficients of the covariance matrix of the observations for the pair of frequencies $(a,b)$ is computed empirically as follows:
\bea\label{eq:r}
\widehat{\tens{R}}_{ab}(p) = \sum_{p'\in \mathcal{D}(p)} x_a(p')\,x_b(p'),
\eea
where $\mathcal{D}(p)$ is a domain of pixels centred around the pixel $p$. In this work, the pixel domain, $\mathcal{D}(p)$, is defined by smoothing the product map $x_i(p)\,x_j(p)$ with a Gaussian window in pixel space. 

\subsection{Nuisance covariance matrix}\label{subsec:covar}

 We define the ``nuisance'' covariance matrix, $\tens{R}_{\rm N}$, as the sum of the CIB covariance matrix, the CMB covariance matrix, and the noise covariance matrix:
\bea\label{eq:rn}
\tens{R}_{\rm N} = \tens{R}_{\rm CIB} + \tens{R}_{\rm CMB}  + \tens{R}_{\rm noise}.
\eea
Therefore,
\bea\label{eq:rn2}
\tens{R} = \tens{R}_{\bdf} + \tens{R}_{\rm N}.
\eea
Our aim is to obtain an estimate, $\widehat{\tens{R}}_{\rm N}$, of the nuisance covariance matrix. In combination with the estimate of the full covariance matrix, $\widehat{\tens{R}}$ (Eq.~\ref{eq:r}), the nuisance covariance matrix, $\widehat{\tens{R}}_{\rm N}$, allows us to constrain the Galactic signal subspace.

 The noise covariance matrix, $\tens{R}_{\rm noise}$, can be estimated from the covariance of the half-difference maps between the HM1 and the HM2 half-mission surveys of \Planck\ \citep{planck2014-a09}, since the sky emission cancels out in the difference while the noise does not. The half-difference map for frequency channel $a$ is
\bea\label{eq:half-diff}
n_a(p) = {x_a^{HM1}(p) - x_a^{HM2}(p) \over 2},
\eea
where $x_a^{HM1}$ and $x_a^{HM2}$ are the \Planck\ HM1 map and the \Planck\ HM2 map, respectively. Similarly to the full covariance matrix in Eq.~(\ref{eq:r}), the noise covariance matrix is then estimated empirically:
\bea\label{eq:rnoise}
\left(\widehat{\tens{R}}_{\rm noise}\right)_{ab}(p) = \sum_{p'\in \mathcal{D}(p)} n_a(p')\,n_b(p').
\eea

The \Planck\ CMB best-fit ${\rm \Lambda}$CDM model $\mathcal{C}_\ell$ \citep{planck2013-p08} and the \Planck\ CIB best-fit $\mathcal{C}^{axb}_\ell$ models \citep{planck2013-pip56} are used as priors to estimate: (i) the CMB covariance matrix, $\tens{R}_{\rm CMB}$; and (ii) the CIB covariance matrix, $\tens{R}_{\rm CIB}$, in pixel space.

We first simulate a Gaussian CMB map, $\widetilde{y}(p)$, having a power spectrum given by the \Planck\ CMB best-fit $\mathcal{C}_\ell$, and we scale it across frequencies through the known CMB spectral distribution, $\bda$. In harmonic space, the cross-power spectrum of the simulated CMB maps is given by
\bea
\langle a_a\widetilde{y}_{\ell m}a_b\widetilde{y}_{\ell m}^{~*}\rangle = \mathcal{C}_\ell a_aa_b,
\eea
where $\mathcal{C}_\ell$ is the \Planck\ CMB best-fit.
From the simulated CMB maps, $\bda\widetilde{y}(p)$, we are able to compute the CMB covariance matrix in pixel space, $\widehat{\tens{R}}_{\rm CMB}(p)$, in the same way as in Eqs.~(\ref{eq:r}) and (\ref{eq:rnoise}).

From the \Planck\ CIB best-fit cross-and auto-power spectra, ${\mathcal{C}_{\rm CIB}^{a\times b}}(\ell)$, we simulate $N_{\rm ch}$ correlated Gaussian maps,\footnote{It could be argued that the non-Gaussianity of the CIB anisotropies should be taken into account in the statistics, in order to make the separation of Galactic foregrounds and CIB more accurate. However, the non-Gaussianity of the CIB is negligible compared to the non-Gaussianity of the Galactic foregrounds, so that Gaussian statistics is a sufficient approximation for separating the thermal dust and the CIB.} $\widetilde{z}_a(p)$, having a correlation matrix in harmonic space given by
\bea
\langle \widetilde{z}_a(\ell,m)\widetilde{z}_b^{~*}(\ell, m)\rangle = {\mathcal{C}_{\rm CIB}^{a\times b}}(\ell),
\eea
where ${\mathcal{C}_{\rm CIB}^{a\times b}}(\ell)$ are the \Planck\ CIB best-fits.
From the simulated correlated CIB maps, $\widetilde{z}_a(p)$, we can compute the CIB covariance matrix in pixel space, $\widehat{\tens{R}}_{\rm CIB}(p)$, in the same way as in Eqs.~(\ref{eq:r}) and (\ref{eq:rnoise}). Note that, for all $a,b < 143$\,GHz, the coefficients of $\widehat{\tens{R}}_{{\rm CIB}~ab}$ are set to zero, because there is no \Planck\ measurement of the CIB at those low frequencies, and therefore they are considered negligible.

\subsection{Determining the Galactic signal subspace with a constrained PCA}\label{subsec:cpca}

Once both the observation covariance matrix, $\widehat{\tens{R}}$, and the nuisance covariance matrix, $\widehat{\tens{R}}_{\rm N}$, have been computed, we can ``whiten'' the \Planck\ data by making the transformation
\bea
\bdx \leftarrow \widehat{\tens{R}}_{\rm N}^{-1/2} \bdx,
\eea 
such that the covariance matrix of the transformed \Planck\ observations is now given by
\bea
\widehat{\tens{R}}_{\rm N}^{-1/2} \tens{R} \widehat{\tens{R}}_{\rm N}^{-1/2}.
\eea
Assuming that the prior CIB covariance matrix, $\widehat{\tens{R}}_{\rm CIB}$, is close the real CIB covariance matrix, $\tens{R}_{\rm CIB}$, we have that
\bea
\widehat{\tens{R}}_{\rm N}^{-1/2} \tens{R}_{\rm N} \widehat{\tens{R}}_{\rm N}^{-1/2} \approx \tens{I},
\eea
where $\tens{I}$ is the identity matrix.
In this way, from Eq.~(\ref{eq:rn2}) the covariance matrix of the transformed \Planck\ observations becomes
\bea\label{eq:wcov}
\widehat{\tens{R}}_{\rm N}^{-1/2} \tens{R} \widehat{\tens{R}}_{\rm N}^{-1/2} \approx \widehat{\tens{R}}_{\rm N}^{-1/2} \tens{R}_{\bdf} \widehat{\tens{R}}_{\rm N}^{-1/2} + \tens{I},
\eea
such that the power of all the nuisance contamination, including CIB, CMB, and noise, is encoded in the matrix $\tens{I}$, which is close to an identity matrix. Typically, the coefficients of the transformed observation covariance matrix (Eq.~\ref{eq:wcov}) provide the signal-to-noise ratio over the sky, i.e., the power of the Galactic signal divided by the overall power of the CIB-plus-CMB-plus-noise.

Therefore, by diagonalizing the transformed observation covariance matrix (Eq.~\ref{eq:wcov}), we obtain the following eigenstructure:
\bea\label{eq:diag}
\widehat{\tens{R}}_{\rm N}^{-1/2} \widehat{\tens{R}} \widehat{\tens{R}}_{\rm N}^{-1/2} = \left[
\begin{array}{c|c}
  &  \\
 \tens{U}_{\rm S}  & \tens{U}_{\rm N} \\
 & 
\end{array}
 \right]
 \cdot \left[
\begin{array}{lll|c}
\lambda_1+1 & & &  \\
 & ...  & & \\
 & &  \lambda_m+1 &  \\
\hline
 & & & \tens{I}  \\
\end{array}
\right]
\cdot \left[
\begin{array}{c}
  \\
\tens{U}_{\rm S}^{\sf{T}} \\
\hline
  \\
\tens{U}_{\rm N}^{\sf{T}} \\
\end{array}
 \right].\nonumber
\\
\eea
In such a representation, the eigenvalues of the covariance matrix $\widehat{\tens{R}}_{\rm N}^{-1/2} \widehat{\tens{R}} \widehat{\tens{R}}_{\rm N}^{-1/2}$ that are close to unity do not contain any relevant power of the Galactic signal and the signal is dominated by the CIB, the CMB, and the noise. The corresponding eigenvectors span the nuisance subspace characterized by the number of degrees of freedom, $N_{\rm dof}$, of the CIB plus CMB plus noise only. Conversely, the subset of eigenvectors collected in the $N_{\rm ch}\times m$ matrix $\tens{U}_{\rm S}$, for which the eigenvalues of $\widehat{\tens{R}}_{\rm N}^{-1/2} \widehat{\tens{R}} \widehat{\tens{R}}_{\rm N}^{-1/2}$ significantly depart from unity, span the Galactic signal subspace (principal components). The number of eigenvalues, $m$, that are much larger than unity corresponds to the dimension of the Galactic signal subspace, i.e., the number of independent (unphysical) templates contributing to the Galactic signal. 
This is a constrained principal component analysis (PCA), in the sense that the PCA is driven by the local signal-to-noise ratio.

\begin{figure}
  \begin{center}
\includegraphics[width=\columnwidth]{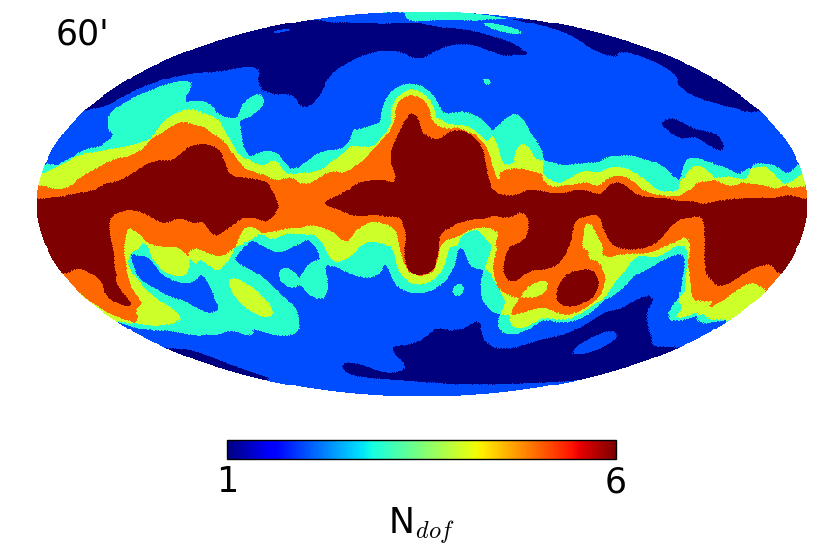}~\\
\includegraphics[width=\columnwidth]{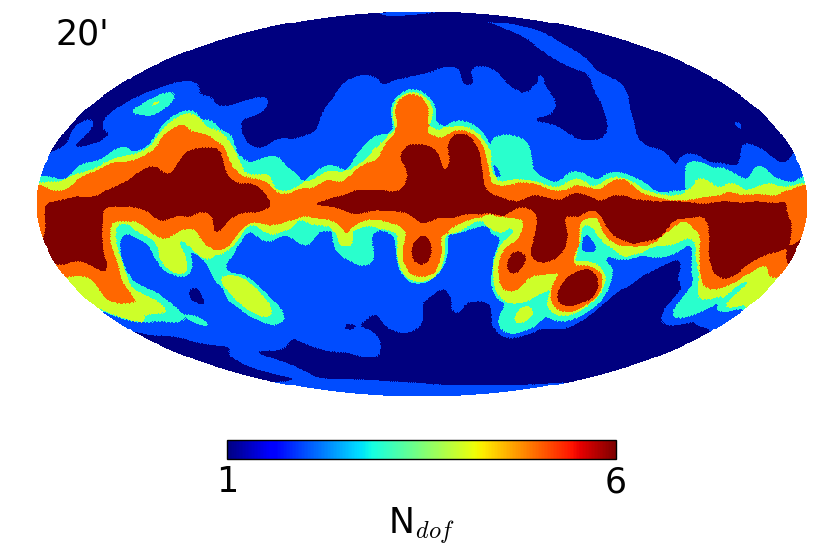}~\\
 \includegraphics[width=\columnwidth]{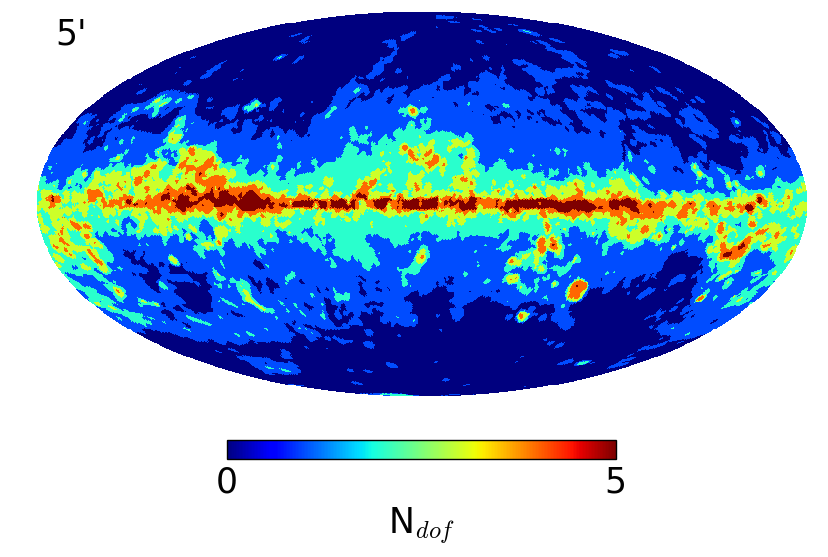}~
\end{center}
\caption{Local number of Galactic foreground degrees of freedom selected by the AIC criterion at one degree angular scale ({\it top}), $20'$ scale ({\it middle}), and $5'$ scale ({\it bottom}). The number of Galactic degrees of freedom decreases at high latitude and small angular scales.}
  \label{Fig:dof}
\end{figure}

\begin{figure}
  \begin{center}
\includegraphics[width=\columnwidth]{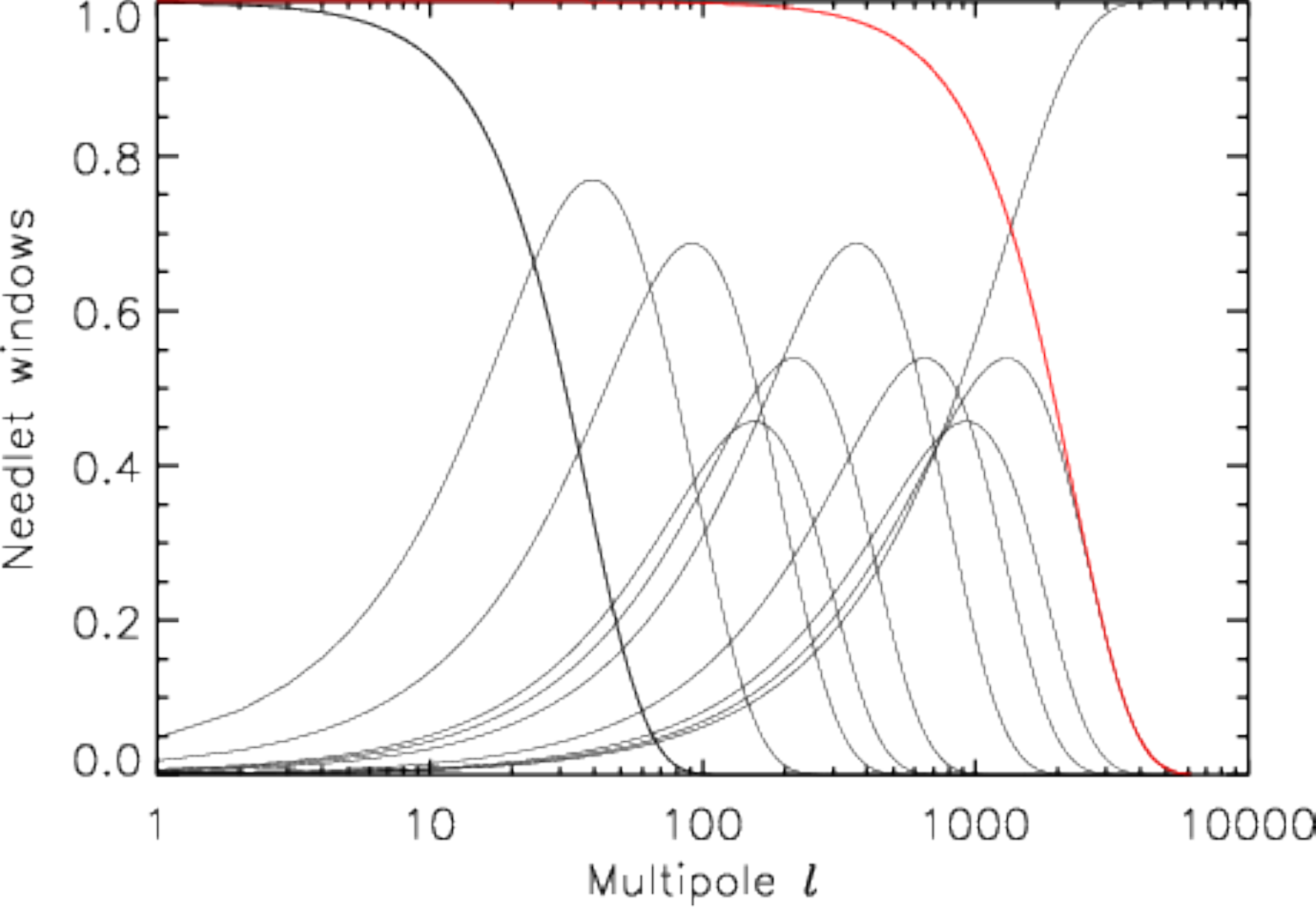}~
\end{center}
\caption{Needlet windows acting as bandpass filters in harmonic space (\emph{black lines}), with the $5'$ beam transfer function overplotted (\emph{red line}).}
  \label{Fig:bands}
\end{figure}

The diagonalization of the transformed \Planck\ covariance matrix of Eq.~(\ref{eq:diag}) can be written in a compact form as
\bea\label{eq:compact}
\widehat{\tens{R}}_{\rm N}^{-1/2} \widehat{\tens{R}} \widehat{\tens{R}}_{\rm N}^{-1/2} = \tens{U}_{\rm S}\tens{D}_{\rm S}\tens{U}_{\rm S}^{\sf{T}}+\tens{U}_{\rm N}\tens{U}_{\rm N}^{\sf{T}},
\eea
where 
\bea
\tens{D}_{\rm S} = {\rm diag} [\lambda_1+1,...,\lambda_m+1]
\eea
 is a $m\times m$ diagonal matrix and $[\tens{U}_{\rm S}|\tens{U}_{\rm N}]$ is an $N_{\rm ch}\times N_{\rm ch}$ orthonormal matrix collecting all the eigenvectors of $\widehat{\tens{R}}_{\rm N}^{-1/2} \widehat{\tens{R}} \widehat{\tens{R}}_{\rm N}^{-1/2}$. Using Eq.~(\ref{eq:compact}) and the orthonormality condition, $\tens{U}_{\rm S}\tens{U}_{\rm S}^{\sf{T}}+\tens{U}_{\rm N}\tens{U}_{\rm N}^{\sf{T}}=\tens{I}$, the covariance matrix of the Galactic signal can be written as
\bea\label{eq:mle}
\widehat{\tens{R}}_{\bdf} &=& \widehat{\tens{R}} - \widehat{\tens{R}}_{\rm N}\cr
          &=& \widehat{\tens{R}}_{\rm N}^{1/2}\,\left(\widehat{\tens{R}}_{\rm N}^{-1/2} \widehat{\tens{R}} \widehat{\tens{R}}_{\rm N}^{-1/2} - \tens{I}\right)\,\widehat{\tens{R}}_{\rm N}^{1/2}\cr
          &=& \widehat{\tens{R}}_{\rm N}^{1/2}\,\left( \tens{U}_{\rm S}\,\left(\tens{D}_{\rm S} - \tens{I}\right)\,\tens{U}_{\rm S}^{\sf{T}}\right)\,\widehat{\tens{R}}_{\rm N}^{1/2}.
\eea
This is the power that ``best matches'' what would be observed by the instrument in the absence of CIB, CMB, and noise. 
Therefore, we estimate the Galactic signal, $\bdf$, by
\bea
\widehat \bdf(p) = \widehat{\tens{F}} \,\bdt,
\eea
where 
\bea\label{eq:F-mle}
\widehat{\tens{F}} = \widehat{\tens{R}}_{\rm N}^{1/2} \tens{U}_{\rm S}
\eea
is an $N_{\rm ch} \times m$ mixing matrix estimate, and $\bdt$ is an $m\times 1$ vector of independent templates whose covariance matrix is given by
\bea
\widehat{\tens{R}}_{\bdt} = \tens{D}_{\rm S} - \tens{I}.
\eea
The estimate, $\widehat{\tens{F}}$, is the only useful information for implementing the multi-dimensional ILC filter of Eq.~(\ref{qqq:forfilter}). It can be different from the true mixing matrix, $\tens{F}$, of the Galactic signal as long as the column space is the same: let $\tens{T}$ be some invertible $m \times m$ matrix and consider the transformed matrices $\widetilde{\tens{F}} = \widehat{\tens{F}}\tens{T}^{-1}$ and $\widetilde{\tens{R}}_{\bdt} = \tens{T} \widehat{\tens{R}}_{\bdt} \tens{T}^{\sf{T}}$. These transformed matrices are an alternate factorization of the covariance matrix of the Galactic signal, $\widehat{\tens{R}}_{\bdf}$, but they are equivalent, since by construction, $\widetilde{\tens{F}} \widetilde{\tens{R}}_{\bdt} \widetilde{\tens{F}}^{\sf{T}} = \widehat{\tens{F}} \widehat{\tens{R}}_{\bdt} \widehat{\tens{F}}^{\sf{T}}$. The ILC weights of Eq.~(\ref{qqq:forfilter}) are unchanged under right multiplication by an invertible matrix. Therefore, the $m$ independent templates, $\bdt$, can be replaced by any other linear combination, $\tens{T} \bdt$, as long as we are interested in reconstructing the overall Galactic signal, $\bdf$.

The number, $m$, of principal components in Eq.~(\ref{eq:diag}), or the effective dimension of the Galactic signal subspace, can vary over the sky, depending on the local signal-to-noise ratio. In particular, at high Galactic latitudes this number decreases because the contributions of the CIB, the CMB, and the noise starts dominating the Galactic signal (Fig.~\ref{Fig:dof}). The dimension of the Galactic signal subspace is also expected to vary with angular scale, because at small angular scales the power of the Galactic signal becomes dominated by the power of the noise and the CIB. Therefore, for the accurate separation of the Galactic and CIB signals we find it useful to estimate the dimension of the Galactic subspace locally both in space and in scale. This is achieved by decomposing the data on a wavelet frame. 

In this work, the analysis is performed on a needlet frame (see e.g., \citet{Delabrouille2009}, \citet{Remazeilles2013} for the use of needlets in component separation). Basically, the spherical harmonic transforms of the maps, $a_{\ell m}$, are bandpass filtered in harmonic space, then transformed back into real space, therefore conserving a specific range of angular scales in the map. The result is called a needlet map, characterized by a given range of angular scales. The multi-dimensional ILC Eq.~(\ref{eq:2}) is performed on the needlet maps independently for each range of scale, and the synthesized map is obtained by co-adding the ILC estimates at the various scales.
The wavelet decomposition allows for estimating the dimension, $m$, of the Galactic signal subspace locally over the sky and over the angular scales, depending on the local conditions of contamination. In this work, the needlet bandpass windows (Fig.~\ref{Fig:bands}) are defined in harmonic space from the difference of successive Gaussian beam transfer functions:
\bea
h^{(1)}(\ell) &=& \sqrt{b_1(\ell)^2},\cr
h^{j}(\ell)& =& \sqrt{b_{j+1}(\ell)^2 - b_{j}(\ell)^2},\cr
h^{(10)}(\ell)& =& \sqrt{1 - b_{10}(\ell)^2},
\eea
where 
\bea
b_{j}(\ell) = \exp{\left(-\ell(\ell+1)\sigma_j^2/2\right)}
\eea
and
\bea
\sigma_j = \left({1\over\sqrt{8\ln 2}}\right)\left({\pi\over 180\times 60}\right){\rm FWHM}[j]
\eea
with ${\rm FWHM} = [300',120',60',45',30',15',10',7.5',5']$. In this way we have
\bea
\sum_{j=1}^{10} \left(h^{j}(\ell)\right)^2 = 1,
\eea
such that there is no effective filtering of any power at any scale in the final maps synthesized from the different needlet scales after component separation.

\subsection{Model selection with the Akaike information criterion}\label{subsec:aic}

 In \citet{Remazeilles2011b}, the effective number, $m$, of Galactic components in each needlet domain was estimated by rejecting the eigenvalues in Eq.~(\ref{eq:diag}) that are smaller than $1.25$, i.e., for which the ``noise'' contributes to the observation by more than $80\%$. This criterion is somewhat arbitrary. In the present work, we propose instead to use a statistical criterion to discriminate between the ``large'' eigenvalues, tracing the Galactic signal, and the ``noisy'' eigenvalues ($\approx 1$) to be rejected; the effective rank of the covariance matrix of the Galactic signal is estimated by statistical model selection through the Akaike information criterion \citep{Akaike1974}.

For a given dimension, or model, $m$, if we assume that the data, $\bdx$, are independent and identically distributed according to the Gaussian distribution $\mathcal{N}(0,\tens{R}(m))$, with $\tens{R}(m) = \tens{R}_{\bdf}(m)+\tens{R}_{\rm N}$, then the likelihood reads as
\bea
\mathcal{L}\left(\{\bdx_k\}_k\vert\tens{R}(m)\right) = \prod_{k=1}^{n}{1\over \sqrt{2\pi\det \tens{R}(m)}}\exp{\left\{-{1\over 2}\bdx^{\sf{T}}_k\tens{R}(m)^{-1}\bdx_k\right\}},\nonumber
\\
\eea
where $n$ is the number of modes in the (needlet) domain considered. The log-likelihood can be written as
\bea
-2\log \mathcal{L} &=& \sum_{k=1}^{n} \bdx^{\sf{T}}_k\tens{R}(m)^{-1}\bdx_k-\log\det\left(\tens{R}(m)^{-1}\right)+{\rm constant}(m)\cr
                   &=&  n\,K\left(\widehat{\tens{R}},\tens{R}(m)\right)+{\rm constant}(m),
\eea
where $K\left(\widehat{\tens{R}},\tens{R}(m)\right)$ is the Kullback-Leibler divergence \citep{Kullback1968}, measuring the spectral mismatch between the model covariance matrix, $\tens{R}(m)$, and the data covariance matrix, $\widehat{\tens{R}}$:
\bea\label{eq:kl}
K\left(\widehat{\tens{R}},\tens{R}(m)\right) = \mbox{Tr}\left(\widehat{\tens{R}}\tens{R}(m)^{-1}\right)-\log \det \left(\widehat{\tens{R}}\tens{R}(m)^{-1}\right)-N_{\rm ch}.
\eea

At this stage, it is interesting to note that the estimate of the Galactic covariance matrix, $\widehat{\tens{R}}_{\bdf}(m)$ computed in Eq.~(\ref{eq:mle}), is nothing other than the maximum likelihood estimate, i.e., the minimizer of the Kullback-Leibler divergence of Eq.~(\ref{eq:kl}), as in {\tt SMICA} \citep{Delabrouille2003,Cardoso2008}. The proof is given in Sect. \ref{subsec:mle-proof}.

In the region of the sky and the range of angular scales considered, we select the best rank value, $m^{*}$, among the class of models, $m$, by minimizing the AIC
\bea
A(m) &=& 2\,n\,m-2\log \left(\mathcal{L}_{\rm max}(m)\right).
\eea
 Through the penalty, $2\,n\,m$, the AIC makes a trade-off between the goodness of fit and the complexity of the model. Let us denote $\widehat{\tens{R}}_{\rm N}^{-1/2}\widehat{\tens{R}}\widehat{\tens{R}}_{\rm N}^{-1/2}=\tens{U}\tens{D}\tens{U}^{\sf{T}}$ the diagonalization of the transformed data covariance matrix, where 
\bea
\tens{U} = [\tens{U}_{\rm S}|\tens{U}_{\rm N}] \quad \textrm{ and } \quad\tens{D} = \left[\begin{array}{cc} \widehat{\tens{D}}_{\rm S}& 0\\ 0& \widehat{\tens{D}}_{\rm N}\end{array}\right].
\eea
Since the maximum likelihood is reached when $\widehat{\tens{R}}_{\rm N}^{-1/2}\widehat{\tens{R}}(m)\widehat{\tens{R}}_{\rm N}^{-1/2} = \tens{U}_{\rm S}\,\left(\tens{D}_{\rm S} - \tens{I}\right)\,\tens{U}_{\rm S}^{\sf{T}} + \tens{I}$,  we have 
\bea
& & -2\log \left(\mathcal{L}_{\rm max}(m)\right)\cr
\,
&=& nK\left(\widehat{\tens{R}}_{\rm N}^{-1/2}\widehat{\tens{R}}\widehat{\tens{R}}_{\rm N}^{-1/2},\tens{U}_{\rm S}\left(\tens{D}_{\rm S}-\tens{I}\right)\tens{U}_{\rm S}^{\sf{T}}  + \tens{I}\right)\cr
\,
&=& nK\left(\tens{D},\tens{U}^{\sf{T}}\left[\tens{U}_{\rm S}\left(\tens{D}_{\rm S}-\tens{I}\right)\tens{U}_{\rm S}^{\sf{T}}  + \tens{I}\right]\tens{U}\right)\cr
\,
&=& nK\left(\left[\begin{array}{cc} \tens{D}_{\rm S}& 0\\ 0& \tens{D}_{\rm N}\end{array}\right], \left[\begin{array}{cc} \tens{D}_{\rm S}& 0\\ 0& \tens{I}\end{array}\right]\right)\cr
\,
&=& n\left\{ \mbox{Tr}\left(\left[\begin{array}{cc} \tens{I}& 0\\ 0& \tens{D}_{\rm N}\end{array}\right]\right)-\log \det \left(\left[\begin{array}{cc} \tens{I}& 0\\ 0& \tens{D}_{\rm N}\end{array}\right]\right) - N_{\rm ch}   \right\}\cr
\,
&=& n\left(m + \sum_{i=m+1}^{N_{\rm ch}} \mu_i-\sum_{i=m+1}^{N_{\rm ch}} \log \mu_i - N_{\rm ch}\right)\cr
&=& n\left(m + \sum_{i=m+1}^{N_{\rm ch}} f(\mu_i)\right),
\eea
where $\mu_i$ are the $(N_{\rm ch}-m)$ eigenvalues of the transformed data covariance matrix, $\tens{R}_{\rm N}^{-1/2}\widehat{\tens{R}}\tens{R}_{\rm N}^{-1/2}$, collected in the matrix $\tens{D}_{\rm N}$, and 
\bea
f(\mu) = \mu - \log \mu -1.
\eea
This convex function is minimum for $\mu=1$. 

Therefore, the AIC criterion reduces to the simple analytical form:
\bea
\label{eq:aic}
 A(m) = n\,\left(2\,m+\sum_{i=m+1}^{N_{\rm ch}} \left[ \mu_i - \log\mu_i -1 \right]\right),
\eea
and the dimension of the Galactic signal subspace is estimated by minimizing Eq.~(\ref{eq:aic}) in each region considered:
\bea
m^{*} = {\rm argmin}\, \left[A(m)\right].
\eea

Minimizing the AIC criterion (Eq.~\ref{eq:aic}) in each needlet domain (i.e., in each region of the sky and each range of angular scales) allows for estimating the effective dimension of the Galactic signal subspace locally, given the level of contamination by noise, CIB, and CMB in this region. The multi-dimensional ILC then adapts the filtering to the effective local dimension of the Galactic signal. In this respect, the {\tt GNILC} method goes beyond the {\tt SMICA} method by relaxing any prior assumption on the number of Galactic components. 

Figure \ref{Fig:dof} shows the local dimension of the Galactic signal subspace over the sky, for three different ranges of angular scales, which we have estimated by minimizing the AIC criterion (Eq.~\ref{eq:aic}) on the \Planck\ data. In practice, the dimension of the Galactic signal subspace (equivalently, the number of Galactic degrees of freedom) clearly depends on the local signal-to-noise ratio, where the ``noise'' here includes the instrumental noise, the CIB, and the CMB signals. As expected, at high latitude and small angular scales (bottom panel in Fig.~\ref{Fig:dof}) this number decreases because the CIB and the instrumental noise become dominant in the observations.

\subsection{Minimization of the Kullback-Leibler divergence}
\label{subsec:mle-proof}

To close this appendix, we show that the {\tt GNILC} estimates of the covariance matrix (Eq.~\ref{eq:mle}) and the mixing matrix (Eq.~\ref{eq:F-mle}) are the maximum likelihood solution that minimizes the Kullback-Leibler divergence (Eq.~\ref{eq:kl}). For the sake of simplicity, we will assume that the data have been whitened so that the nuisance covariance matrix is represented by an identity matrix, $\tens{I}$. 

Let us consider a foreground model with a fixed number, $m$, of independent templates in the domain being considered. The foreground covariance matrix can therefore be modelled as a rank--$m$ matrix
\bea
\tens{R}_{\bdf}(m) = \tens{A}_m\tens{\Lambda}_m\tens{A}_m^{\sf{T}},
\eea
and the full data covariance matrix as
\bea
\tens{R}(m) = \tens{A}_m\tens{\Lambda}_m\tens{A}_m^{\sf{T}}  + \tens{I},
\eea
where $\tens{\Lambda}_m$ is an $m\times m$ diagonal matrix collecting the $m$ non-null eigenvalues of $\tens{R}_{\bdf}(m)$ and $\tens{A}_m$ is an $N_{\rm ch}\times m$ matrix collecting the $m$ corresponding eigenvectors. In other words, the eigendecomposition of the foreground covariance matrix is given by
\bea
{\tens{R}_{\bdf}(m)=[ \tens{A}_{m} \tens{A}_{N_{\rm ch}-m} ] 
  \left[\begin{array}{cc} \tens{\Lambda}_{m}& 0\\ 0& 0\end{array}\right]
  [ \tens{A}_{m} \tens{A}_{N_{\rm ch}-m} ] ^{\sf{T}}}.
\eea

The maximum likelihood estimate, $\tens{\Lambda}_m$, minimizing the Kullback-Leibler divergence, must satisfy for all $(i,j) \in \left[1,m\right]^2$
\bea
& & {\partial K\left(\widehat{\tens{R}},\tens{R}(m)\right)\over \partial (\tens{\Lambda}_m)_{ij}} \cr
& = & -\mbox{Tr}\left(\widehat{\tens{R}}\tens{R}(m)^{-1}{\partial\tens{R}(m)\over\partial(\tens{\Lambda}_m)_{ij}}\tens{R}(m)^{-1}\right)+\mbox{Tr}\left(\tens{R}(m)^{-1}{\partial\tens{R}(m)\over\partial(\tens{\Lambda}_m)_{ij}}\right)\cr 
& = &  -{\left(\tens{A}_m^{\sf{T}} \widehat{\tens{R}} \tens{A}_m\right)_{ij}\over \left((\tens{\Lambda}_m)_{ii}+1\right)\left((\tens{\Lambda}_m)_{jj}+1\right)}+{\delta_{ij}\over \left((\tens{\Lambda}_m)_{ii}+1\right)}\cr
& = & 0.
\eea
Therefore, the maximum likelihood estimate, $\{\tens{A}_m, \tens{\Lambda}_m\}$, is solution of
\bea\label{qqq:maxlkl}
\left(\tens{A}_m^{\sf{T}} \widehat{\tens{R}} \tens{A}_m\right) = \left(\tens{\Lambda}_m+\tens{I}\right).
\eea
where $\tens{I}$ is an $m\times m$ identity matrix. If we consider the eigenvalue decomposition of the full data covariance matrix as
\bea
\widehat{\tens{R}} = \tens{U}\tens{D}\tens{U}^{\sf{T}} = [ \tens{U}_{\rm S} \tens{U}_{\rm N} ] 
  \left[\begin{array}{cc} \tens{D}_{\rm S}& 0\\ 0& \tens{D}_{\rm N}\end{array}\right]
  [ \tens{U}_{\rm S} \tens{U}_{\rm N} ]^{\sf{T}}
\eea
where $\tens{D}_{\rm S}$ is the $m\times m$ diagonal matrix collecting the $m$ largest eigenvalues of $\widehat{\tens{R}}$, and $\tens{U}_{\rm S}$ is an $N_{\rm ch} \times m$ matrix collecting the $m$ corresponding eigenvectors, then the maximum likelihood solution of Eq.~(\ref{qqq:maxlkl}) is given by
\bea\label{qqq:sol}
\tens{\Lambda}_m &= &\tens{D}_{\rm S}-\tens{I},\cr
\tens{A}_m &= &\tens{U}_{\rm S}.
\eea

\subsection{Test on \Planck\ simulations }
\label{subsec:simu}

We apply the {\tt GNILC} algorithm to the public\footnote{\href{http://crd.lbl.gov/cmb-data}{http://crd.lbl.gov/cmb-data}} \Planck\ full focal plane simulations (FFP8) from \citet{planck2014-a14} in order to validate the component-separation method. 

We still adopt as a prior for {\tt GNILC} the \Planck\ 2013 CIB best-fit power spectra, although the simulations of the CIB components in FFP8 do not follow the exact same statistics, since the FFP8 CIB maps are generated across frequencies from simulated dark matter shells in \citet{planck2014-a14}. The imperfect agreement between the prior and the actual CIB power spectrum enables us to test the robustness of {\tt GNILC} when the knowledge of the CIB power spectrum is not perfect.

\begin{figure*}
\begin{center}
\includegraphics[width=0.3\textwidth]{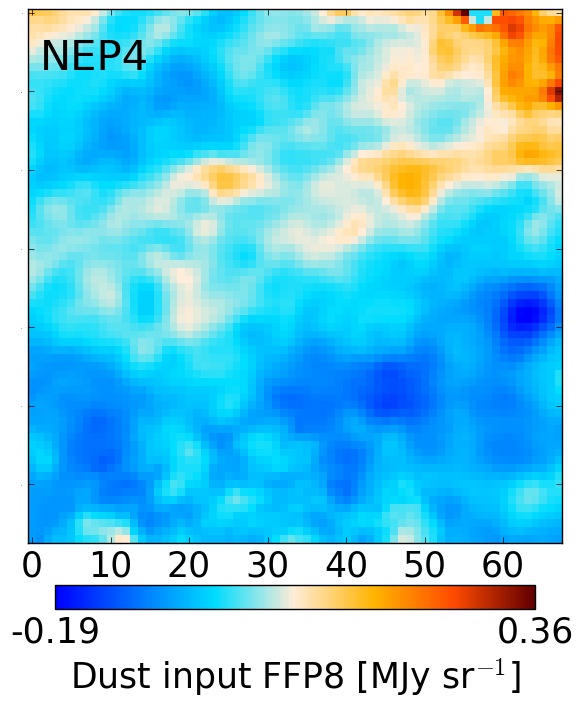}~
\includegraphics[width=0.3\textwidth]{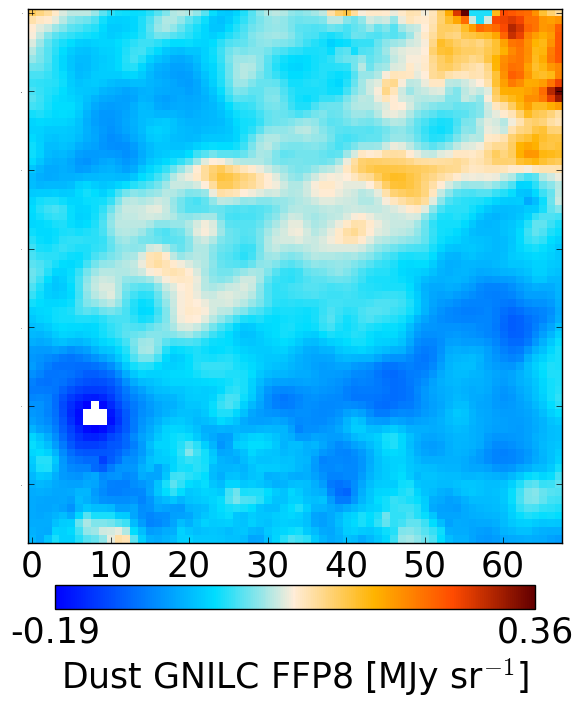}~
\includegraphics[width=0.3\textwidth]{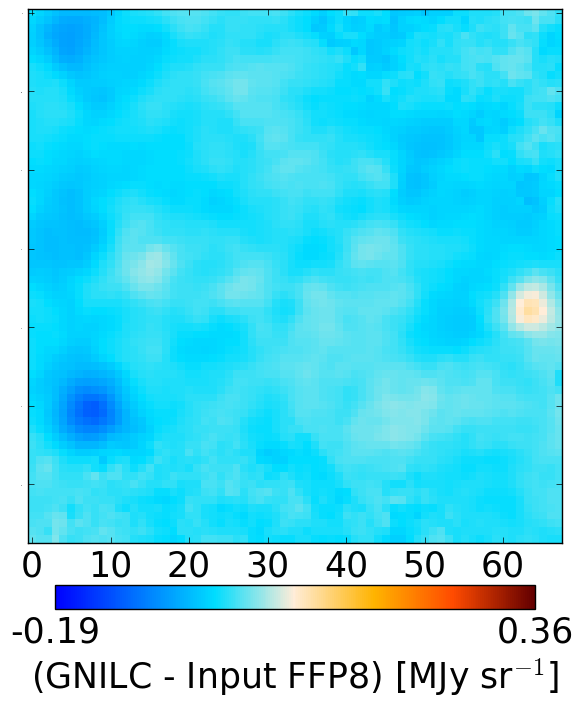}~\\
\includegraphics[width=0.3\textwidth]{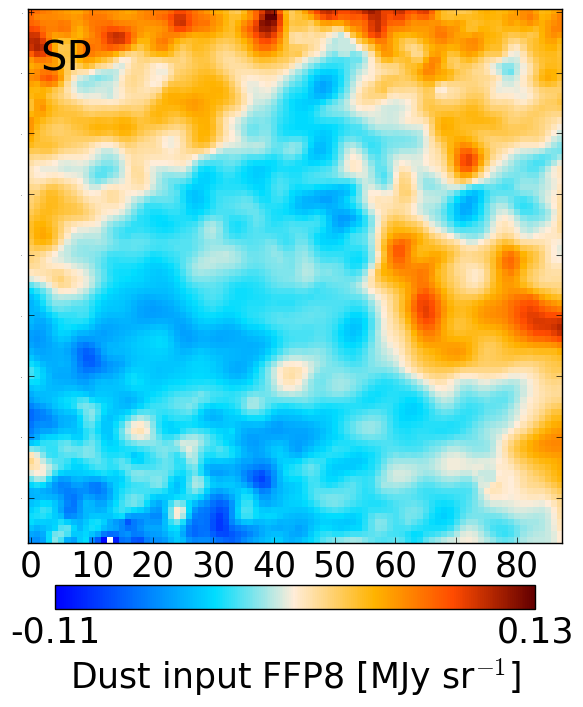}~
\includegraphics[width=0.3\textwidth]{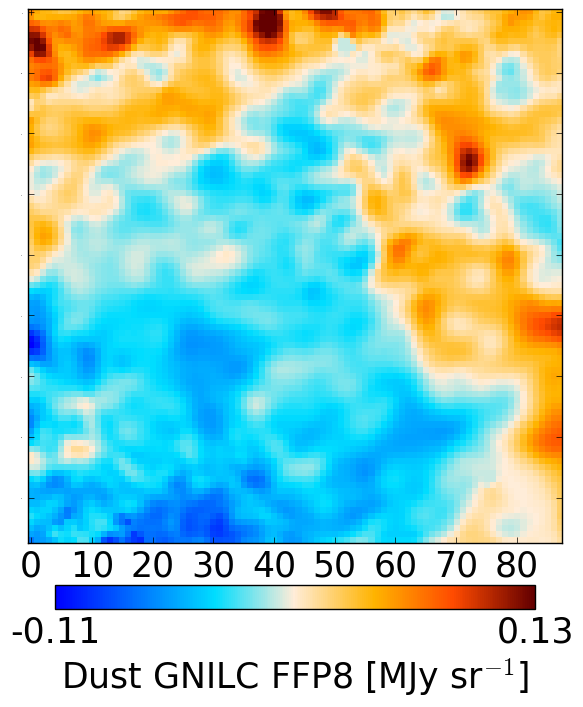}~
\includegraphics[width=0.3\textwidth]{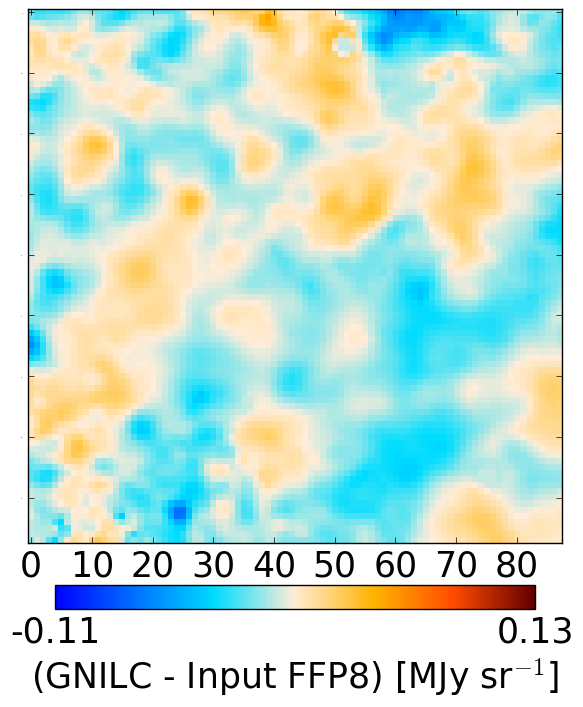}~\\
\end{center}
  \caption{Thermal dust maps from the \Planck\ FFP8 simulations. Input dust FFP8 (\emph{left}), {\tt GNILC} dust FFP8 (\emph{middle}), difference ({\tt GNILC} - input) (\emph{right}). The fields are: NEP4 (\emph{top}) and SP (\emph{bottom}).}
  \label{Fig:ffp8dust}
\end{figure*}

\begin{figure*}
\begin{center}
\includegraphics[width=0.3\textwidth]{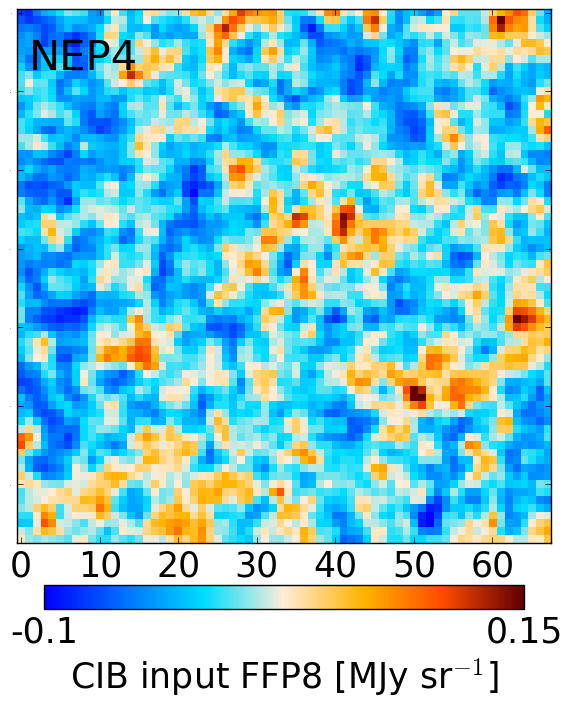}~
\includegraphics[width=0.3\textwidth]{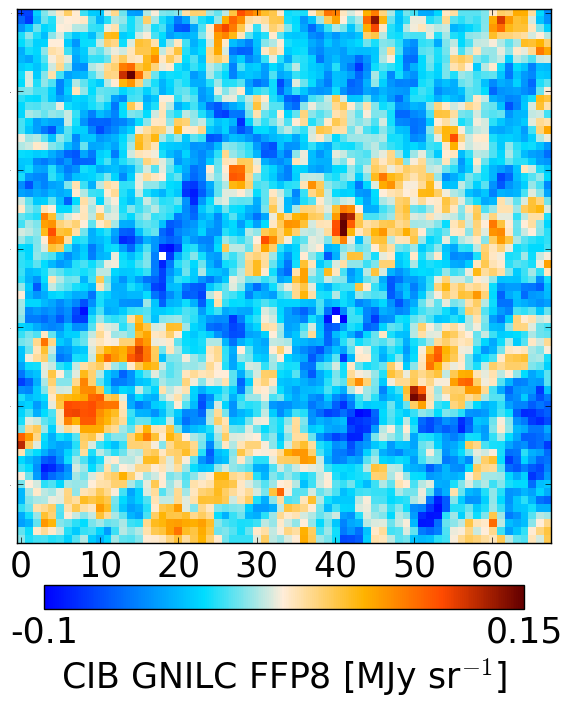}~
\includegraphics[width=0.3\textwidth]{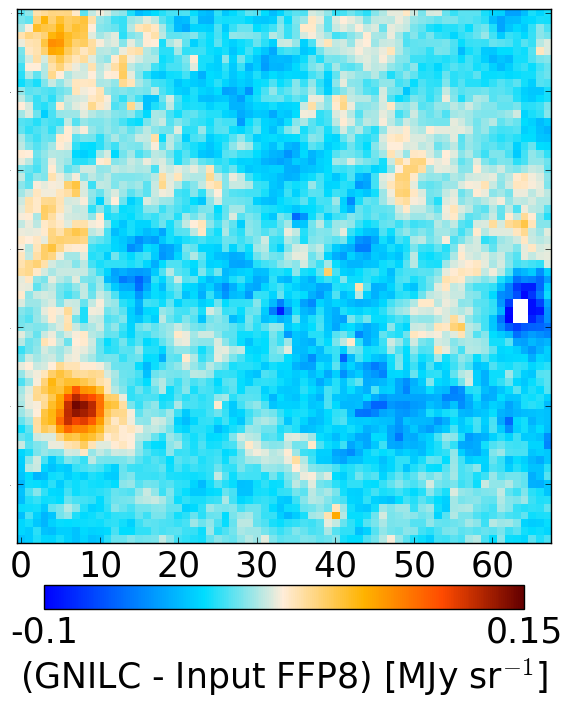}~\\
\includegraphics[width=0.3\textwidth]{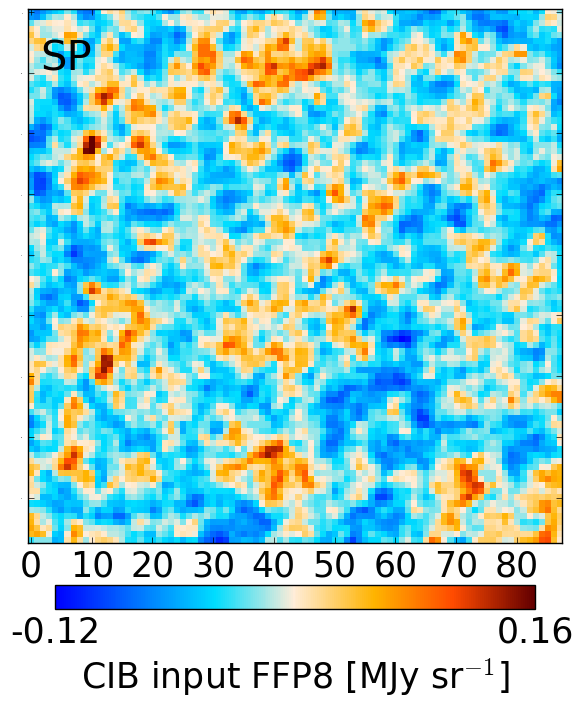}~
\includegraphics[width=0.3\textwidth]{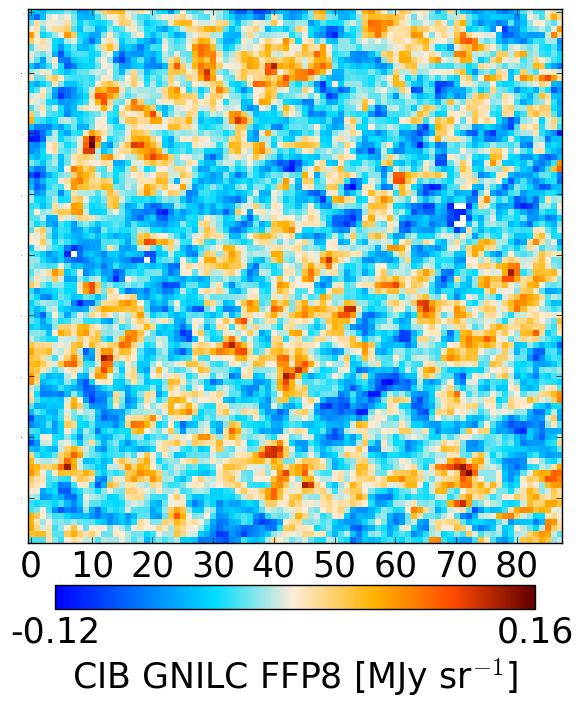}~
\includegraphics[width=0.3\textwidth]{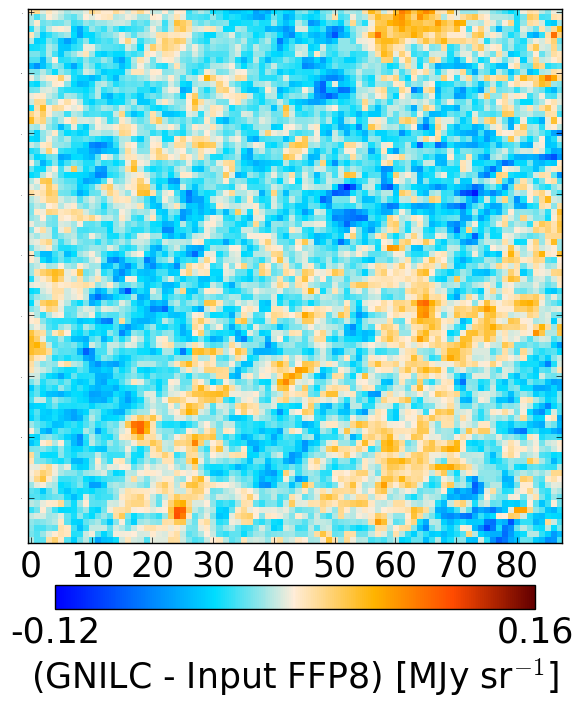}~\\
\end{center}
  \caption{CIB maps from the FFP8 simulations. Input CIB FFP8 (\emph{left}), {\tt GNILC} CIB FFP8 (\emph{middle}), difference ({\tt GNILC} - input) (\emph{right}). The fields are: NEP4 (\emph{top}) and SP (\emph{bottom}).}
  \label{Fig:ffp8cib}
\end{figure*}

In Fig.~\ref{Fig:ffp8dust} we compare the reconstructed {\tt GNILC} dust map with the input FFP8 dust map within some small CIB fields of \citet{planck2013-pip56}, where the dust is faint, namely NEP4 (part of the GHIGLS NEP field) and SP. The difference between the {\tt GNILC} dust map and the input dust map is also shown. In Fig.~\ref{Fig:ffp8cib}, we compare the reconstructed {\tt GNILC} CIB map with the input FFP8 CIB map. The difference between the {\tt GNILC} CIB and the input CIB is shown in the third column of Fig.~\ref{Fig:ffp8cib}.

\begin{figure*}
\begin{center}
\includegraphics[width=0.9\columnwidth]{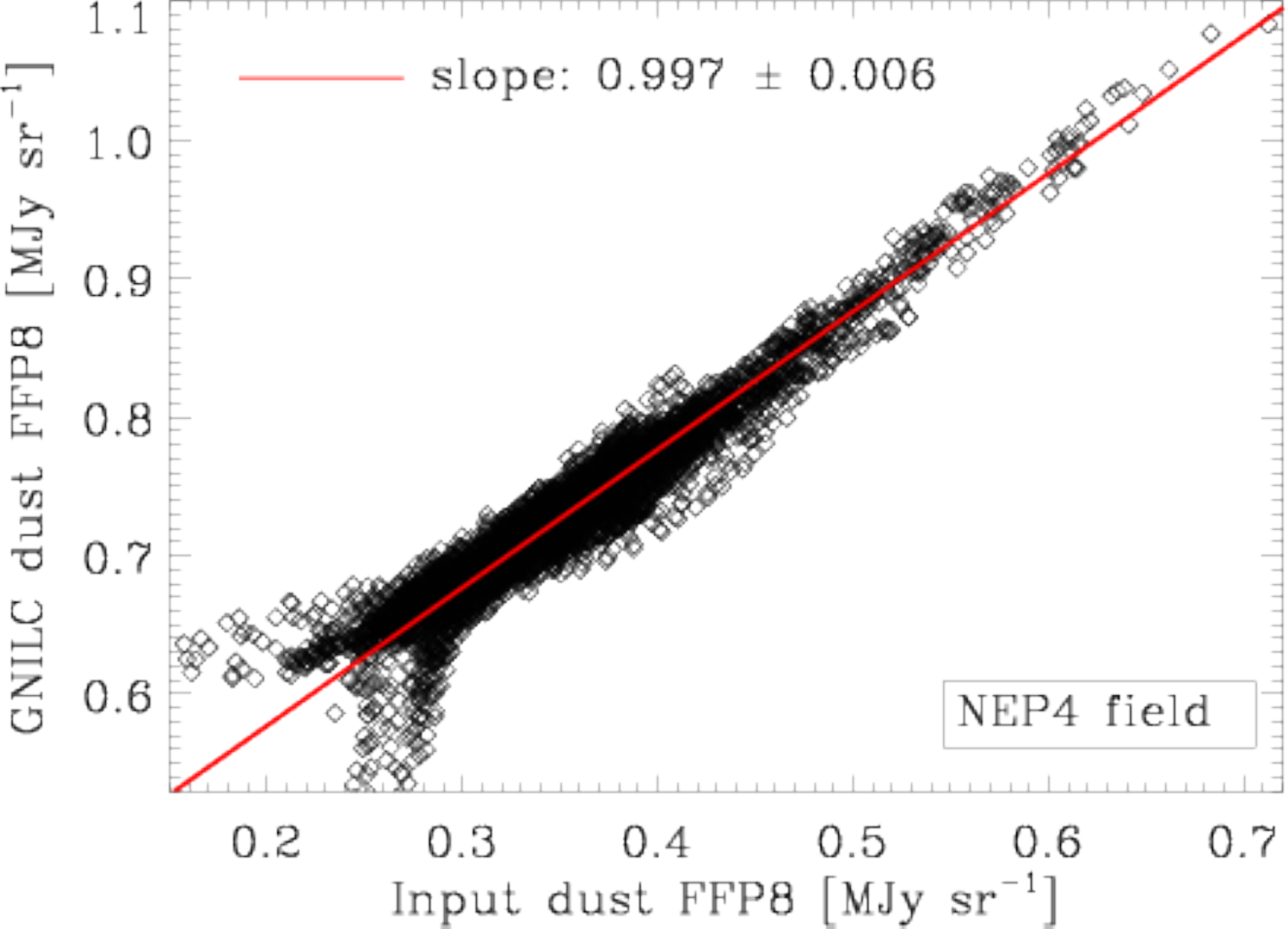}~
\includegraphics[width=0.9\columnwidth]{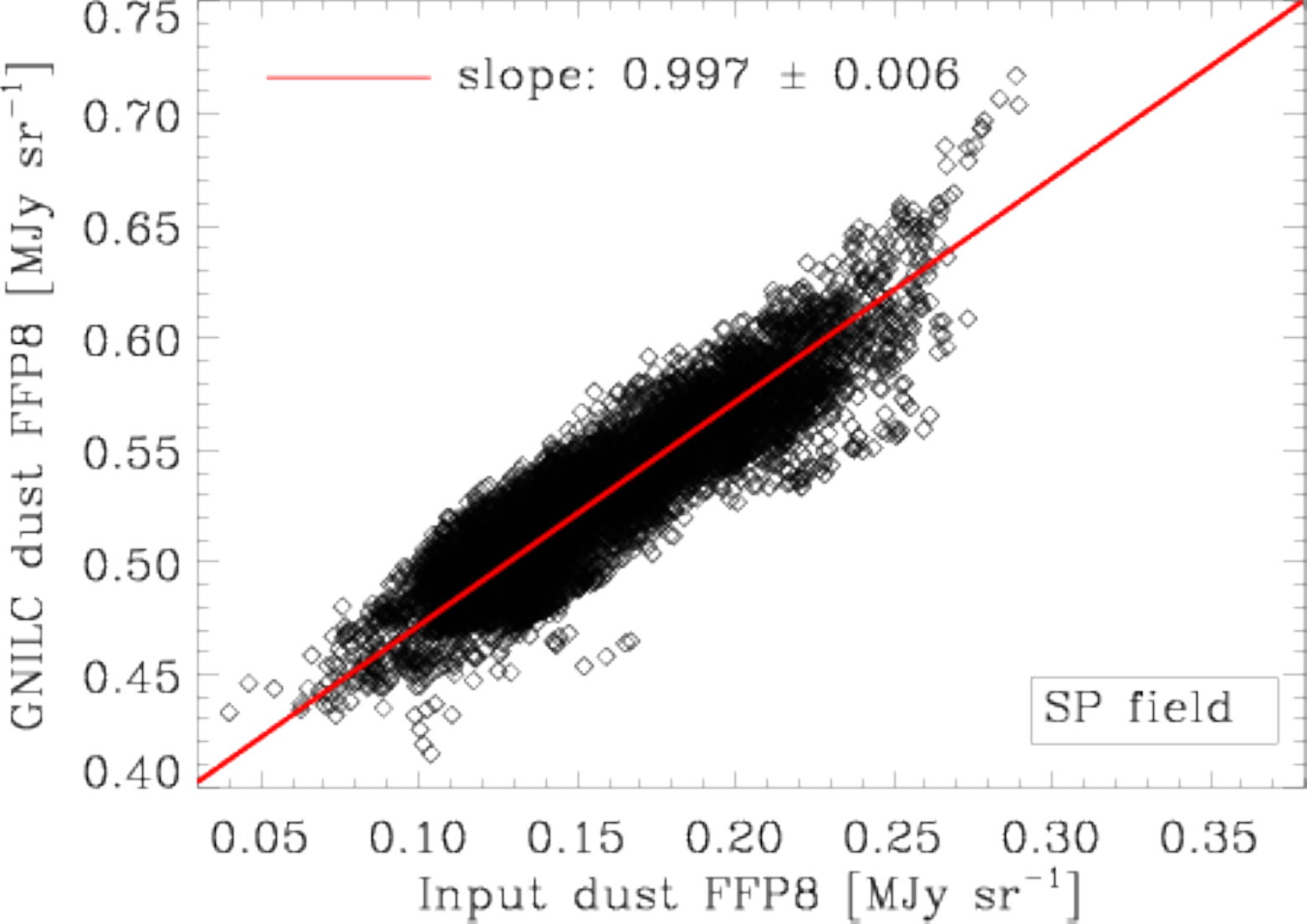}~
\end{center}
  \caption{$T$--$T$ scatter plots of the \Planck\ FFP8 simulations between {\tt GNILC} and the input thermal dust maps in the NEP4 field (\emph{left}) and the SP field (\emph{right}).}
  \label{Fig:ffp8dust_tt}
\end{figure*}

\begin{figure*}
\begin{center}
\includegraphics[width=0.9\columnwidth]{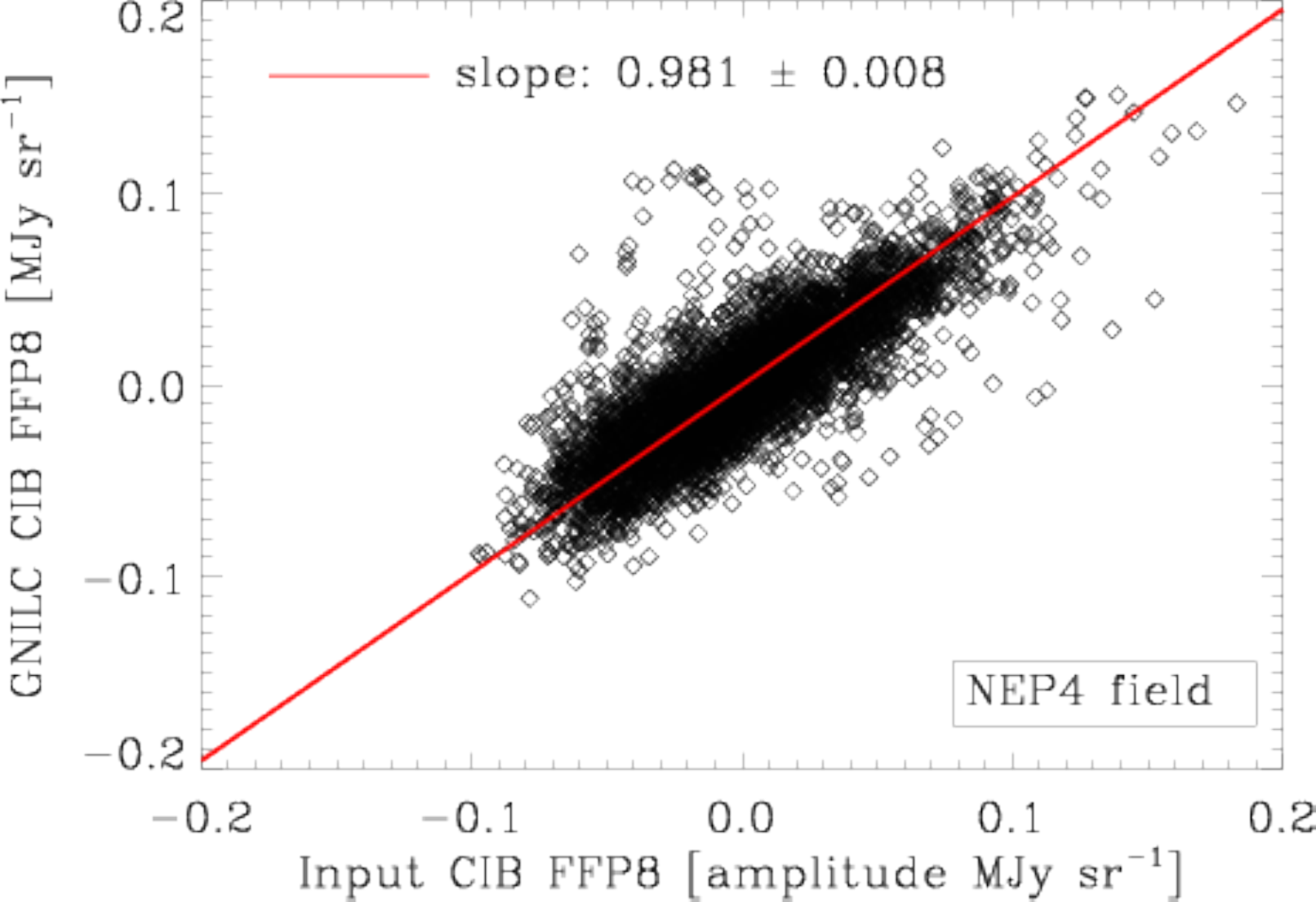}~
\includegraphics[width=0.9\columnwidth]{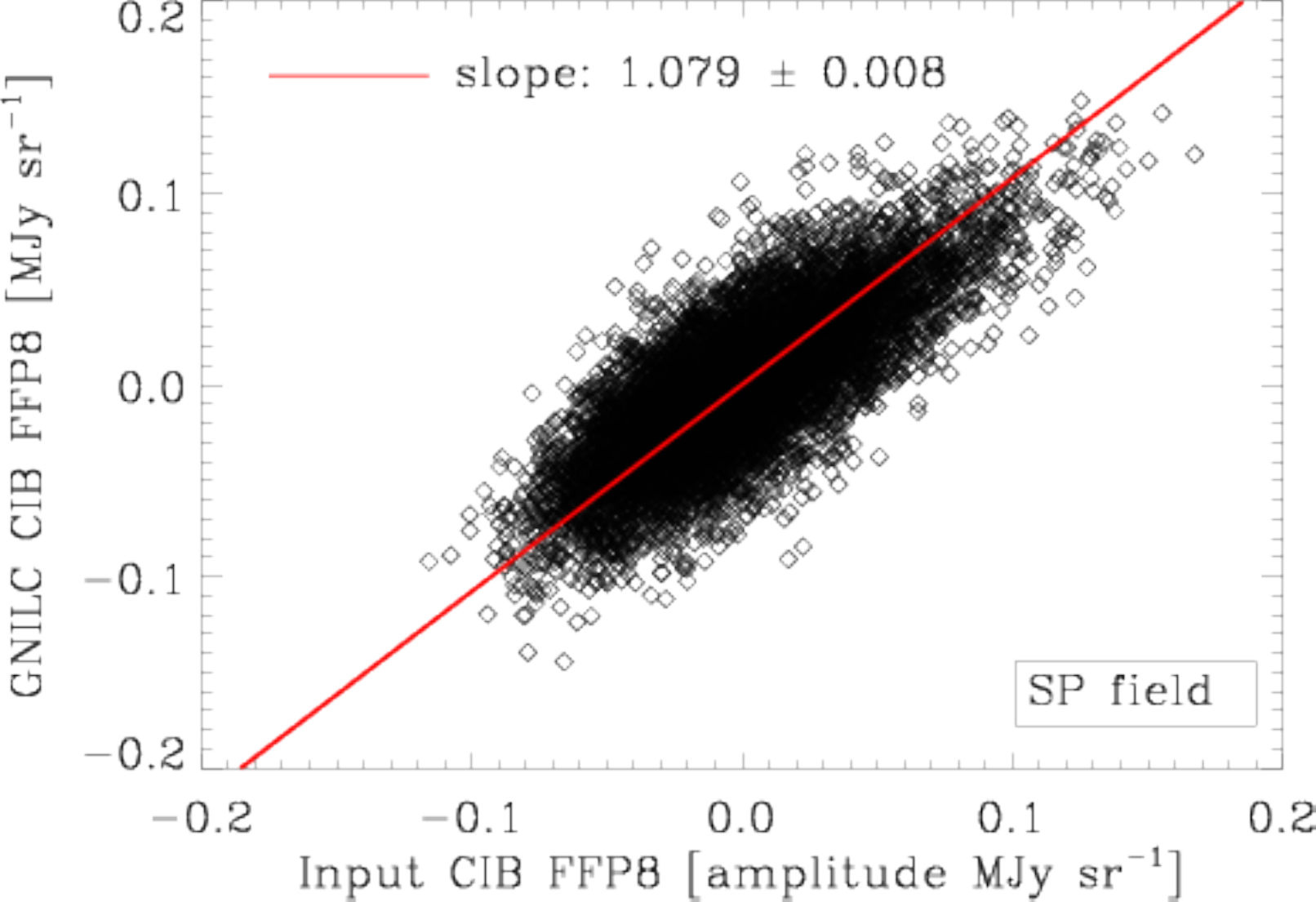}~
\end{center}
  \caption{$T$--$T$ scatter plots of the \Planck\ FFP8 simulations between {\tt GNILC} and input CIB maps in the NEP4 field (\emph{left}) and the SP field (\emph{right}).}
  \label{Fig:ffp8cib_tt}
\end{figure*}

The $T$--$T$ correlation between the {\tt GNILC} output dust and the FFP8 input dust is plotted in Fig.~\ref{Fig:ffp8dust_tt}. The $T$--$T$ correlation between the {\tt GNILC} output CIB and the FFP8 input CIB is plotted in Fig.~\ref{Fig:ffp8cib_tt}. 
In the NEP4 field, the slope of the $T$--$T$ scatter plots is $0.997\pm 0.006$ for the dust and $0.981\pm 0.008$ for the CIB. In the SP field, the slope of the $T$--$T$ scatter plots is $0.997\pm 0.006$ for the dust and $1.079\pm 0.008$ for the CIB. 
In both fields, the Pearson correlation coefficient between the input and the {\tt GNILC} reconstruction is found larger than $0.9$ for the dust and larger than $0.8$ for the CIB. 

The successful reconstruction of dust and CIB shows that the {\tt GNILC} method is robust to imperfect prior assumptions on the CIB angular power spectra. 
As shown in \citet{Olivari2016}, it is not the exact morphology of the CIB power spectrum but the overall slope that matters for enabling {\tt GNILC} to discriminate between dust and CIB. The prior power spectrum in {\tt GNILC} is only intended to estimate the dimension of the dust and CIB component subspaces (Sect. \ref{subsec:cpca}) not the amplitude of these components (Sect. \ref{subsec:ilc}).



\end{document}